\documentclass{emulateapj}
\usepackage{graphicx}
\usepackage{rotating}
\usepackage{pdfpages}
\usepackage{morefloats}
\usepackage{threeparttable}
\usepackage{longtable}
\usepackage{enumerate}
\usepackage{epstopdf}
\usepackage{natbib}
\usepackage{amsmath}
\usepackage{afterpage}

\newcommand{\zabs}{$z_{\rm abs}$}

\newcommand{\hone}{H~\textsc{i}}
\newcommand{\oone}{O~\textsc{i}}
\newcommand{\cfour}{C~\textsc{iv}}
\newcommand{\cfoureq}{{\rm C}~\textsc{iv}}
\newcommand{\ctwo}{C~\textsc{ii}}
\newcommand{\sifour}{Si~\textsc{iv}}
\newcommand{\sithree}{Si~\textsc{iii}}
\newcommand{\sitwo}{Si~\textsc{ii}}

\newcommand{\osix}{O~\textsc{vi}}
\newcommand{\lya}{Ly$\alpha$~}
\newcommand{\om}{$\Omega_{\rm C~\textsc{IV}}$}
\newcommand{\omeq}{\Omega_{\rm C~\textsc{IV}}}
\newcommand{\Dz}{$\Delta z~$}
\newcommand{\DX}{$\Delta X$}
\newcommand{\dNdX}{$d\mathcal{N}/dX~$}
\newcommand{\dNdz}{$d\mathcal{N}/dz$}
\newcommand{\limEW}{W$_{\rm lim}$}
\newcommand{\nhi}{$N(H~\textsc{i})$}
\newcommand{\ncfour}{$\mathcal{N}_{\rm abs}(N({\rm C}~\textsc{iv}))$}
\newcommand{\cfourcol}{$N({\text C}~\textsc{iv})$}
\newcommand{\logcfourcol}{log $N({\text C}~\textsc{iv})$}
\newcommand{\cfourb}{$b({\text C}~\textsc{iv})$}
\newcommand{\cfourcoleq}{N({\text C}~\textsc{iv})}
\newcommand{\ncfoureq}{\mathcal{N}_{\rm abs}(N({\rm C}~\textsc{iv}))}

\newcommand{\fn}{$f(N({\rm C}~\textsc{iv}))$}

\newcommand{\beq}{\begin{equation}}
\newcommand{\eeq}{\end{equation}}
\newcommand{\lam}{$\lambda$}
\newcommand{\dv}{$\delta v$}
\newcommand{\cmt}{{\rm cm}^{-2}}
\def\sci#1{{\; \times \; 10^{#1}}}
\begin{document}
\title{A Deep Search For Faint Galaxies Associated With Very Low-redshift C IV Absorbers: II. Program Design, Absorption-line Measurements, and Absorber Statistics}
\author{Joseph N. Burchett\altaffilmark{1}$^,$\altaffilmark{2}, Todd M. Tripp\altaffilmark{2}, J. Xavier Prochaska\altaffilmark{3}, Jessica K. Werk\altaffilmark{3}, Jason Tumlinson\altaffilmark{4}, John M. O'Meara\altaffilmark{5}, Rongmon Bordoloi\altaffilmark{4}, Neal Katz\altaffilmark{2}, and C. N. A. Willmer\altaffilmark{6}}
\altaffiltext{1}{Email: jburchet@astro.umass.edu}
\altaffiltext{2}{Department of Astronomy, University of Massachusetts, 710 North Pleasant Street, Amherst, MA 01003-9305}
\altaffiltext{3}{UCO/Lick Observatory, University of California, Santa Cruz, CA}
\altaffiltext{4}{Space Telescope Science Institute, Baltimore, MD 21218}
\altaffiltext{5}{Department of Chemistry and Physics, Saint MichaelÕs College, One Winooski Park, Colchester, VT 05439}
\altaffiltext{6}{Steward Observatory, University of Arizona, Tucson, AZ, 85721}

\begin{abstract}To investigate the evolution of metal-enriched gas over recent cosmic epochs as well as to characterize the diffuse, ionized, metal-enriched circumgalactic medium (CGM), we have conducted a blind survey for C IV absorption systems in 89 QSO sightlines observed with the Hubble Space Telescope (HST) Cosmic Origins Spectrograph (COS).  We have identified 42 absorbers at z $<$ 0.16, comprising the largest uniform blind sample size to date in this redshift range.  Our measurements indicate an increasing C IV absorber number density per comoving path length (\dNdX = $7.5 \pm 1.1$) and modestly increasing mass density relative to the critical density of the Universe (\om ~= $10.0 \pm 1.5 \times 10^{-8}$ ) from $z\sim1.5$ to the present epoch, consistent with predictions from cosmological hydrodynamical simulations.  Furthermore, the data support a functional form for the column density distribution function that deviates from a single power-law, also consistent with independent theoretical predictions.  As the data also probe heavy element ions in addition to \cfour\ at the same redshifts, we identify, measure, and search for correlations between column densities of these species where components appear aligned in velocity. Among these ion-ion correlations, we find evidence for tight correlations between \ctwo~and \sitwo, \ctwo~and \sithree, and \cfour~and \sifour, suggesting that these pairs of species arise in similar ionization conditions. However, the evidence for correlations decreases as the difference in ionization potential increases. Finally, when controlling for observational bias, we find only marginal evidence for a correlation (86.8\% likelihood) between the Doppler line width \cfourb~and column density \cfourcol.
\end{abstract}

\section{Introduction}

Within the first spectra of quasars \citep{Burbidge:1966lr,Lynds:1966fk}, astronomers detected absorption lines from intervening, enriched gas with properties distinct from the dense, neutral gas characteristic of the interstellar medium (ISM).  This gas contains neutral hydrogen column densities \nhi\ that are several orders-of-magnitude lower than galactic disks and have associated high-ion absorption (e.g., \cfour, \sifour) that is suggestive of a diffuse, ionized medium. Together, these data inspired predictions that the absorption arises in a `halo gas' that surrounds distant galaxies \citep{Bahcall:1969qy}.  Decades of subsequent research has confirmed this basic interpretation \citep[e.g.,][]{Bergeron:1987jk,Morris:1993kq,Bowen:1995fj,Lanzetta:1995rt,Bowen:1996yu,Chen:2001lr,Prochaska:2011yq}, and dedicated surveys of metal-line transitions have yielded statistical constraints on the cosmic distribution and mass density of heavy elements \citep[e.g.,][]{Sargent:1979lr,Sargent:1988sf}.

Ironically, the most rapid progress occurred first for the high-$z$ universe owing to the construction of 10 m-class, ground-based telescopes. These facilities could access redshifted far-UV transitions of, e.g., Mg~\textsc{ii} and \cfour\ in the spectra of $z>2$ quasars.   Indeed, the first such spectra recorded with Keck/HIRES revealed a remarkably high incidence of \cfour\ absorbers from gas with $\nhi \lesssim 10^{15}~\cmt$ \citep{Cowie:1995lr}, and statistical techniques indicated significant \cfour\ enrichment also for gas with $\nhi \lesssim 10^{14}~\cmt$  \citep{Cowie:1998qv,Ellison:2000wd}.  Specifically,  the cosmic incidence of \cfour\ systems,  \dNdX \citep[with $dX$ the differential absorption path length; ][]{Bahcall:1969lr}, at column densities $12 \leq {\rm log} \cfourcoleq \leq 15\ \cmt$ was $\sim14$ at $z \approx 3$ \citep{DOdorico:2010fj}, exceeding theoretical predictions \citep{Cen:2011zl} in the $\Lambda$CDM cosmology.  Also, cosmological simulations \citep{Booth:2012nr} reproducing the observed \cfour\ optical depth relative to \hone\ \citep{Schaye:2003rr} require that the gas between galaxies, the intergalactic medium (IGM), was enriched by ejecta from low-mass haloes at early times.

A simple integration of the observed \cfour\ column densities \cfourcol\ normalized by the total survey path $\Delta X$ yields the cosmic mass density in  \cfour, \om.  In principle, this quantity assesses the enrichment of intergalactic gas, and researchers have measured \om\ across cosmic time to track chemical evolution  \citep[e.g,][]{Songaila:2001qy,Cooksey:2010lr,DOdorico:2010fj}. In practice, however, \om\ may be dominated by metals from gas surrounding galaxies (the so-called circumgalactic medium or CGM) and may have little correspondence to the enrichment of the IGM. Furthermore, the \cfour\ ion may not be the dominant ionization state of C in diffuse gas at any epoch, and the  \cfour/C ratio likely evolves with redshift in a complex fashion \citep{Oppenheimer:2012qy, Cen:2011zl}.  Nevertheless, an evaluation of \om\ with cosmic time does offer a unique constraint on the enrichment history of the universe \citep{Oppenheimer:2006uq}.  Analysis at $z>2$ yields a relatively constant \om\ value from $z \sim 2-5$  \citep{Songaila:2001qy,Pettini:2003vn}, although \citet{Cooksey:2013lr} and \citet{DOdorico:2010fj} show a modest smooth decrease in \om\ with increasing $z$, and then a steep decline at higher $z$  \citep{Ryan-Weber:2009fk,Simcoe:2011rt}. The latter may suggest a decline in enrichment at early times (although \citet{DOdorico:2013fk} do not find a sharp decline at $z>5$) while the former has been interpreted as evidence for ongoing enrichment \citep{Simcoe:2011ys}.

Progress on such research at $z \ll 2$ has been stymied by technical limitations. At these redshifts, the key (far-UV) transitions for diffuse gas shift to observed wavelengths $\lambda_{\rm obs} < 3000~$\AA, requiring space-borne UV spectrometers. Furthermore, the expansion of the universe alone implies fewer detections per \AA\ of spectrum.  Indeed, a statistical survey at low-$z$ requires nearly an order-of-magnitude more sightlines than at $z \approx 3$.   This has resulted in relatively slow progress at $z<1$ despite the many years of observations with UV spectrometers on {\it HST}.

Over time, however, the well-maintained {\it HST} archive has eventually enabled such analysis.   Drawing on the GHRS and STIS high-dispersion datasets, \citet{Cooksey:2010lr} and \citet{Cooksey:2011bh} examined the incidence and mass density of  \cfour\ and \sifour, respectively, at $z<1$. Furthermore, the $z \lesssim 0.15$ regime has been studied by \citet{Tilton:2013lr} and \citet{Shull:2014zr}.  These four studies all agree that the frequency of strong \cfour\ absorption and, therefore, the related mass density have increased since early cosmic time ($z\sim5$),  suggesting that the diffuse gas surrounding and between galaxies has been continuously enriched.  However, some discrepancy remains in the very-low-$z$ regime whether the \cfour\ mass density as traced by \om\ has experienced a sudden upturn \citep{Tilton:2013lr}.

The installation of COS has lead to a resurgence of quasar spectroscopy with {\it HST} and, subsequently, a statistically powerful archival dataset covering  $\lambda \approx 1150-1700$ \AA.  We therefore recognized the potential for two major advances as regards \cfour:

\begin{enumerate}[i.]
  \item improved statistics in the present-day universe ($z<0.1$);
  \item the opportunity to examine the physical association of this enriched gas to galaxies at unprecedented levels.
\end{enumerate}

As stated above, additional motivation for conducting an absorber survey at low redshift comes from the feasible opportunity to conduct deep, high spatial resolution, multiwavelength studies of the galaxy environments near the absorbers with high {\it spectroscopic} completeness; indeed, rich publicly available data already exist.  The papers in this series present the results of our survey that combines HST/COS UV QSO spectroscopy covering the $\lambda \lambda$ 1548.2, 1550.8 \AA ~\cfour\ doublet and many other heavy element ion transitions down to $z\sim0$ with corresponding deep galaxy spectroscopy and imaging in these QSO fields.  Our survey aims for unprecedented spectroscopic completeness to faint galaxies on the order of $0.01~L*$ in the environments along the QSO lines of sight, once again enabled by the low-redshift nature of the absorber sample. The first paper in this series \citep[Paper I, ][]{Burchett:2013qy} focused on one such absorber environment.  The current paper (Paper II) presents the parent \cfour\ absorber sample and focuses on analyses of the UV absorption data, including the integrated cosmic enrichment in the most recent epoch as traced by the cosmic mass density mentioned above. Subsequent papers will present analyses of galaxy-absorber CGM relationships leveraging the galaxy survey data from public sources (Paper III) and our own ongoing ground-based observational campaign (Paper IV).

This paper is organized as follows: In Sections 2 and 3, we describe our data sources and measurements, respectively.  Section 4 presents our calculated \cfour~evolutionary statistics and discusses them in context with previous work. Section 5 examines relationships between the various metal ions measured in our QSO spectroscopy, and Section 6 focuses on the possible relationship between \cfour\ column density and Doppler line width.  We summarize our results in Section 7.  Throughout, we assume a cosmology of $H_0 = 70$ km/s Mpc$^{-1}$, $\Omega_M=0.3$, and $\Omega_{\Lambda}=0.7$.

\section{Data}

\subsection{HST/COS Spectroscopy}
Our \cfour~sample is composed of systems detected in 89 sightlines targeted by three Hubble Space Telescope (HST) programs using the Cosmic Origins Spectrograph \citep[COS;][]{Green:2012qy}: COS-Halos \citep{Tumlinson:2013cr,Werk:2012qy}, COS-Dwarfs \citep{Bordoloi:2014lr}, and the COS Absorption Survey of Baryon Harbors \citep[CASBaH, ][]{Tripp:2011wd, Meiring:2013fj}.  Information about the QSOs included in this study is presented in Table \ref{tab:QSOinfo}, and we will generally refer to them with abbreviated versions of their names in this table.  All of these data use the G130M and G160M gratings, covering the wavelength range 1100-1800 \AA, and were reduced as descrbed by \citet{Meiring:2011fj}.  Due to the differing goals of each survey, the spectra possess various signal-to-noise ratios (S/N), and the S/N varies greatly across the wavelength range of each spectrum (see Section \ref{sec:path}); the COS-Halos and COS-Dwarfs data have typical S/N values of $\sim$11 per $\sim$18 km/s resolution element, while CASBaH spectra have S/N $\sim$30 per resolution element (also $\sim$18 km/s).  The COS-Halos survey targeted 42 QSO sightlines that pass within 150 kpc of $L\sim L*$ galaxies with various stellar masses, star-formation rates, and impact parameters at redshifts $z=0.15-0.35$ that bring the  \osix~$\lambda \lambda$ 1031.7, 1037.8 \AA~ doublet and  \lya~1215.7 \AA~lines into the G130M/G160M bandpasses.  The COS-Dwarfs survey similarly targeted sightlines that pass near known galaxies with selected star-formation properties and masses, but the 43 galaxies selected were specifically $L < 0.1~L*$ galaxies.  Also, the galaxies selected for COS-Dwarfs are at a lower redshift range, $z_{gal}$ = 0.02-0.08, ideal for detecting the \cfour~doublet with COS.  Lastly, the highest S/N spectra in our dataset comes from the CASBaH survey (PI: Tripp), which targeted higher-redshift QSOs to measure transitions from high ions such as Ne~\textsc{VIII}, Mg~\textsc{X}, and Si~\textsc{XII}.  These 9 sightlines were not targeted based on preselected proximal galaxies, but deep follow-up galaxy environment data is currently being obtained around them \citep{Meiring:2011fj}.  

\begin{longtable}{lrrc} 
\tabletypesize{\scriptsize} 
\tablewidth{0pt} 
\tablecaption{QSO Sample for \cfour\ survey.} 
\tablehead{\colhead{QSO Name}  & \colhead{$\alpha$ (J2000)} & \colhead{$\delta$ (J2000)} & \colhead{$z_{\rm qso}$} \\ \ & \multicolumn{2}{c}{(degrees)} & \ }
SDSS J001224.01-102226.5 & 3.1001 & $-10.3740$ & 0.228 \\ 
SDSS J004222.29-103743.8 & 10.5929 & $-10.6288$ & 0.424 \\ 
SDSS J015530.02-085704.0 & 28.8751 & $-8.9511$ & 0.165 \\ 
SDSS J021218.32-073719.8 & 33.0764 & $-7.6222$ & 0.174 \\ 
SDSS J022614.46+001529.7 & 36.5603 & $0.2583$ & 0.615 \\ 
PHL 1337 & 38.7808 & $-4.0349$ & 1.437 \\ 
SDSS J024250.85-075914.2 & 40.7119 & $-7.9873$ & 0.377 \\ 
SDSS J025937.46+003736.3 & 44.9061 & $0.6268$ & 0.534 \\ 
SDSS J031027.82-004950.7 & 47.6159 & $-0.8308$ & 0.080 \\ 
SDSS J040148.98-054056.5 & 60.4541 & $-5.6824$ & 0.570 \\ 
FBQS 0751+2919 & 117.8013 & $29.3273$ & 0.915 \\ 
SDSS J080359.23+433258.4 & 120.9968 & $43.5496$ & 0.449 \\ 
SDSS J080908.13+461925.6 & 122.2839 & $46.3238$ & 0.657 \\ 
SDSS J082024.21+233450.4 & 125.1009 & $23.5807$ & 0.470 \\ 
SDSS J082633.51+074248.3 & 126.6396 & $7.7134$ & 0.311 \\ 
SDSS J084349.49+411741.6 & 130.9562 & $41.2949$ & 0.990 \\ 
SDSS J091029.75+101413.6 & 137.6240 & $10.2371$ & 0.463 \\ 
SDSS J091235.42+295725.4 & 138.1476 & $29.9571$ & 0.305 \\ 
SDSS J091440.38+282330.6 & 138.6683 & $28.3918$ & 0.735 \\ 
SDSS J092554.43+453544.4 & 141.4768 & $45.5957$ & 0.329 \\ 
SDSS J092554.70+400414.1 & 141.4779 & $40.0706$ & 0.471 \\ 
SDSS J092837.98+602521.0 & 142.1583 & $60.4225$ & 0.296 \\ 
SDSS J092909.79+464424.0 & 142.2908 & $46.7400$ & 0.240 \\ 
SDSS J093518.19+020415.5 & 143.8258 & $2.0710$ & 0.649 \\ 
SDSS J094331.61+053131.4 & 145.8817 & $5.5254$ & 0.564 \\ 
SDSS J094621.26+471131.3 & 146.5886 & $47.1920$ & 0.230 \\ 
SDSS J094733.21+100508.7 & 146.8884 & $10.0858$ & 0.139 \\ 
SDSS J094952.91+390203.9 & 147.4705 & $39.0344$ & 0.365 \\ 
SDSS J095000.73+483129.3 & 147.5031 & $48.5248$ & 0.589 \\ 
SDSS J095915.65+050355.1 & 149.8152 & $5.0653$ & 0.162 \\ 
SDSS J100102.55+594414.3 & 150.2606 & $59.7373$ & 0.746 \\ 
SDSS J100902.06+071343.8 & 152.2586 & $7.2289$ & 0.456 \\ 
SDSS J101622.60+470643.3 & 154.0942 & $47.1120$ & 0.822 \\ 
SDSS J102218.99+013218.8 & 155.5791 & $1.5386$ & 0.789 \\ 
PG1049-005 & 162.9643 & $-0.8549$ & 0.359 \\ 
SDSS J105945.23+144142.9 & 164.9385 & $14.6953$ & 0.631 \\ 
SDSS J105958.82+251708.8 & 164.9951 & $25.2858$ & 0.662 \\ 
SDSS J110312.93+414154.9 & 165.8039 & $41.6986$ & 0.402 \\ 
SDSS J110406.94+314111.4 & 166.0289 & $31.6865$ & 0.434 \\ 
SDSS J111239.11+353928.2 & 168.1630 & $35.6578$ & 0.636 \\ 
SDSS J111754.31+263416.6 & 169.4763 & $26.5713$ & 0.421 \\ 
SDSS J112114.22+032546.7 & 170.3092 & $3.4297$ & 0.152 \\ 
SDSS J112244.89+575543.0 & 170.6870 & $57.9286$ & 0.906 \\ 
SDSS J113327.78+032719.1 & 173.3658 & $3.4553$ & 0.525 \\ 
SDSS J113457.62+255527.9 & 173.7401 & $25.9244$ & 0.710 \\ 
PG1148+549 & 177.8353 & $54.6259$ & 0.975 \\ 
SDSS J115758.72-002220.8 & 179.4947 & $-0.3725$ & 0.260 \\ 
PG1202+281 & 181.1754 & $27.9033$ & 0.165 \\ 
SDSS J120720.99+262429.1 & 181.8375 & $26.4081$ & 0.324 \\ 
PG1206+459 & 182.2417 & $45.6765$ & 1.163 \\ 
SDSS J121037.56+315706.0 & 182.6565 & $31.9517$ & 0.389 \\ 
SDSS J121114.56+365739.5 & 182.8107 & $36.9610$ & 0.171 \\ 
SDSS J122035.10+385316.4 & 185.1463 & $38.8879$ & 0.376 \\ 
SDSS J123304.05-003134.1 & 188.2669 & $-0.5262$ & 0.471 \\ 
SDSS J123335.07+475800.4 & 188.3962 & $47.9668$ & 0.382 \\ 
SDSS J123604.02+264135.9 & 189.0168 & $26.6933$ & 0.209 \\ 
SDSS J124154.02+572107.3 & 190.4751 & $57.3520$ & 0.583 \\ 
SDSS J124511.25+335610.1 & 191.2969 & $33.9361$ & 0.711 \\ 
SDSS J132222.68+464535.2 & 200.5945 & $46.7598$ & 0.375 \\ 
SDSS J132704.13+443505.0 & 201.7672 & $44.5847$ & 0.331 \\ 
SDSS J133045.15+281321.4 & 202.6881 & $28.2226$ & 0.417 \\ 
SDSS J133053.27+311930.5 & 202.7220 & $31.3252$ & 0.242 \\ 
PG1338+416 & 205.2533 & $41.3872$ & 1.214 \\ 
SDSS J134206.56+050523.8 & 205.5274 & $5.0900$ & 0.266 \\ 
SDSS J134231.22+382903.4 & 205.6301 & $38.4843$ & 0.172 \\ 
SDSS J134246.89+184443.6 & 205.6954 & $18.7455$ & 0.383 \\ 
SDSS J134251.60-005345.3 & 205.7150 & $-0.8959$ & 0.326 \\ 
SDSS J135625.55+251523.7 & 209.1065 & $25.2566$ & 0.164 \\ 
SDSS J135712.61+170444.1 & 209.3026 & $17.0789$ & 0.150 \\ 
PG1407+265 & 212.3496 & $26.3059$ & 0.940 \\ 
SDSS J141910.20+420746.9 & 214.7925 & $42.1297$ & 0.873 \\ 
SDSS J143511.53+360437.2 & 218.7980 & $36.0770$ & 0.429 \\ 
SDSS J143726.14+504555.8 & 219.3589 & $50.7655$ & 0.783 \\ 
LBQS 1435-0134 & 219.4512 & $-1.7863$ & 1.308 \\ 
SDSS J144511.28+342825.4 & 221.2970 & $34.4737$ & 0.697 \\ 
SDSS J145108.76+270926.9 & 222.7865 & $27.1575$ & 0.064 \\ 
SDSS J151428.64+361957.9 & 228.6194 & $36.3328$ & 0.695 \\ 
SDSS J152139.66+033729.2 & 230.4153 & $3.6248$ & 0.126 \\ 
PG1522+101 & 231.1022 & $9.9748$ & 1.328 \\ 
SDSS J154553.48+093620.5 & 236.4729 & $9.6057$ & 0.665 \\ 
SDSS J155048.29+400144.9 & 237.7012 & $40.0291$ & 0.497 \\ 
SDSS J155304.92+354828.6 & 238.2705 & $35.8079$ & 0.722 \\ 
SDSS J155504.39+362848.0 & 238.7683 & $36.4800$ & 0.714 \\ 
SDSS J161649.42+415416.3 & 244.2059 & $41.9046$ & 0.440 \\ 
SDSS J161711.42+063833.4 & 244.2976 & $6.6426$ & 0.229 \\ 
SDSS J161916.54+334238.4 & 244.8189 & $33.7107$ & 0.471 \\ 
PG1630+377 & 248.0047 & $37.6306$ & 1.476 \\ 
SDSS J225738.20+134045.4 & 344.4092 & $13.6793$ & 0.594 \\ 
SDSS J234500.43-005936.0 & 356.2518 & $-0.9933$ & 0.789 \\ 
\label{tab:QSOinfo} 
\end{longtable}

\section{Measurements}

\subsection{Line Identification}
We used a multi-step visual identification process to search for \cfour~systems within our 89 spectra observed with the G130M and G160M gratings, which enable coverage of the \cfour\ doublet at $z\lesssim0.16$.  To aid in the search for the \cfour\ doublet, we created a user interface to scroll through a spectrum, select a candidate $\lambda$ 1548 feature, and view the alignment of the 1548 and 1550 features in velocity space and in their apparent column density profiles \citep{Sembach:1992rt,Savage:1991vn}.  The apparent column density (as a function of velocity) is defined as follows:
\beq
N(v) = \frac{m_e c}{\pi e^2 f \lambda} \tau(v)
\eeq
where $m_e$ is the electron mass, $c$ is the speed of light, $\lambda$ is the wavelength, and $f$ is the line oscillator strength.  $\tau(v)$ is the apparent optical depth at a given velocity and is defined as
\beq
\tau(v) = \text{ln} \frac{I_c(v)}{I(v)}
\eeq
where $I_c(v)$ is the continuum intensity and $I(v)$ is the observed intensity.  

\begin{figure}[!t]
\centering
\includegraphics[width=1.0\linewidth]{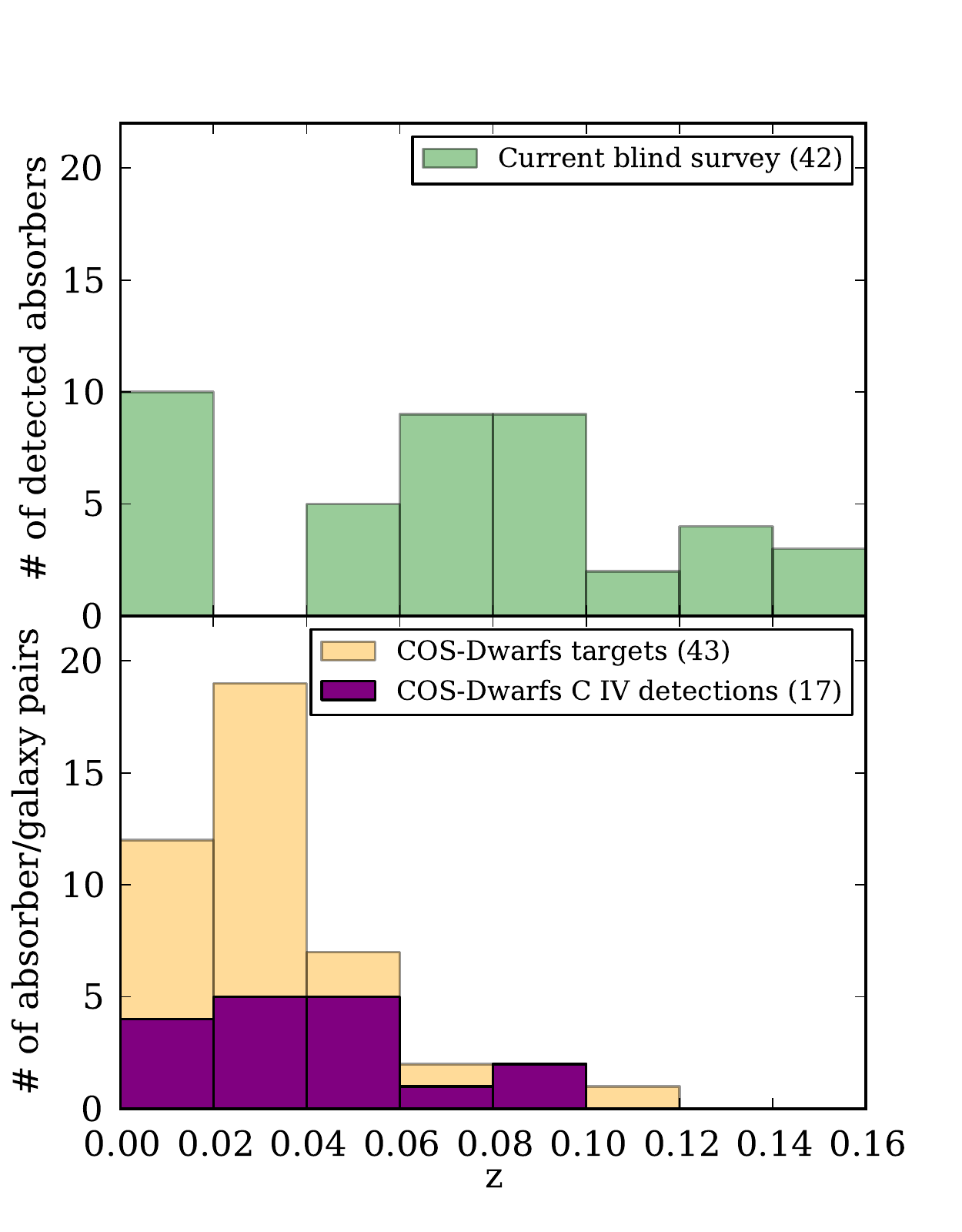}
\caption{(Top) Our blind survey sample of 42 \cfour~absorbers distributed by redshift. (Bottom) Galaxy-selected targets from the COS-Dwarfs survey \citep{Bordoloi:2014lr} and the subset for which they detected \cfour. To achieve a purely blind sample, absorbers associated with the COS-Dwarfs target galaxies were excluded from the analyses in this paper and are thus not counted in the top panel.}
\label{fig:civhist}
\end{figure}

If neither of the lines is blended with interloping absorption from another species and if the lines are not saturated, the apparent column density profiles should align.  However, if the apparent column density corresponding to the 1550 line appears greater over a portion of the profile, the 1548 line may be saturated.  If the 1548 line apparent column density appears greater than that of the 1550 line, the 1548 line may be blended with an interloper; otherwise, the candidate absorber may be rejected because $f_{1548}\lambda_{1548}/f_{1550}\lambda_{1550} \sim 2$, and a greater apparent column density for the 1548 line is unphysical.  In cases of possible blending, we accepted the candidate as a detection if the two apparent column density profiles were aligned in regions of the profile unaffected by a conspicuous interloper.

To further scrutinize the candidate systems, we searched for other common lines, such as \lya, \ctwo, and \sithree, within $\pm$400 km/s of the `systemic redshift' determined by the \cfour\ absorption.  However, we emphasize that the presence of these other lines \textit{was not a necessary criterion} for inclusion in our sample; the purpose was merely to offset some doubt initially held about the system, such as the apparent column density profiles being imprecisely aligned, which could occur merely due to known errors in the COS wavelength calibration \citep[up to $\pm$40 km/s according to][]{Wakker:2015rf}.    Then, each system was checked for significance of detection, where we required $3\sigma$ detections for both the 1548 and 1550 lines.  If a suspected \cfour\ doublet was possibly blended with an interloping line, we used Voigt profile fitting to first fit the contaminant and then assess whether enough optical depth remained in the line profile to account for the presence of \cfour.  The candidates passing these tests were then checked for common misidentifications, notably interloping Lyman series lines and higher-$z$ O~\textsc{i} 1302, Si~\textsc{ii} 1304 pairs, which have a similar wavelength separation to the \cfour~doublet.  

Once we had established our independent sample of \cfour~candidates, we flagged those systems that were at the redshifts of the galaxies targeted in the original COS-Halos and COS-Dwarfs surveys, as those targeted systems would not compose a blind sample and were thus not included in the analyses presented here.  Although none were detected, the survey design also called for omitting absorbers within 5000 km/s of the QSO redshifts, based on the \citet{Tripp:2008lr} finding of a proximate \osix\ absorber overabundance within 5000 km/s of observed QSO redshifts. Lastly, we compared our identifications with those from the COS-Halos and COS-Dwarfs studies for verification. A previous detection by those studies was not required because their analysis was primarily focused on absorption associated with their targeted galaxies.  Finally, we were left with a 42-absorber sample (see Figure~ \ref{fig:civhist}).

\subsection{Absorption Line Measurements}
For our equivalent width (EW) and column density measurements, we fit local continua within $\pm$ 500 km/s of each line center using Legendre polynomials whose order was determined by an F-test \citep{Sembach:1992rt}, typically 3rd or 4th order.  We then measured the equivalent width of each line and calculated the integrated apparent column density, assuming the line was unsaturated and, therefore, on the linear part of the curve of growth.  

We then normalized the spectrum flux by the fitted continuum and fit Voigt profiles to the absorption lines to measure column densities, Doppler parameters, and velocity centroids using our own software based on a Levenberg-Marquardt optimization algorithm. COS possesses a unique line spread function \citep{COSLSF} with large wings, and we accounted for this in the profile fitting process. In several cases, multiple absorption components were evident for a given species, and we attempted to fit the minimum number of components to account for the optical depth in each profile, including any apparent asymmetry, upon visual inspection.  Doublet components were fit simultaneously as were interloping lines where blending from other species was evident.  This blending could be observed, for example, by odd features in the (mostly) aligned apparent column density profiles of the doublet or by large asymmetries in the line profiles.  In certain cases, line profiles were partially or completely obscured by the geocoronal O~\textsc{I} $\lambda$ 1302 \AA~emission lines, preventing their measurement.  

The resulting Voigt profile fitting measurements are presented in Table 1, and plots of the absorber sample spectra along with our profile fits are presented in the Appendix. Likely due in part to the resolution of our data, certain lines yield invalid Doppler $b$-value measurements, and these these values are omitted from the table.  However, measurements of \cfourb$ < 10$ km/s may also be unreliable.  For components of species with no unsaturated lines that would yield reliable Voigt profile fitting measurements, we report the integrated apparent column density profiles as lower limits.
\clearpage
\LongTables 
\begin{deluxetable*}{lllllll} 
\tabletypesize{\scriptsize} 
\tablewidth{0pt} 
\tablecaption{Voigt profile fitting measurements by absorber component} 
\tablehead{\colhead{QSO} & \colhead{$z_{abs}$} & \colhead{Ion}  & \colhead{Lines in profile fit} & \colhead{$\rm{log }~N\tablenotemark{c}$ (cm$^{-2})$} &\colhead{$b$ (km/s)} & \colhead{$v$ (km/s)}} 
\startdata 
J0155-0857 & 0.00547 & C IV & 1548.2, 1550.8 & 13.78 $\pm$ 0.07 & 34 $\pm$ 9 & -38 $\pm$ 6 \\ 
~ & ~ & C IV & 1548.2, 1550.8 & 13.02 $\pm$ 0.32 & 16 $\pm$ 19 & 27 $\pm$ 11 \\ 
~ & ~ & H I &  & $>$14.35 &  -  &  -  \\ 
~ & 0.05416 & C IV & 1548.2, 1550.8 & 13.75 $\pm$ 0.29 & 7 $\pm$ 5 & 40 $\pm$ 2 \\ 
~ & ~ & H I &  & $>$14.15 &  -  &  -  \\ 
J0242-0759 & 0.00477 & C IV & 1548.2, 1550.8 & 13.81 $\pm$ 0.08 & 39 $\pm$ 10 & -23 $\pm$ 6 \\ 
~ & ~ & H I &  & $>$14.42 &  -  &  -  \\ 
~ & ~ & H I & 1215.7 & 13.79 $\pm$ 0.12 & 22 $\pm$ 7 & -148 $\pm$ 5 \\ 
~ & ~ & Si IV & 1393.8, 1402.8 & 13.04 $\pm$ 0.07 & 30 $\pm$ 8 & -6 $\pm$ 5 \\ 
J0925+4004 & 0.00261 & C IV & 1548.2, 1550.8 & 14.19 $\pm$ 0.07 & 34 $\pm$ 6 & 9 $\pm$ 4 \\ 
~ & ~ & H I &  & $>$14.13 &  -  &  -  \\ 
J0928+6025 & 0.01494 & C IV & 1548.2, 1550.8 & 13.77 $\pm$ 0.03 & 25 $\pm$ 3 & -23 $\pm$ 2 \\ 
~ & ~ & H I &  & $<$14.19 &  -  &  -  \\ 
J0950+4831 & 0.08049 & C IV & 1548.2, 1550.8 & 14.08 $\pm$ 0.12 & 22 $\pm$ 6 & -8 $\pm$ 4 \\ 
~ & ~ & H I & 1215.7 & 13.84 $\pm$ 0.06 & 26 $\pm$ 4 & -26 $\pm$ 3 \\ 
J1001+5944 & 0.15037 & C II & 1334.5 & 13.41 $\pm$ 0.12 & 29 $\pm$ 13 & -29 $\pm$ 8 \\ 
~ & ~ & C IV\tablenotemark{a} & 1548.2, 1550.8 & 14.25 $\pm$ 0.40 & 9 $\pm$ 3 & -16 $\pm$ 2 \\ 
~ & ~ & H I &  & $>$14.30 &  -  &  -  \\ 
~ & ~ & N V & 1238.8, 1242.8 & 13.59 $\pm$ 0.16 & 6 $\pm$ 3 & -6 $\pm$ 1 \\ 
~ & ~ & Si III & 1206.5 & 12.47 $\pm$ 0.08 & 22 $\pm$ 7 & 6 $\pm$ 4 \\ 
J1009+0713 & 0.05860 & C II & 1334.5 & 13.79 $\pm$ 0.09 & 13 $\pm$ 5 & 7 $\pm$ 3 \\ 
~ & ~ & C IV & 1548.2, 1550.8 & 14.08 $\pm$ 0.10 & 29 $\pm$ 7 & -1 $\pm$ 5 \\ 
~ & ~ & H I &  & $>$14.61 &  -  &  -  \\ 
~ & ~ & Si II & 1304.4 & 13.59 $\pm$ 0.21 & 16 $\pm$ 14 & -14 $\pm$ 8 \\ 
~ & ~ & Si III\tablenotemark{a} & 1206.5 & 13.37 $\pm$ 0.29 & 18 $\pm$ 7 & 4 $\pm$ 4 \\ 
~ & ~ & Si IV & 1393.8, 1402.8 & 13.09 $\pm$ 0.09 & 16 $\pm$ 6 & 9 $\pm$ 4 \\ 
~ & 0.11401 & C II &  & $>$14.61 &  -  &  -  \\ 
~ & ~ & C IV & 1548.2, 1550.8 & 13.57 $\pm$ 0.29 & 9 $\pm$ 10 & 37 $\pm$ 5 \\ 
~ & ~ & C IV & 1548.2, 1550.8 & 13.86 $\pm$ 0.17 & 52 $\pm$ 26 & -32 $\pm$ 18 \\ 
~ & ~ & Fe II\tablenotemark{a} & 1608.5 & 14.82 $\pm$ 0.46 & 23 $\pm$ 12 & -6 $\pm$ 7 \\ 
~ & ~ & H I\tablenotemark{a} & 1215.7 & 20.71 $\pm$ 0.01 &  -  & -23 $\pm$ 14 \\ 
~ & ~ & N I &  & $>$14.24 &  -  &  -  \\ 
~ & ~ & Si II &  & $>$14.22 &  -  &  -  \\ 
~ & ~ & Si III &  & $>$13.42 &  -  &  -  \\ 
J1059+1441 & 0.00242 & C II & 1334.5 & 13.46 $\pm$ 0.11 & 10 $\pm$ 6 & -18 $\pm$ 3 \\ 
~ & ~ & C IV & 1548.2, 1550.8 & 14.06 $\pm$ 0.04 & 20 $\pm$ 2 & -6 $\pm$ 1 \\ 
~ & ~ & H I &  & $>$14.15 &  -  &  -  \\ 
~ & 0.13291 & C IV & 1548.2, 1550.8 & 13.57 $\pm$ 0.11 & 42 $\pm$ 14 & 10 $\pm$ 10 \\ 
~ & ~ & H I &  & $>$14.93 &  -  &  -  \\ 
~ & ~ & O VI & 1031.9, 1037.6 & 14.26 $\pm$ 0.04 & 72 $\pm$ 9 & 36 $\pm$ 7 \\ 
J1059+2517 & 0.07884 & C II & 1334.5 & 13.97 $\pm$ 0.05 & 51 $\pm$ 7 & 5 $\pm$ 5 \\ 
~ & ~ & C IV & 1548.2, 1550.8 & 13.48 $\pm$ 0.22 & 9 $\pm$ 8 & -21 $\pm$ 4 \\ 
~ & ~ & C IV & 1548.2, 1550.8 & 13.87 $\pm$ 0.14 & 35 $\pm$ 16 & 32 $\pm$ 9 \\ 
~ & ~ & H I &  & $>$14.55 &  -  &  -  \\ 
~ & ~ & Si IV & 1393.8, 1402.8 & 13.38 $\pm$ 0.05 & 53 $\pm$ 8 & 11 $\pm$ 5 \\ 
~ & 0.11884 & C II & 1334.5 & 13.59 $\pm$ 0.10 & 22 $\pm$ 9 & -3 $\pm$ 5 \\ 
~ & ~ & C IV & 1548.2, 1550.8 & 14.16 $\pm$ 0.06 & 25 $\pm$ 3 & -8 $\pm$ 2 \\ 
~ & ~ & H I\tablenotemark{a} & 1215.7, 1025.7 & 14.43 $\pm$ 0.13 & 38 $\pm$ 4 & -33 $\pm$ 3 \\ 
~ & ~ & O VI & 1031.9, 1037.6 & 13.95 $\pm$ 0.11 & 37 $\pm$ 13 & -10 $\pm$ 9 \\ 
~ & ~ & Si III & 1206.5 & 12.91 $\pm$ 0.06 & 23 $\pm$ 4 & -28 $\pm$ 3 \\ 
J1104+3141 & 0.06223 & C IV & 1548.2, 1550.8 & 14.28 $\pm$ 0.09 & 18 $\pm$ 2 & -24 $\pm$ 2 \\ 
~ & ~ & H I &  & $>$14.25 &  -  &  -  \\ 
~ & ~ & H I & 1215.7 & 13.53 $\pm$ 0.09 & 36 $\pm$ 9 & -137 $\pm$ 7 \\ 
J1112+3539 & 0.01726 & C IV & 1548.2, 1550.8 & 14.05 $\pm$ 0.10 & 44 $\pm$ 11 &  -  \\ 
~ & ~ & C IV & 1548.2, 1550.8 & 13.68 $\pm$ 0.50 &  -  & -34 $\pm$ 7 \\ 
~ & ~ & H I & 1215.7 & 13.48 $\pm$ 0.27 & 10 $\pm$ 6 & -49 $\pm$ 3 \\ 
~ & ~ & H I & 1215.7 & 13.36 $\pm$ 0.13 & 16 $\pm$ 8 & 13 $\pm$ 5 \\ 
~ & ~ & H I & 1215.7 & 13.15 $\pm$ 0.18 & 35 $\pm$ 21 & 144 $\pm$ 14 \\ 
J1117+2634 & 0.04758 & C II & 1334.5 & 14.05 $\pm$ 0.08 & 20 $\pm$ 5 & -71 $\pm$ 4 \\ 
~ & ~ & C II & 1334.5 & 14.51 $\pm$ 0.19 & 15 $\pm$ 4 & -25 $\pm$ 2 \\ 
~ & ~ & C IV & 1548.2, 1550.8 & 13.91 $\pm$ 0.08 & 18 $\pm$ 5 & -65 $\pm$ 4 \\ 
~ & ~ & C IV & 1548.2, 1550.8 & 14.12 $\pm$ 0.08 & 19 $\pm$ 4 & -8 $\pm$ 3 \\ 
~ & ~ & H I &  & $>$14.52 &  -  &  -  \\ 
~ & ~ & Si II & 1193.3, 1190.4 & 13.44 $\pm$ 0.04 & 22 $\pm$ 3 & -19 $\pm$ 3 \\ 
~ & ~ & Si III & 1206.5 & 13.30 $\pm$ 0.23 & 16 $\pm$ 6 & -14 $\pm$ 6 \\ 
~ & ~ & Si III & 1206.5 & 12.99 $\pm$ 0.24 & 24 $\pm$ 11 & -57 $\pm$ 13 \\ 
~ & ~ & Si IV & 1393.8, 1402.8 & 13.43 $\pm$ 0.04 & 19 $\pm$ 2 & -16 $\pm$ 2 \\ 
J1122+5755 & 0.00640 & C IV\tablenotemark{a} & 1548.2, 1550.8 & 14.33 $\pm$ 0.07 & 24 $\pm$ 3 & 21 $\pm$ 2 \\ 
~ & ~ & H I &  & $>$14.03 &  -  &  -  \\ 
~ & ~ & H I & 1215.7 & 12.97 $\pm$ 0.22 &  -  & -77 $\pm$ 4 \\ 
~ & ~ & H I & 1215.7 & 13.86 $\pm$ 0.15 & 24 $\pm$ 6 & -162 $\pm$ 4 \\ 
~ & 0.05316 & C IV & 1548.2, 1550.8 & 13.73 $\pm$ 0.11 & 30 $\pm$ 10 & 0 $\pm$ 7 \\ 
~ & ~ & H I\tablenotemark{a} & 1215.7 & 15.32 $\pm$ 2.33 & 19 $\pm$ 12 & -10 $\pm$ 3 \\ 
~ & ~ & Si III & 1206.5 & 12.72 $\pm$ 0.10 & 16 $\pm$ 6 & 16 $\pm$ 4 \\ 
J1210+3157 & 0.05991 & C II & 1334.5 & 14.10 $\pm$ 0.06 & 19 $\pm$ 3 & -42 $\pm$ 2 \\ 
~ & ~ & C IV\tablenotemark{a} & 1548.2, 1550.8 & 14.38 $\pm$ 0.08 & 21 $\pm$ 2 & -30 $\pm$ 2 \\ 
~ & ~ & H I &  & $>$14.15 &  -  &  -  \\ 
~ & ~ & Si II & 1260.4, 1190.4 & 13.21 $\pm$ 0.07 & 43 $\pm$ 8 & -59 $\pm$ 6 \\ 
~ & ~ & Si III & 1206.5 & 13.09 $\pm$ 0.13 & 20 $\pm$ 7 & -48 $\pm$ 4 \\ 
~ & ~ & Si IV & 1402.8, 1393.8 & 13.29 $\pm$ 0.05 & 19 $\pm$ 4 & -37 $\pm$ 5 \\ 
~ & 0.07374 & C IV & 1548.2, 1550.8 & 13.75 $\pm$ 0.05 & 25 $\pm$ 4 & -10 $\pm$ 3 \\ 
~ & ~ & H I\tablenotemark{b} &-& -  &  -  &  -  \\ 
~ & 0.07818 & C IV & 1548.2, 1550.8 & 13.66 $\pm$ 0.07 & 9 $\pm$ 2 & -12 $\pm$ 2 \\ 
~ & ~ & H I & 1215.8 & 13.89 $\pm$ 0.12 & 20 $\pm$ 5 & -23 $\pm$ 3 \\ 
~ & ~ & H I & 1215.7 & 13.84 $\pm$ 0.11 & 22 $\pm$ 5 & -119 $\pm$ 4 \\ 
~ & 0.14959 & C II & 1036.3, 1334.5 & 14.19 $\pm$ 0.05 & 41 $\pm$ 6 & 10 $\pm$ 4 \\ 
~ & ~ & C IV\tablenotemark{a} & 1548.2, 1550.8 & 14.33 $\pm$ 0.12 & 37 $\pm$ 9 & -8 $\pm$ 7 \\ 
~ & ~ & H I &  & $>$14.90 &  -  &  -  \\ 
~ & ~ & O VI & 1031.9, 1037.6 & 14.65 $\pm$ 0.06 & 45 $\pm$ 6 & -16 $\pm$ 4 \\ 
~ & ~ & Si II & 1190.4, 1193.3, 1260.4 & 13.04 $\pm$ 0.04 & 40 $\pm$ 5 & -4 $\pm$ 21 \\ 
~ & ~ & Si III\tablenotemark{a} & 1206.5 & 13.68 $\pm$ 0.07 & 38 $\pm$ 3 & 4 $\pm$ 2 \\ 
~ & ~ & Si IV & 1393.8, 1402.8 & 13.49 $\pm$ 0.07 & 42 $\pm$ 10 & -2 $\pm$ 7 \\ 
J1211+3657 & 0.07777 & C IV & 1548.2, 1550.8 & 14.11 $\pm$ 0.07 & 24 $\pm$ 4 & -5 $\pm$ 2 \\ 
~ & ~ & H I &  & $>$14.16 &  -  &  -  \\ 
J1233-0031 & 0.00392 & C IV & 1548.2, 1550.8 & 13.59 $\pm$ 0.09 & 13 $\pm$ 4 & -22 $\pm$ 3 \\ 
~ & ~ & H I &  & $>$14.18 &  -  &  -  \\ 
J1241+5721 & 0.14728 & C II & 1334.5 & 13.96 $\pm$ 0.06 & 40 $\pm$ 7 & -47 $\pm$ 5 \\ 
~ & ~ & C IV\tablenotemark{a} & 1548.2, 1550.8 & 13.64 $\pm$ 0.20 & 29 $\pm$ 19 & -13 $\pm$ 12 \\ 
~ & ~ & H I &  & $>$17.86 &  -  &  -  \\ 
~ & ~ & O VI & 1031.9, 1037.6 & 14.45 $\pm$ 0.04 & 67 $\pm$ 7 & -14 $\pm$ 5 \\ 
~ & ~ & Si II & 1260.4 & 12.79 $\pm$ 0.06 & 28 $\pm$ 7 & -33 $\pm$ 4 \\ 
~ & ~ & Si III & 1206.5 & 13.21 $\pm$ 0.04 & 44 $\pm$ 5 & -45 $\pm$ 4 \\ 
~ & ~ & Si III & 1206.5 & 12.65 $\pm$ 0.10 & 33 $\pm$ 11 & 75 $\pm$ 7 \\ 
J1342+0505 & 0.13993 & C II & 1036.3, 1334.5 & 14.26 $\pm$ 0.02 & 32 $\pm$ 2 & -148 $\pm$ 1 \\ 
~ & ~ & C II & 1036.3, 1334.5 & 13.85 $\pm$ 0.04 & 30 $\pm$ 5 & 42 $\pm$ 3 \\ 
~ & ~ & C IV & 1548.2, 1550.8 & 13.85 $\pm$ 0.15 & 19 $\pm$ 8 & -177 $\pm$ 6 \\ 
~ & ~ & C IV\tablenotemark{a} & 1548.2, 1550.8 & 14.30 $\pm$ 0.08 & 65 $\pm$ 13 & -9 $\pm$ 9 \\ 
~ & ~ & H I &  & $>$14.67 &  -  &  -  \\ 
~ & ~ & O VI & 1031.9, 1037.6 & 14.60 $\pm$ 0.03 &  -  & -1 $\pm$ 6 \\ 
~ & ~ & O VI & 1031.9, 1037.6 & 14.43 $\pm$ 0.04 &  -  & -156 $\pm$ 9 \\ 
~ & ~ & Si II & 1190.4, 1193.3 & 13.41 $\pm$ 0.08 & 43 $\pm$ 8 & -164 $\pm$ 6 \\ 
~ & ~ & Si III &  & $>$13.05 &  -  &  -  \\ 
~ & ~ & Si III & 1206.5 & 13.03 $\pm$ 0.07 & 18 $\pm$ 3 & 37 $\pm$ 2 \\ 
~ & ~ & Si IV & 1393.8, 1402.8 & 13.22 $\pm$ 0.07 & 31 $\pm$ 7 & 32 $\pm$ 5 \\ 
~ & ~ & Si IV & 1393.8, 1402.8 & 13.52 $\pm$ 0.04 & 57 $\pm$ 7 & -173 $\pm$ 5 \\ 
J1342+1844 & 0.08474 & C IV & 1548.2, 1550.8 & 13.35 $\pm$ 0.05 & 24 $\pm$ 5 & 24 $\pm$ 3 \\ 
~ & ~ & H I\tablenotemark{a} & 1215.8 & 17.91 $\pm$ 0.27 & 15 $\pm$ 9 & 46 $\pm$ 16 \\ 
~ & ~ & H I\tablenotemark{a} & 1215.7 & 17.50 $\pm$ 0.50 & 22 $\pm$ 2 & 214 $\pm$ 2 \\ 
J1342-0053 & 0.07174 & C II & 1334.5 & 14.43 $\pm$ 0.11 & 18 $\pm$ 3 & -114 $\pm$ 1 \\ 
~ & ~ & C II & 1334.5 & 13.85 $\pm$ 0.28 &  -  & -39 $\pm$ 2 \\ 
~ & ~ & C II & 1334.5 & 13.87 $\pm$ 0.11 & 34 $\pm$ 9 & -11 $\pm$ 9 \\ 
~ & ~ & C IV & 1548.2, 1550.8 & 14.09 $\pm$ 0.04 & 39 $\pm$ 5 & -17 $\pm$ 3 \\ 
~ & ~ & C IV & 1548.2, 1550.8 & 13.54 $\pm$ 0.10 & 16 $\pm$ 6 & -114 $\pm$ 4 \\ 
~ & ~ & H I &  & $>$14.40 &  -  &  -  \\ 
~ & ~ & N II & 1084.0 & 14.24 $\pm$ 0.14 & 19 $\pm$ 8 & -113 $\pm$ 5 \\ 
~ & ~ & N II & 1084.0 & 13.41 $\pm$ 0.62 &  -  & -49 $\pm$ 15 \\ 
~ & ~ & N II & 1084.0 & 13.96 $\pm$ 0.20 & 25 $\pm$ 21 & 5 $\pm$ 13 \\ 
~ & ~ & O I & 1302.2 & 14.25 $\pm$ 0.08 & 13 $\pm$ 4 & -110 $\pm$ 2 \\ 
~ & ~ & Si II & 1190.4, 1193.3, 1260.4, 1304.4, 1526.7 & 13.97 $\pm$ 0.09 & 10 $\pm$ 1 & -124 $\pm$ 1 \\ 
~ & ~ & Si II & 1190.4, 1193.3, 1260.4, 1304.4, 1526.7 & 13.28 $\pm$ 0.08 &  -  & -43 $\pm$ 3 \\ 
~ & ~ & Si II & 1260.4 & 12.77 $\pm$ 0.11 &  -  & 8 $\pm$ 2 \\ 
~ & ~ & Si III & 1206.5 & 12.79 $\pm$ 0.09 & 11 $\pm$ 4 & -35 $\pm$ 2 \\ 
~ & ~ & Si III & 1206.5 & 13.29 $\pm$ 0.07 & 21 $\pm$ 2 & -113 $\pm$ 1 \\ 
~ & ~ & Si III & 1206.5 & 12.90 $\pm$ 0.15 & 9 $\pm$ 4 & 0 $\pm$ 2 \\ 
~ & 0.08795 & C IV & 1548.2, 1550.8 & 13.86 $\pm$ 0.08 & 22 $\pm$ 6 & 17 $\pm$ 4 \\ 
~ & ~ & H I &  & $>$14.29 &  -  &  -  \\ 
~ & ~ & Si III & 1206.5 & 12.86 $\pm$ 0.12 & 13 $\pm$ 5 & 6 $\pm$ 3 \\ 
J1357+1704 & 0.09779 & C II & 1334.5 & 13.99 $\pm$ 0.04 & 27 $\pm$ 3 & 25 $\pm$ 2 \\ 
~ & ~ & C IV & 1548.2, 1550.8 & 13.61 $\pm$ 0.17 & 18 $\pm$ 8 & -37 $\pm$ 5 \\ 
~ & ~ & C IV & 1548.2, 1550.8 & 13.60 $\pm$ 0.19 & 39 $\pm$ 22 & 21 $\pm$ 16 \\ 
~ & ~ & H I &  & $>$14.59 &  -  &  -  \\ 
~ & ~ & Si II & 1260.4 & 12.78 $\pm$ 0.04 & 22 $\pm$ 4 & 25 $\pm$ 2 \\ 
~ & ~ & Si III\tablenotemark{a} & 1206.5 & 13.58 $\pm$ 0.13 & 21 $\pm$ 3 & 25 $\pm$ 1 \\ 
~ & ~ & Si III & 1206.5 & 12.90 $\pm$ 0.05 & 17 $\pm$ 3 & -44 $\pm$ 2 \\ 
~ & ~ & Si IV & 1402.8, 1393.8 & 13.16 $\pm$ 0.09 & 33 $\pm$ 9 & -36 $\pm$ 6 \\ 
~ & ~ & Si IV & 1393.8, 1402.8 & 13.23 $\pm$ 0.07 & 23 $\pm$ 5 & 29 $\pm$ 4 \\ 
~ & 0.08366 & C II & 1334.5 & 13.70 $\pm$ 0.05 & 33 $\pm$ 5 & -2 $\pm$ 4 \\ 
~ & ~ & C IV & 1548.2, 1550.8 & 13.95 $\pm$ 0.07 & 20 $\pm$ 4 & 0 $\pm$ 3 \\ 
~ & ~ & C IV & 1548.2, 1550.8 & 13.60 $\pm$ 0.11 & 32 $\pm$ 12 & -71 $\pm$ 8 \\ 
~ & ~ & H I &  & $>$14.67 &  -  &  -  \\ 
~ & ~ & Si III & 1206.5 & 13.22 $\pm$ 0.17 & 15 $\pm$ 4 & -1 $\pm$ 2 \\ 
~ & ~ & Si III & 1206.5 & 12.85 $\pm$ 0.07 & 37 $\pm$ 8 & -170 $\pm$ 5 \\ 
~ & ~ & Si IV & 1402.8, 1393.8 & 13.20 $\pm$ 0.04 & 18 $\pm$ 3 & 6 $\pm$ 2 \\ 
J1437+5045 & 0.12971 & C II & 1334.5 & 14.21 $\pm$ 0.15 & 71 $\pm$ 27 & -13 $\pm$ 19 \\ 
~ & ~ & C IV &  & $>$14.62 &  -  &  -  \\ 
~ & ~ & H I &  & $>$14.46 &  -  &  -  \\ 
~ & ~ & N V\tablenotemark{a} & 1238.8, 1242.8 & 14.13 $\pm$ 0.20 & 36 $\pm$ 21 & 15 $\pm$ 16 \\ 
~ & ~ & O VI\tablenotemark{a} & 1031.9, 1037.6 & 14.61 $\pm$ 0.20 & 37 $\pm$ 14 & 10 $\pm$ 10 \\ 
~ & ~ & Si III &  & $>$13.06 &  -  &  -  \\ 
J1445+3428 & 0.00549 & C IV\tablenotemark{a} & 1548.2, 1550.8 & 14.15 $\pm$ 0.13 & 17 $\pm$ 4 & 13 $\pm$ 3 \\ 
~ & ~ & H I\tablenotemark{b} &-& -  &  -  &  -  \\ 
~ & ~ & N V & 1242.8, 1238.8 & 13.94 $\pm$ 0.09 & 29 $\pm$ 8 & 17 $\pm$ 5 \\ 
J1521+0337 & 0.09674 & C IV\tablenotemark{a} & 1548.2, 1550.8 & 14.21 $\pm$ 0.12 & 40 $\pm$ 11 & -4 $\pm$ 8 \\ 
~ & ~ & H I &  & $>$14.56 &  -  &  -  \\ 
J1553+3548 & 0.08291 & C II & 1334.5 & 14.52 $\pm$ 0.05 & 27 $\pm$ 2 & -13 $\pm$ 1 \\ 
~ & ~ & C IV & 1548.2, 1550.8 & 14.01 $\pm$ 0.07 & 41 $\pm$ 8 & -7 $\pm$ 7 \\ 
~ & ~ & Fe II & 1144.9, 1143.2 & 14.08 $\pm$ 0.15 & 9 $\pm$ 4 & -5 $\pm$ 2 \\ 
~ & ~ & H I\tablenotemark{a} & 1215.7 & 18.67 $\pm$ 0.55 & 25 $\pm$ 25 & 157 $\pm$ 68 \\ 
~ & ~ & H I\tablenotemark{a} & 1215.7 & 19.53 $\pm$ 0.10 &  -  & -42 $\pm$ 21 \\ 
~ & ~ & N II & 1084.0 & 14.28 $\pm$ 0.06 & 32 $\pm$ 6 & -8 $\pm$ 4 \\ 
~ & ~ & O I & 1302.2 & 14.72 $\pm$ 0.08 & 21 $\pm$ 3 & -13 $\pm$ 2 \\ 
~ & ~ & Si II & 1190.4, 1193.3, 1260.4, 1526.7 & 14.14 $\pm$ 0.08 & 19 $\pm$ 1 & -19 $\pm$ 3 \\ 
~ & ~ & Si III\tablenotemark{a} & 1206.5 & 13.42 $\pm$ 0.09 & 28 $\pm$ 6 & 4 $\pm$ 4 \\ 
~ & ~ & Si IV & 1393.8, 1402.8 & 13.30 $\pm$ 0.07 & 30 $\pm$ 7 & -5 $\pm$ 4 \\ 
J1619+3342 & 0.09637 & C II & 1334.5 & 14.06 $\pm$ 0.02 & 24 $\pm$ 2 & -104 $\pm$ 1 \\ 
~ & ~ & C II & 1334.5 & 14.35 $\pm$ 0.06 & 15 $\pm$ 1 & 6 $\pm$ 1 \\ 
~ & ~ & C IV & 1548.2, 1550.8 & 13.82 $\pm$ 0.05 & 15 $\pm$ 2 & -114 $\pm$ 1 \\ 
~ & ~ & C IV & 1548.2, 1550.8 & 13.70 $\pm$ 0.07 & 74 $\pm$ 16 & -7 $\pm$ 11 \\ 
~ & ~ & Fe II & 1144.9 & 13.94 $\pm$ 0.05 & 14 $\pm$ 3 & -4 $\pm$ 2 \\ 
~ & ~ & Fe III & 1122.5 & 13.82 $\pm$ 0.10 & 40 $\pm$ 13 & 5 $\pm$ 8 \\ 
~ & ~ & H I\tablenotemark{a} & 1215.7 & 20.53 $\pm$ 0.01 & 126 $\pm$ 11 & 10 $\pm$ 6 \\ 
~ & ~ & N I & 1199.5, 1200.2, 1200.7 & 13.99 $\pm$ 0.08 & 8 $\pm$ 1 & -8 $\pm$ 1 \\ 
~ & ~ & N II & 1084.0 & 15.94 $\pm$ 0.30 &  -  & -16 $\pm$ 1 \\ 
~ & ~ & O I & 1039.2, 1302.2 & 14.55 $\pm$ 0.03 & 17 $\pm$ 2 & -1 $\pm$ 1 \\ 
~ & ~ & O VI & 1037.6 & 14.80 $\pm$ 0.07 & 42 $\pm$ 8 & -154 $\pm$ 6 \\ 
~ & ~ & P II & 1152.8 & 13.03 $\pm$ 0.19 & 12 $\pm$ 12 & -7 $\pm$ 7 \\ 
~ & ~ & S II & 1250.6, 1259.5 & 14.93 $\pm$ 0.07 &  -  & -10 $\pm$ 2 \\ 
~ & ~ & Si II & 1193.3, 1260.4, 1526.7 & 14.04 $\pm$ 0.08 & 13 $\pm$ 1 & -13 $\pm$ 2 \\ 
~ & ~ & Si III & 1206.5 & 13.22 $\pm$ 0.04 & 27 $\pm$ 3 & 7 $\pm$ 2 \\ 
~ & ~ & Si IV & 1393.8, 1402.8 & 13.22 $\pm$ 0.03 & 13 $\pm$ 2 & 0 $\pm$ 1 \\ 
PG1202+281 & 0.08026 & C IV & 1548.2, 1550.8 & 13.73 $\pm$ 0.30 &  -  & 0 $\pm$ 5 \\ 
~ & ~ & H I\tablenotemark{b} &-& -  &  -  &  -  \\ 
FBQS 0751+2919 & 0.06029 & C IV & 1548.2, 1550.8 & 13.62 $\pm$ 0.03 & 15 $\pm$ 2 & -30 $\pm$ 1 \\ 
~ & ~ & C IV & 1548.2, 1550.8 & 13.21 $\pm$ 0.06 & 15 $\pm$ 4 & 17 $\pm$ 2 \\ 
~ & ~ & H I &  & $>$14.70 &  -  &  -  \\ 
LBQS 1435-0134 & 0.13849 & C IV & 1548.2, 1550.8 & 13.40 $\pm$ 0.04 & 20 $\pm$ 3 & -6 $\pm$ 2 \\ 
~ & ~ & H I & 1025.7, 1215.7 & 14.70 $\pm$ 0.02 & 27 $\pm$ 0 & 2 $\pm$ 1 \\ 
~ & ~ & O VI & 1031.9, 1037.6 & 13.79 $\pm$ 0.03 & 21 $\pm$ 3 & 9 $\pm$ 2 \\ 
PG1148+549 & 0.00349 & C IV & 1548.2, 1550.8 & 13.66 $\pm$ 0.03 & 11 $\pm$ 1 & -10 $\pm$ 1 \\ 
~ & ~ & H I &  & $>$14.17 &  -  &  -  \\ 
~ & ~ & H I & 1215.7 & 13.50 $\pm$ 0.02 & 29 $\pm$ 2 & 132 $\pm$ 1 \\ 
PG1407+265 & 0.07227 & C IV & 1550.8, 1548.2 & 13.47 $\pm$ 0.04 & 22 $\pm$ 3 & -17 $\pm$ 2 \\ 
~ & ~ & H I\tablenotemark{b} &-& -  &  -  &  -  \\ 
PG1522+101 & 0.07523 & C IV & 1548.2, 1550.8 & 13.56 $\pm$ 0.05 & 9 $\pm$ 2 & -6 $\pm$ 1 \\ 
~ & ~ & H I & 1215.7 & 13.87 $\pm$ 0.02 & 26 $\pm$ 1 & -24 $\pm$ 1 \\ 
~ & ~ & N V & 1238.8, 1242.8 & 13.19 $\pm$ 0.08 & 10 $\pm$ 4 & -4 $\pm$ 2 \\ 
\enddata 
\tablecaption{Voigt profile fitting results for our \cfour sample.} 
\tablenotetext{a}{Measurements shown with errors from Voigt profile fitting but may suffer from saturation.} 
\tablenotetext{b}{The presence of the line is apparent but either telluric emission or bad pixels prevent a measurement. } 
\tablenotetext{c}{Column densities expressed as lower limits were measured using the apparent optical method \citep{Savage:1991vn}. } 
\label{table:abslines} 
\end{deluxetable*} 

\clearpage

\begin{figure*}[!h]
\centering
\includegraphics[width=.8\paperwidth]{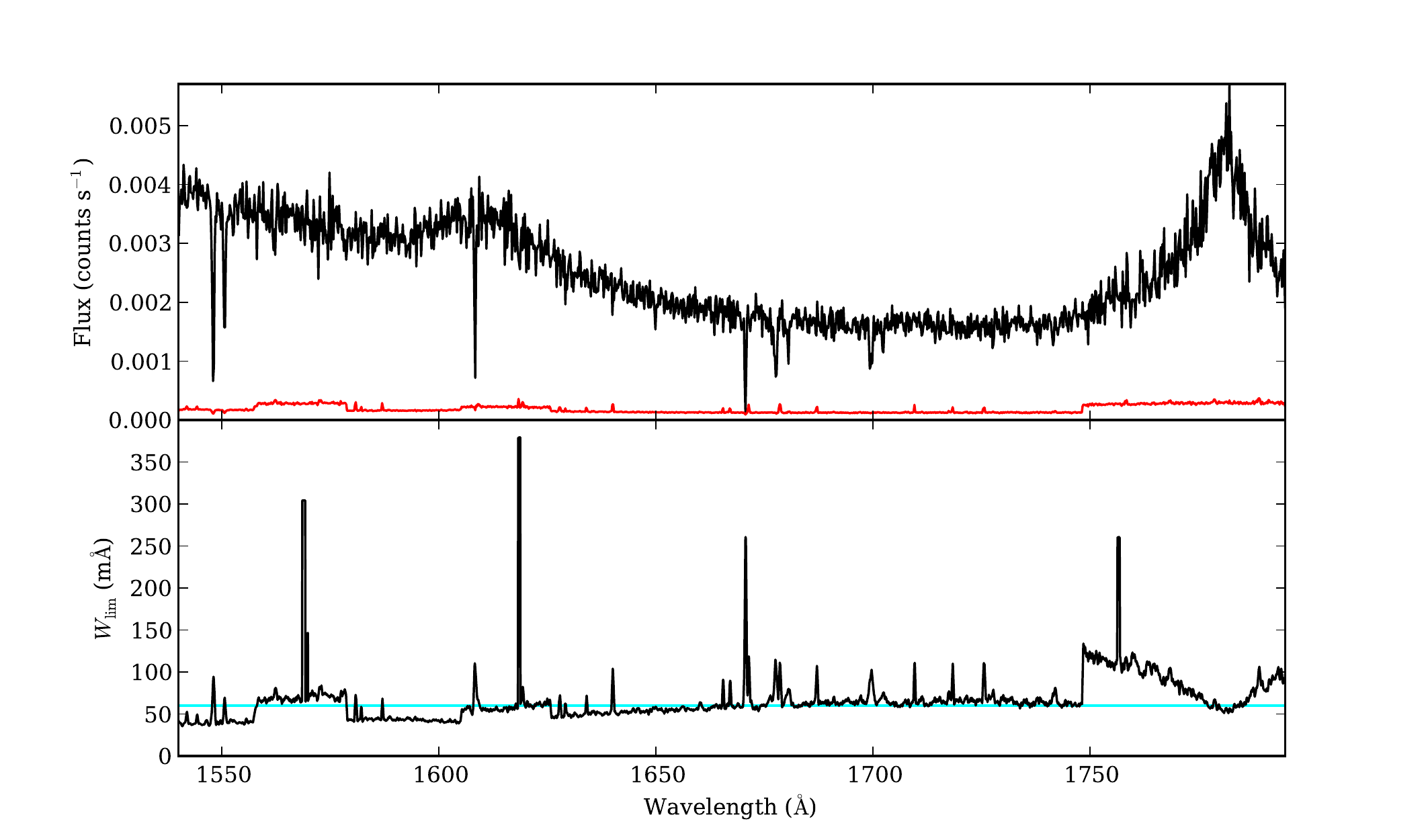}
\caption{The spectral region of one sightline, QSO J1342+1844, over which the COS bandpass covers the \cfour~doublet.  (Top) The QSO spectrum and errors are plotted in black and red, respectively. (Bottom) The limiting equivalent width for a 3-$\sigma$ detection of a given line. The cyan line marks a fiducial 60 m\AA~threshold; if chosen as a limiting equivalent width for \Dz or \DX, all regions of the spectrum with \limEW~ above this value are rejected from the \Dz and \DX\ measurement.  Spikes in the \limEW\ panel that do not coincide with absorption features in the spectra primarily occur due to pixel-scale noise fluctuations.  The discontinuous periods of increased $W_{\rm lim}$ near, e.g., 1560 \AA~correspond to detector gaps at the individual central wavelength settings of the COS G160M grating. }
\label{fig:limEW}
\end{figure*}

\subsection{Path Length Calculation}
\label{sec:path}
Our statistics calculations require measuring the total path over which we may detect the \cfour~doublet.  We express the total path in terms of two quantities: the redshift path length \Dz and the comoving path length \DX \citep{Bahcall:1969lr}.  The redshift path is simply the sum of the redshift ranges over which the doublet is detectable in each spectrum to a limiting equivalent width or column density, but this calculation must account for varying S/N across the spectrum and absorption lines from systems at other redshifts, especially strong \lya lines.  Therefore, we must ignore those segments where the S/N is insufficient to reveal lines of a given strength.  

To account for the decreasing sensitivity to lower column densities, authors using automatic line identification methods may also assume a constant path length across all column densities and employ Monte Carlo completeness corrections \citep[e.g.,][]{Simcoe:2011rt} in the derived statistics.  Our survey comprises a visually identified absorber sample, and the method presented here alternatively measures a variable path length from the data S/N; the two methods should produce consistent results.

Because a varying redshift describes a varying comoving volume, we also employ the comoving path length X($z$), which is defined as follows:
\beq
\text{X}(z) = \frac{2}{3 \Omega_M} \sqrt{\Omega_M \times (1+z)^3 + \Omega_{\Lambda}} - \frac{2 \sqrt{\Omega_{\Lambda}}}{3 \Omega_M} 
\eeq
For the \Dz and \DX\ calculation, we first convert the wavelength at each pixel to $z$ and $X$, respectively, assuming the \cfour~\lam 1550 line were to fall at each location (i.e., $z=1550.77/\lambda -1$).  We then sum over the regions in each spectrum (in terms of $z$ or $X$) where the \lam 1550 would be detectable and finally sum the detectable regions in all spectra.

A line's detectability is a function of the line strength and of the S/N of the data at the line's location.  Therefore, we calculate at each wavelength position the limiting rest equivalent width (\limEW), a threshold above which lines may be detected.  The \limEW ~is defined as follows:
\beq
\text{W}_{\rm lim}(\lambda) = \frac{3 \sigma_{W(\lambda)}}{1+z_{abs}}
\eeq
where $\sigma_{W(\lambda)}$ is the uncertainty of the observed equivalent width summed in quadrature over a number of pixels.  The number of integrated pixels is taken as a typical width (in pixels) of a line with equivalent width \limEW, between 8 and 18 pixels for the absorbers in our sample and wider than the full width at half maximum of the spectra.  The following expression defines $\sigma_{W(\lambda)}$:
\beq
\sigma_{W(\lambda)}^2 = \sum_i \left( \Delta \lambda(i) \left[\frac{\sigma_{I(\lambda_i)}}{I(\lambda_i)} \right] \right)^2
\eeq
where $\Delta \lambda(i)$ is the pixel width (in \AA), $I(\lambda_i)$ is the continuum flux at pixel $i$, and $\sigma_{I(\lambda_i)}$ is the flux uncertainty at pixel $i$. We chose the number of integrated pixels based on typical widths of lines from our data with measured equivalent widths similar to the threshold \limEW, e.g., the average profile of an $\sim$80 m\AA\ line was 12 pixels wide.   An example \limEW~calculation for the spectrum of J1342+1844 is shown in Figure~\ref{fig:limEW}.

\begin{table}[!t] 
\caption{Limiting equivalent widths and path lengths calculated as functions of column density.} 
\begin{center} 
\begin{threeparttable} 
\begin{tabular}{ccccc} 
\toprule 
log N(C~\textsc{iv}) [cm$^{-2}$] & W$_{\rm 1550}$\tnote{a} \ [m\AA] & \Dz  & \DX  & \dNdz\tnote{b} \\ 
\hline 
13.0 & 19 & 0.1 & 0.1 & - \\[4pt]  
13.1 & 24 & 0.5 & 0.5 & - \\[4pt]  
13.2 & 29 & 0.8 & 0.8 & - \\[4pt]  
13.3 & 35 & 1.1 & 1.3 & 7.80$^{+2.4}_{-1.4}$ \\[4pt]  
13.4 & 43 & 1.7 & 1.9 & 6.79$^{+2.1}_{-1.2}$ \\[4pt]  
13.5 & 54 & 1.8 & 2.0 & 6.79$^{+2.1}_{-1.2}$ \\[4pt]  
13.6 & 67 & 2.8 & 3.1 & 5.51$^{+1.6}_{-1.0}$ \\[4pt]  
13.7 & 83 & 4.0 & 4.4 & 4.59$^{+1.4}_{-0.8}$ \\[4pt]  
13.8 & 102 & 4.3 & 4.7 & 3.22$^{+1.1}_{-0.6}$ \\[4pt]  
13.9 & 124 & 5.5 & 6.1 & 2.59$^{+0.9}_{-0.5}$ \\[4pt]  
14.0 & 150 & 7.1 & 7.9 & 2.43$^{+0.8}_{-0.5}$ \\[4pt]  
14.1 & 178 & 8.7 & 9.6 & 1.56$^{+0.7}_{-0.4}$ \\[4pt]  
14.2 & 208 & 9.9 & 11.1 & 0.91$^{+0.6}_{-0.3}$ \\[4pt]  
14.3 & 241 & 10.9 & 12.1 & 0.52$^{+0.5}_{-0.2}$ \\[4pt]  
14.4 & 275 & 11.6 & 12.9 & 0.17$^{+0.4}_{-0.1}$ \\[4pt]  
14.5 & 310 & 12.0 & 13.4 & 0.08$^{+0.3}_{-0.1}$ \\[4pt]  
14.6 & 345 & 12.3 & 13.8 & 0.08$^{+0.3}_{-0.1}$ \\[4pt]  
14.7 & 379 & 12.5 & 14.0 & - \\[4pt]  
14.8 & 412 & 12.7 & 14.2 & - \\[4pt]  
14.9 & 442 & 12.8 & 14.3 & - \\[4pt]  
15.0 & 468 & 12.9 & 14.4 & - \\ 
\hline 
\end{tabular} 
\begin{tablenotes} 
\item[a]{\scriptsize Equivalent width corresponding to \cfourcol\ in Column 1} 
\item[b]{\scriptsize Cumulative number of detected absorbers per redshift path length (see Eq. \ref{eq:dndz})} 
\end{tablenotes} 
\end{threeparttable} 
\end{center} 
\label{tab:EWCD} 
\end{table}

\begin{figure}[!t]
\centering
\includegraphics[width=1.05\linewidth]{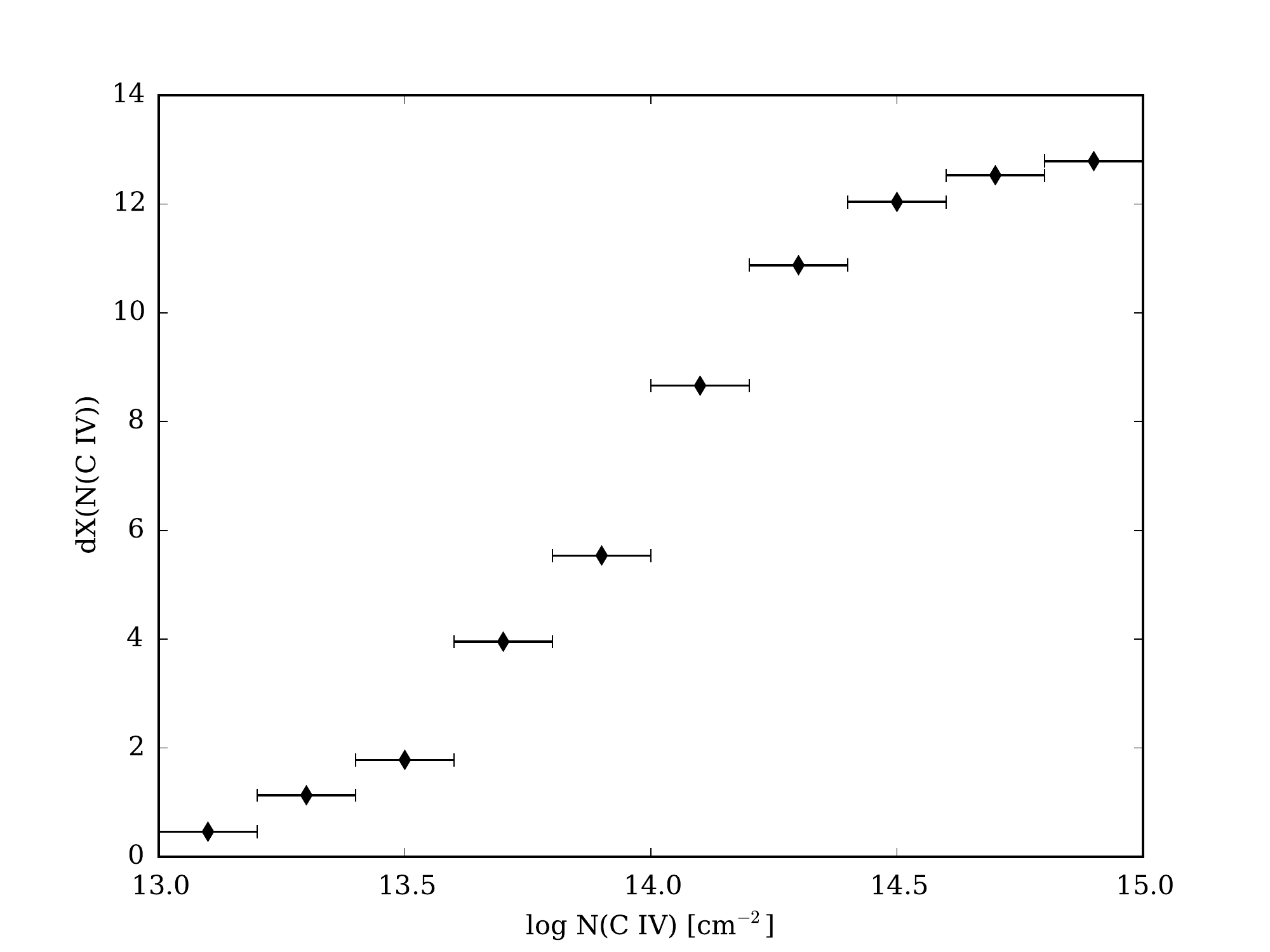}
\caption{The path length as measured from our QSO spectra as a function of column density in the same bins used to calculate the observed column density distribution function, f(N).}
\label{fig:dX}
\end{figure}

The \limEW~was measured using these equations centered at each pixel of every spectrum in our dataset in multiple passes using the number of pixels as described above for \limEW\ values commensurate with the column density bins used to compute the column density distribution function (see Section 4.1).  The absorber column density range we are probing ($13 \lesssim {\rm log}~\cfourcoleq \lesssim 15~\cmt$) falls in the region of the curve of growth where the lines of the doublet begin to saturate.  Therefore, we mapped equivalent width to column density by creating 1000 synthetic \cfour\ doublets with S/N $\sim$ 12 (a typical S/N for our QSO spectra), measuring the resulting equivalent width, and fitting the resulting $W_r$-\logcfourcol~relation from the simulated data with a 4th-order polynomial.  We then measured the path length at each $13 \lesssim {\rm log}~\cfourcoleq \lesssim 15~\cmt$ at intervals of $\Delta$(\logcfourcol) = 0.05 $\cmt$.  In accordance with our blind survey criteria, we did not include regions of the spectra within $\pm$ 500 km/s of the COS-Halos and COS-Dwarfs target galaxy redshifts (the CASBaH survey is intrinsically blind) or  $\pm$ 5000 km/s of the QSO redshift in our path lengths. The path lengths measured for our dataset and employed for the following statistics calculations are shown in Figure \ref{fig:dX} and tabulated in Table~\ref{tab:EWCD}; the limiting equivalent width values corresponding to these column density bins are listed in Table~\ref{tab:EWCD}.  We also list in Table~\ref{tab:EWCD} the cumulative number of absorbers per unit redshift path length as a function of \cfourcol, \dNdz, defined as follows:

\beq
\frac{d\mathcal{N}}{dz}(\cfourcoleq) = \sum_i \frac{\mathcal{N}(N_i(\cfoureq))}{\Delta z(N_i(\cfoureq))}
\label{eq:dndz}
\eeq

\noindent where $\mathcal{N}(N_i(\cfoureq))$ is the number of \cfour\ absorbers in the $i$-th column density bin, $\Delta z(N_i(\cfoureq))$ is the redshift path length for detecting absorbers with $N_i(\cfoureq)$, and the summation is carried out for column density bins with $N_i(\cfoureq) \geq \cfourcoleq$.

\section{Absorber Statistics}

\subsection{Column density distribution function $f(N)$}
We begin our analysis by deriving the column density distribution function, or $f(N)$, of the \cfour\ absorber sample.  A fundamental observable measured for a sample of absorption systems, $f(N)$ is useful for comparisons to theoretical predictions and describing the incidence and mass density of an absorber sample.  One may evaluate
$f(N)$ in discrete column density bins $\Delta N(\cfoureq)$ as follows:
\beq
f(N) = \frac{\ncfoureq}{\Delta \cfourcoleq ~\Delta X(\cfourcoleq)}
\label{eq:fn}
\eeq
where $\Delta X(N(C~\textsc{IV}))$ is the comoving path length corresponding to the threshold set by the individual column density bin, and \ncfour\ is the number of absorbers within the specified column density bin.

\begin{figure}[!t]
\centering
\includegraphics[width=1.0\linewidth]{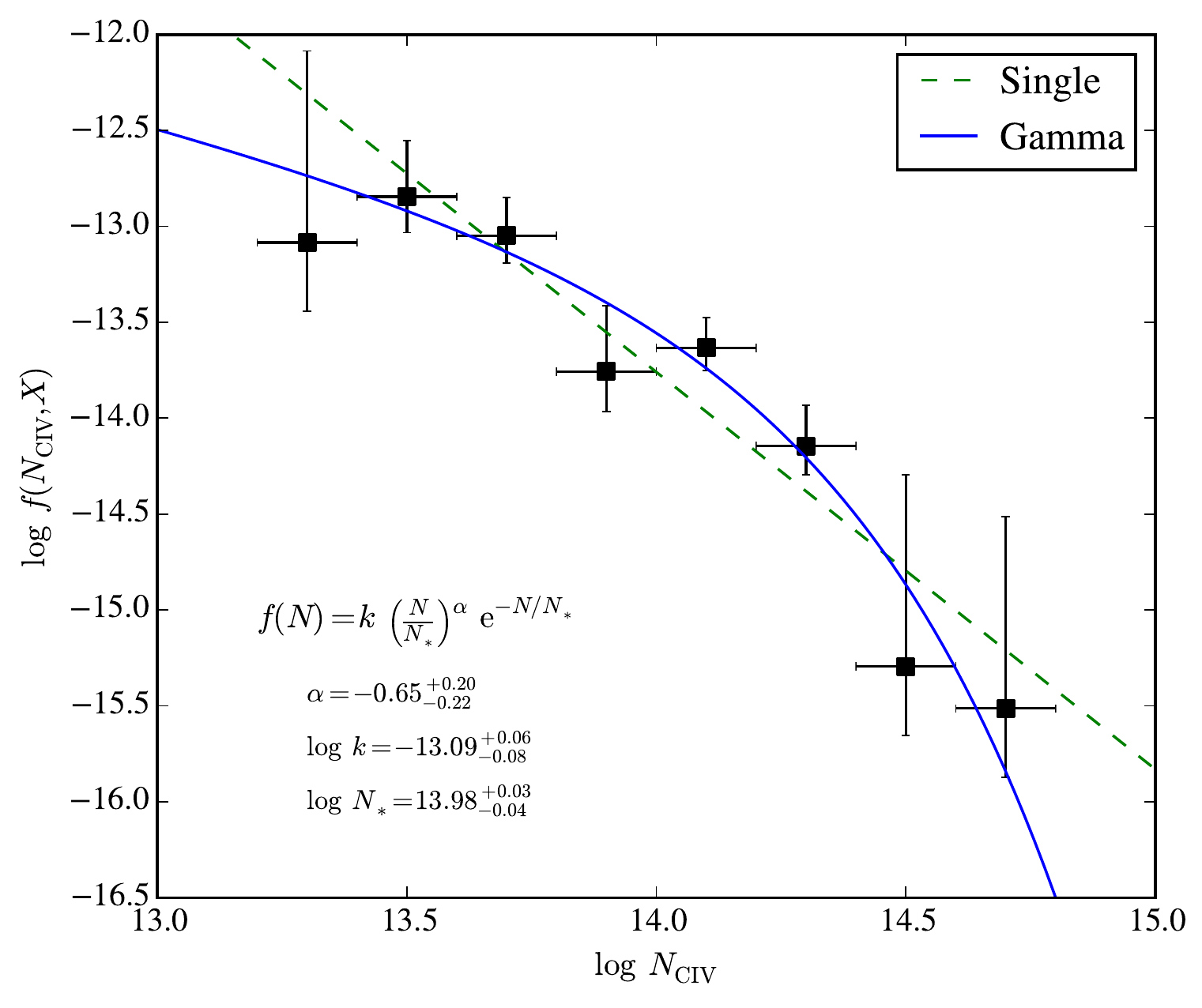}
\caption{The column density distribution function for our 42-absorber \cfour\ sample.  Data points indicate binned evaluations with vertical error bars corresponding to 1-$\sigma$ Poisson errors of the absorber counts in each bin, and the horizontal bars denote the column density bin width, $\Delta$(\logcfourcol) = 0.2.  The overplotted models assume a single power law (green, dashed) and $\Gamma$ function (blue, solid) distributions for \fn.  The latter model offers an excellent description of the observations; the best-fit parameters are labeled in the figure.}
\label{fig:fn}
\end{figure}

Our binned evaluation of
\fn~is presented in Figure~\ref{fig:fn}, assuming 
Poisson uncertainties in \ncfour\ for error estimation.
Similar to previous work, we observe a steep decline in $f(N)$
with increasing \cfourcol.
Earlier studies have frequently modeled $f(N)$ as a single power-law,

\beq
f(N) = k_P  \left( \frac{\cfourcoleq}{10^{14} \, {\rm cm^{-2}}} \right)^{\alpha_P}
\label{eq:fnpower}
\eeq
where $\alpha_P$ is the power law index, and $k_P$ is the normalization constant.
For small samples, this functional form has offered a good description 
for the limited data.
Following \citet{Cooksey:2010lr}, we used a maximum likelihood analysis
to find the best-fit parameters for this model:
$\log k_P = -13.76 \pm 0.07$ and $\alpha_P = -2.07 \pm 0.15$ 
(68\% c.l.).  
For this likelihood analysis, we adopted a saturation limit
$N_{\rm sat} = 10^{14.3} \, \rm cm^{-2}$ and have analyzed the dataset
from $N_{\rm min} = 10^{13.3} \, \rm cm^{-2}$ to
$N_{\rm max} = 10^{15} \, \rm cm^{-2}$.  This model is overplotted on Figure~\ref{fig:fn} and offers a fair
description of the data.

\begin{figure}[!t]
\centering
\includegraphics[width=1.0\linewidth]{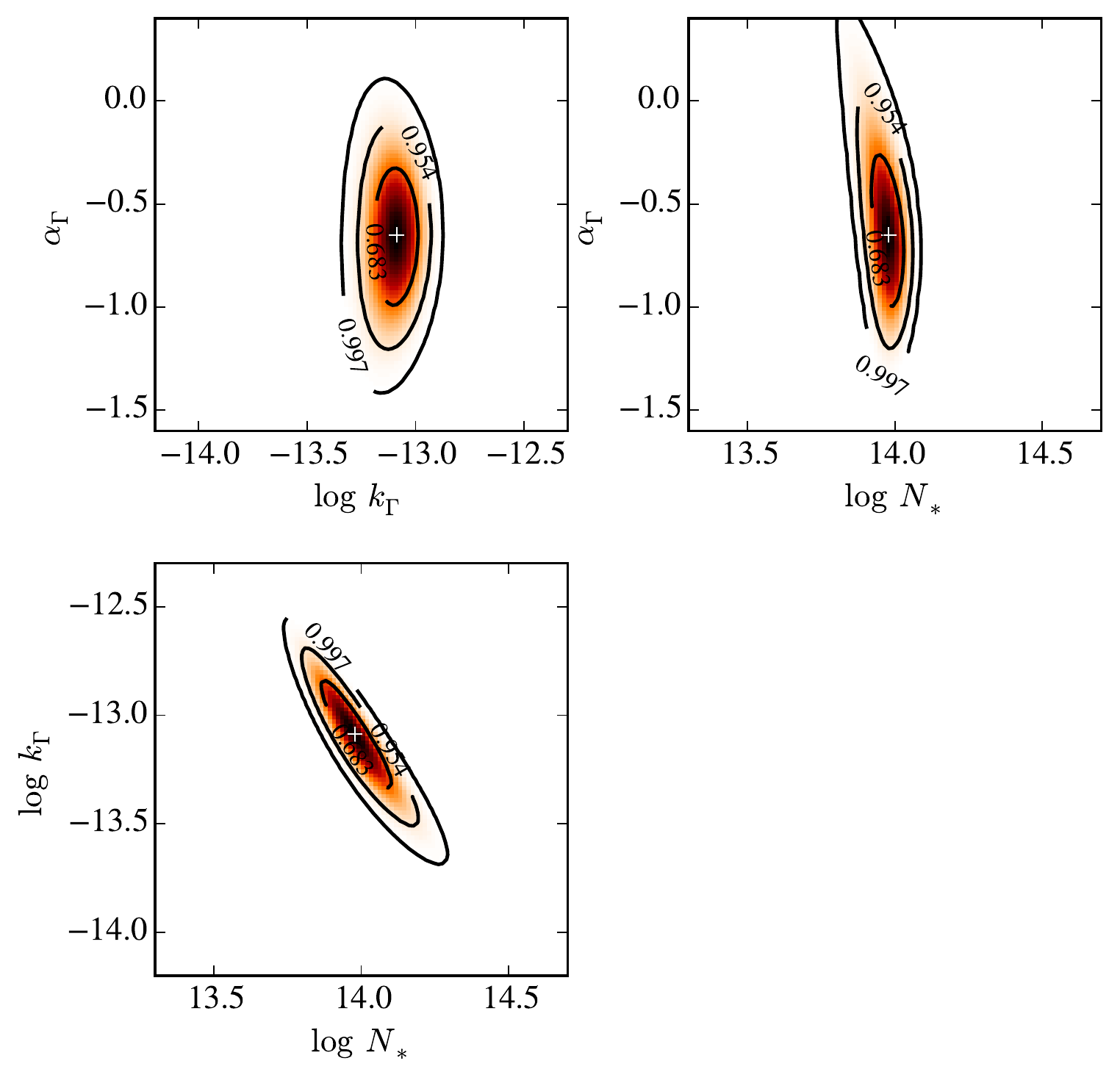}
\caption{
Correlation contours for the three parameters describing the 
$\Gamma$-function model for $f(N)$ at 68.4, 95.4, and 99.7\% confidence (inner to outer).  As expected, $k_\Gamma$
and $N_*$ are highly correlated.  The power-law exponent
is more poorly constrained but indicates a shallower slope
than $\alpha_\Gamma = -1.5$ at $99.7\%$ c.l.
}
\label{fig:corr}
\end{figure}

However, \citet{Cooksey:2013lr} found a steep
(approximately exponential) turn-over in the equivalent width
distribution of strong \cfour\ systems.  Furthermore, \citet{Danforth:2014zr}, whose sample contains more $13.0 \leq {\rm log}~\cfourcoleq \leq 13.5$ absorbers than that presented here, argue that a broken power law better fits their measured \fn\  than a single power law.  We were thus inspired to consider an alternative model, specifically, fitting \fn\ with a $\Gamma$-function

\beq
f(N) = k_\Gamma \left ( \frac{\cfourcoleq}{N_*} \right )^{\alpha_\Gamma}
  \, \exp \left [ \frac{-\cfourcoleq}{N_*} \right ] \;\;\; ,
\label{eq:fngamma}
\eeq
parameterized by a normalization constant\footnote{
	Note that this normalization constant is at least partially 
	degenerate with $N_*^{-\alpha_\Gamma}$.
} $k_\Gamma$, a power-law exponent $\alpha_\Gamma$, and a `break'
column density $N_*$.  
Once again, we perform a maximum likelihood analysis
with $N_{\rm min}$, $N_{\rm max}$, and $N_{\rm sat}$ as above on
a 3-dimenstional grid of $k_\Gamma$, $\alpha_\Gamma$, and $N_*$ values.
The best-fit $\Gamma$-function is overplotted on the binned evaluations
in Figure~\ref{fig:fn} and provides an excellent description of
the observations. 
The correlation in these parameters is illustrated in 
Figure~\ref{fig:corr}; as expected, we find significant degeneracy
between $k_\Gamma$ and $N_*$ although the latter is rather 
tightly constrained.  
The power-law exponent is shallow and constrained at the
$99.7\%$~c.l. to be greater than $\alpha_\Gamma = -1.5$.

\begin{figure}[!t]
\centering
\includegraphics[width=1.05\linewidth]{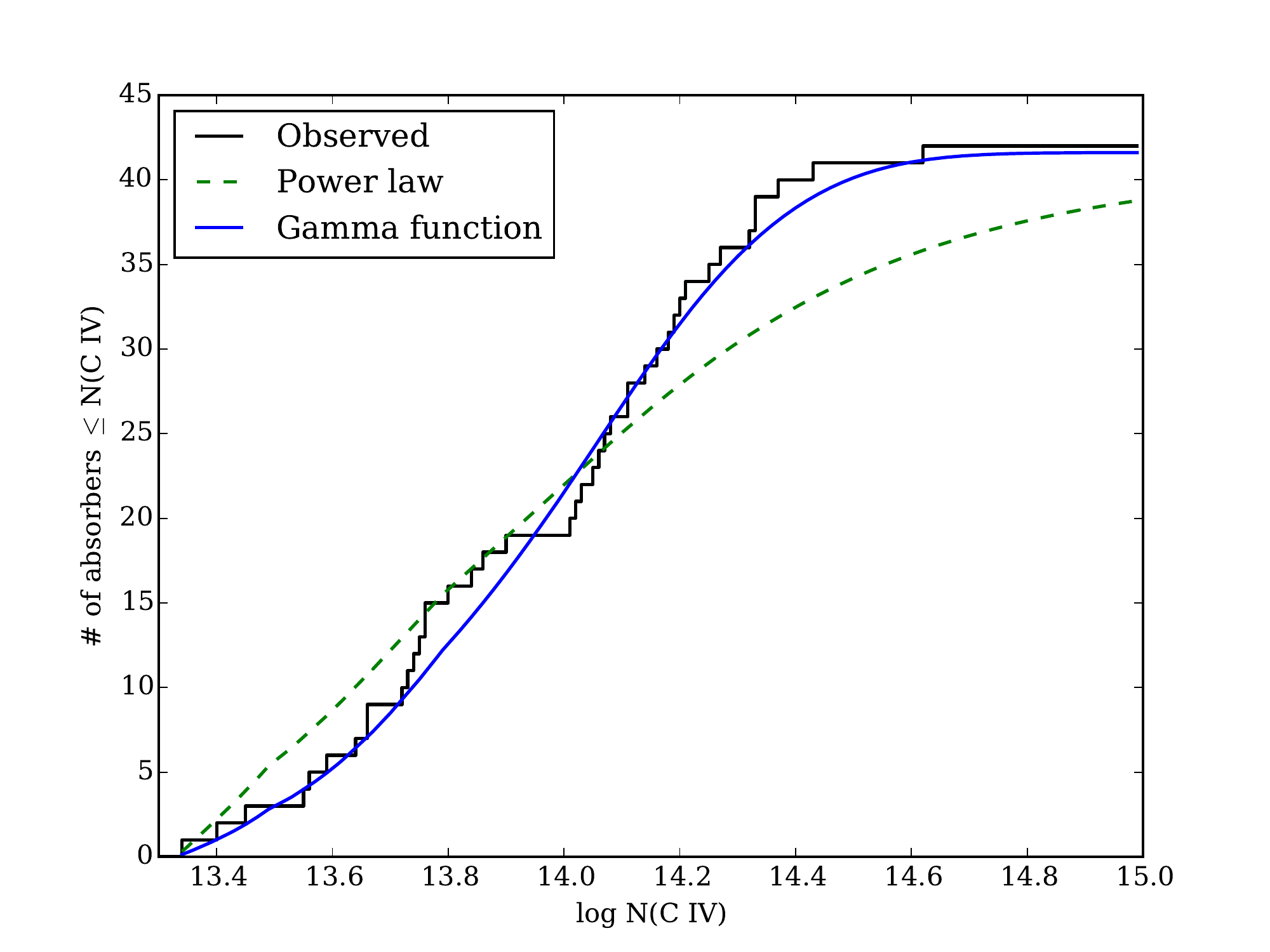}
\caption{
Cumulative distribution as a function of \cfourcol\ of our \cfour\ absorber sample compared with those resulting from the best-fit $\Gamma$- and power law column density distribution functions.  Note that the $\Gamma$ functional form of \fn\ closely reproduces our absorber sample cumulative distribution, while the power-law distribution appears to deviate from the observed distribution at $14.1 < {\rm log}~\cfourcoleq < 14.7$.}
\label{fig:cumabs}
\end{figure}

The best-fit $\Gamma$-function is overplotted on the binned evaluations
in Figure~\ref{fig:fn} and provides an excellent description of
the observed \fn.  To statistically assess the goodness-of-fit between these models, we conducted one-sample Anderson-Darling comparison tests between our absorber sample and each model.  The traditional, but often-employed, implementation of the Kolmogorov-Smirnoff (K-S) and/or Anderson-Darling tests, wherein the so-called `$D$ statistic' follows the K-S distribution (the integrals of which yield p-values for the null hypothesis) do not apply in this situation because the parameters of the models to be tested are derived from the sample distribution to be compared.  Therefore, we produced distributions of the $D$ statistics calculated between the normalized empirical cumulative distribution function of $10^5$ bootstrap resamples of our \cfour\ absorber sample and the continuous cumulative distribution function of each model.  Potentially due to the small numbers of absorbers we have at the lowest and highest column densities, the Anderson-Darling test does not yield a small enough p-value to reject the power-law model for \fn\ with strong statistical confidence.  However, we call attention to Figure \ref{fig:cumabs}, which shows the cumulative distribution of our \cfour\ absorber sample as a function of \cfourcol\ alongside the cumulative distributions following from the $\Gamma$-function and power-law models.  Due to its superior reproduction of our absorber sample, we adopt the $\Gamma$ functional form for \fn\ in the following statistics calculations.  

From any $f(N)$ model, it is trivial to integrate from any
$N_{\rm min}$ value to estimate the incidence of absorption
systems 

\beq
d\mathcal{N}/dX = \int_{N_{\rm min}}^\infty f(N,X) dN
\label{eq:dNdX}
\eeq

For $N_{\rm min} = 10^{13.3} \, \rm cm^{-2}$, we calculate
$dN/dX=7.5$ for our preferred model.  Evaluating the likelihood
function of this model to a $68\%$ confidence limit, we find the
RMS in the resultant $dN/dX$ values to be 1.1 ($\approx 15\%$)
which is consistent with the Poisson uncertainty of an $N=42$
sample.

\subsection{$\Omega_{C~\textsc{IV}}$}
The mass density of triply ionized carbon is quantified by 
the \om~statistic, which is the ratio of the mass density of \cfour\ 
to the critical density of the Universe.  
Formally, \om~may be calculated from $f(N)$ as follows:

\beq
\omeq = \frac{H_0 m_C}{c \rho_{c,0}} \int_{0}^{\infty} 
f(N(\cfour)) N(\cfour) \, dN(\cfour) \;\;\; ,
\eeq

\noindent where $H_0$ is the Hubble constant, $\rho_{c,0} = 3 H_0^2(8\pi G)^{-1}$ is the critical density, $m_C$ is the mass of the carbon atom, and the other symbols have their typical meanings.  
In practice, \om\ is typically estimated over a finite column density
interval $\cfourcoleq = [N_{\rm min},N_{\rm max}]$.
This is required for simple power-law models which will diverge at 
either high or low \cfourcol.  Furthermore, $f(N)$
is generally only constrained over a finite \cfourcol\ interval.

\begin{figure}[!t]
\centering
\includegraphics[width=1.0\linewidth]{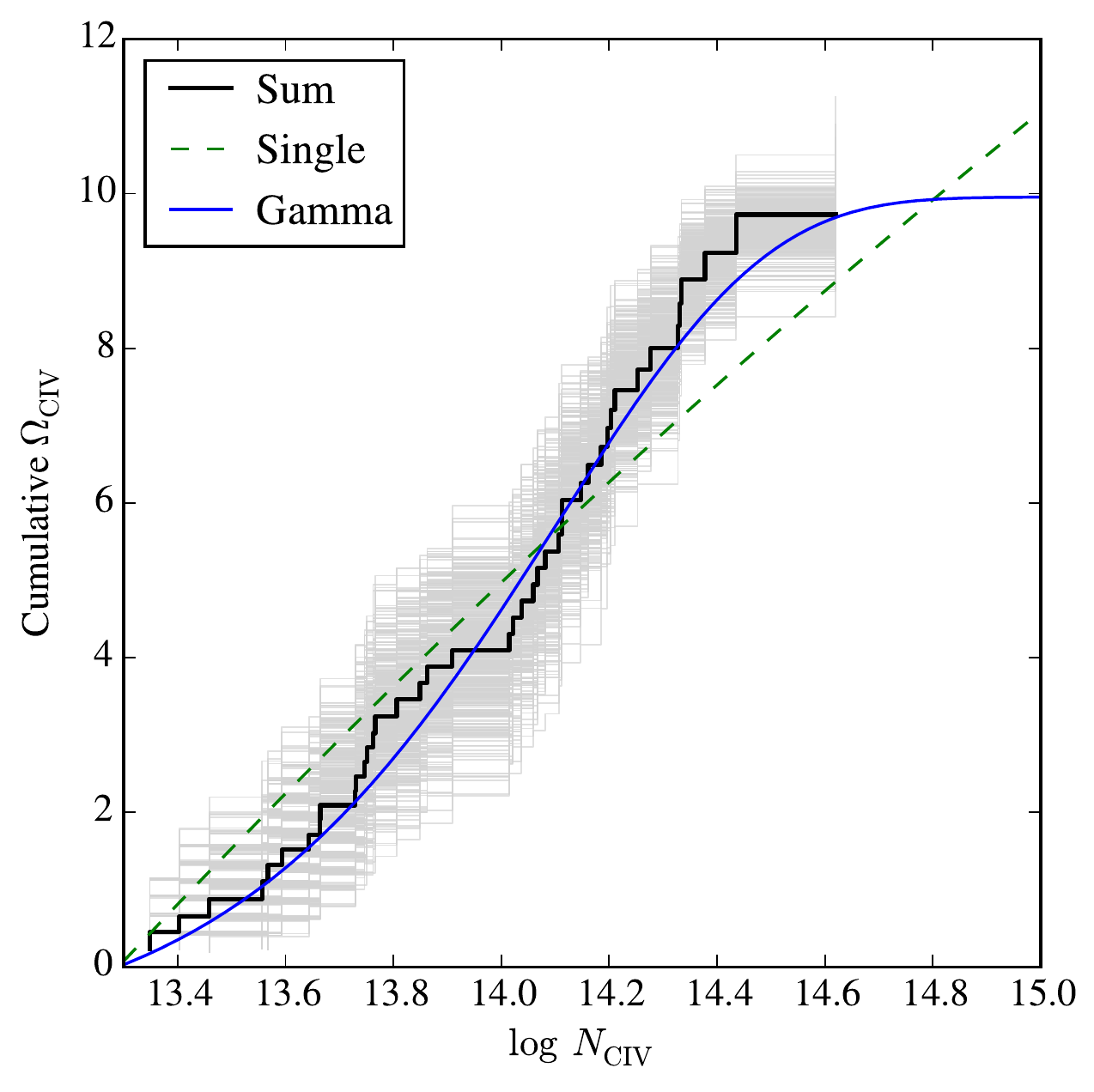}
\caption{Cumulative evaluation of \om\ from a direct summation of our
\cfourcol\ measurements (black histogram; Eq.~\ref{eqn:omsum}).
Overplotted is the integration of our best-fit $f(N)$ model (solid blue curve)
from $N_{\rm min} = 10^{13.3} \, \cmt$.
These lie in excellent agreement.  For an estimate of sample variance 
we have performed a bootstrap analysis of the summed evaluation;  500 
of these trials are shown in light gray on the figure.
Lastly, for comparison we show the results for the 
best-fit power-law model (dashed green curve), 
which systematically overestimates the \om\ value at low and high column densities.
}
\label{fig:cumomciv}
\end{figure}

In Figure~\ref{fig:cumomciv}, we present two evaluations of \om:
(1) the integration of our $\Gamma$-model for $f(N)$ with
$N_{\rm min} = 10^{13.3} \, \cmt$ and 
$N_{\rm max} = 10^{15} \, \cmt$; and 
(2) the summed evaluation of all \cfour\ systems in our 
statistical analysis (which spans from $10^{13.3} \, \cmt$ to
$10^{14.6} \, \cmt$):

\begin{equation}
\omeq = \frac{H_0 m_C}{c \rho_{c,0}} \sum_i
\frac{\cfourcoleq_i}{\Delta X(\cfourcoleq_i)} \;\;\; .
\label{eqn:omsum}
\end{equation}
These two evaluations are in excellent agreement and yield central
values of $10.0\sci{-8}$ and $9.7\sci{-8}$, respectively.
We further emphasize that the shallow power-law $\alpha_\Gamma$
derived from the
$\Gamma$-model and its exponential decrease
at high \cfourcol\ imply that our estimation is rather
insensitive to the choice of $N_{\rm min}$ and $N_{\rm max}$.

We have assessed the uncertainty in \om\ from sample variance through a 
bootstrap estimation of the summed evaluation (Eq.~\ref{eqn:omsum}).  
Specifically, we have
randomly sampled the observed distribution of \cfourcol\ values
with 10,000 trials (allowing for duplications) and evaluated
Eq.~\ref{eqn:omsum} for each trial.  Remarkably, the RMS of the resultant
distribution of \om\ values is small: $\sigma(\omeq) = 0.35\sci{-8}$.
A set of 500 trials are overplotted in gray on Figure~\ref{fig:cumomciv}.
We caution, however, that this bootstrap analysis may not sufficiently
capture the sample variance in the highest \cfourcol\ systems that
contribute $\approx 20\%$ to \om.
Furthermore, this summed evaluation does not correct 
for line-saturation unlike our maximum likelihood analysis
of $f(N)$.  On the other hand, we find excellent agreement between
the \om\ evaluations and have confidence that the effects
of line-saturation are small.
We adopt a $15\%$ uncertainty from systematic error
and report a final estimate of $\omeq = 10.0 \pm 1.5 \sci{-8}$.

Previous analysis of {\it HST} spectral datasets have presented
estimates for \om\ at $z \approx 0$.  
\cite{Cooksey:2010lr} reported $\omeq = 7.0\sci{-8}$ ($\approx 30\%$ error)
from a sample of 19 absorbers at $z<0.6$ discovered in
STIS and GHRS observations.
\cite{Tilton:2013lr} analyzed a larger set of STIS 
spectra and reported an \om\ value over twice as large from 29 absorbers
at $z<0.12$.  That estimation was revised downward by
\cite{Shull:2014zr}, who adopted a different approach to estimating 
\om\ using binned evaluations of $f(N)$.  From their analysis of the
COS linelist posted to HST/MAST by \cite{Danforth:2014zr},
they report $\omeq = 10.1^{+5.2}_{-2.4} \sci{-8}$ (the methodology for
their error estimation was not specified).  
Aside from the original \cite{Tilton:2013lr} estimate, which was later revised, these various \om\ evaluations are in good agreement
with our new analysis.

\begin{figure}[!h]
\centering
\includegraphics[width=1.05\linewidth]{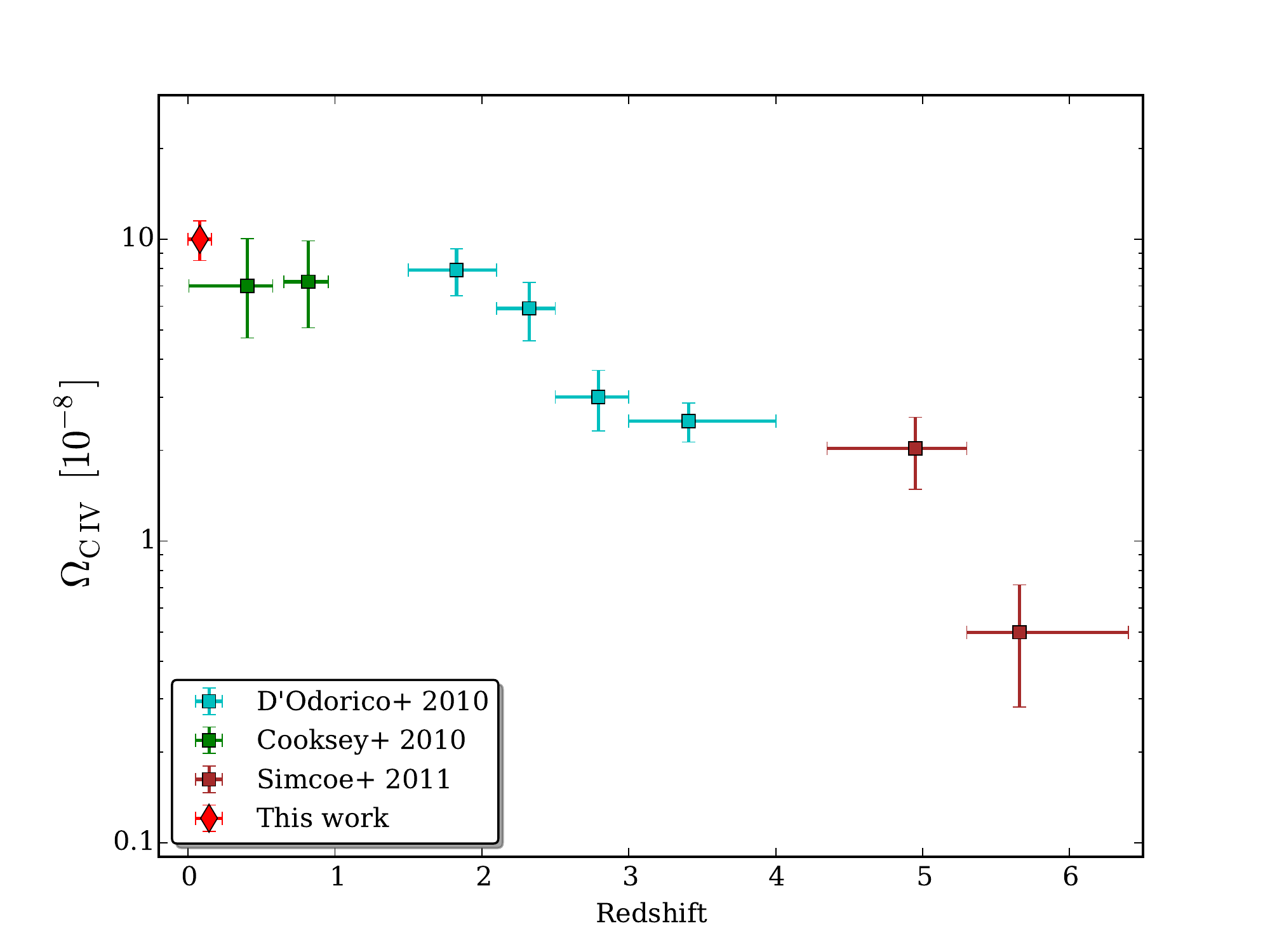}
\caption{Our result for \om\ (red diamond) plotted along with those from \citet{Cooksey:2010lr}, \citet{DOdorico:2010fj}, and \citet{Simcoe:2011rt}.  For visual clarity, we have included only our measurement at $z\sim0$, but the other studies in this redshift regime are discussed in the main text. Our results suggest that only a marginal evolutionary increase in \om~has occurred since $z \sim 2$, but the \cfour~mass density in the present epoch marks a significant increase over that at $z\gtrsim4$. }
\label{fig:omegaevo}
\end{figure}

To place our result in an evolutionary context with previous authors' findings, we convert their \om~values to our adopted integration limits and cosmology per Appendix C of \citet{Cooksey:2010lr} where necessary. Figure~\ref{fig:omegaevo} shows \om~values spanning z$\sim$6 to the present.  
The present ($z=0$) value of \om\ shows a clear
increase over that of earlier cosmic time ($z>4$), 
but the evolution since $z\sim2$ is modest and
not statistically significant.

\subsection{Comparison with simulations}

As we have alluded to above, the measurements we report are key observables that can be predicted by cosmological hydrodynamic simulations, which must reproduce not only key properties of galaxies themselves, such as the stellar mass function and evolution of the star formation density \citep{Madau:1998uq} but also observed properties of the CGM and IGM.  In fact, galaxies, the CGM, and the IGM are intimately related in the simulations, as gas infall and outflows are required to produce the observed global properties of the galaxy population; these processes in turn produce observational signatures in the CGM and IGM \citep{Bordoloi:2011uq,Fumagalli:2011qy,Bouche:2012kx} only probed by absorption line spectroscopy.  As the scale and sophistication of cosmological simulations have progressed, the key factors driving their ability to reproduce actual observations are often encapsulated in `sub-grid' physics, prescriptions for processes such as supernova feedback and galactic winds that are not directly resolved.  We attempt to use the preferred sub-grid variants of the simulations compared below, e.g., the $vzw$ wind model of \citet{Oppenheimer:2012qy}, which is a momentum conserving wind prescription where the velocity of the outflowing particles scale as the internal velocity dispersion of the galaxy and which produces galaxies that more closely match observations than a constant-velocity wind model.

\begin{figure}[t!]
\centering
\includegraphics[width=1.05\linewidth]{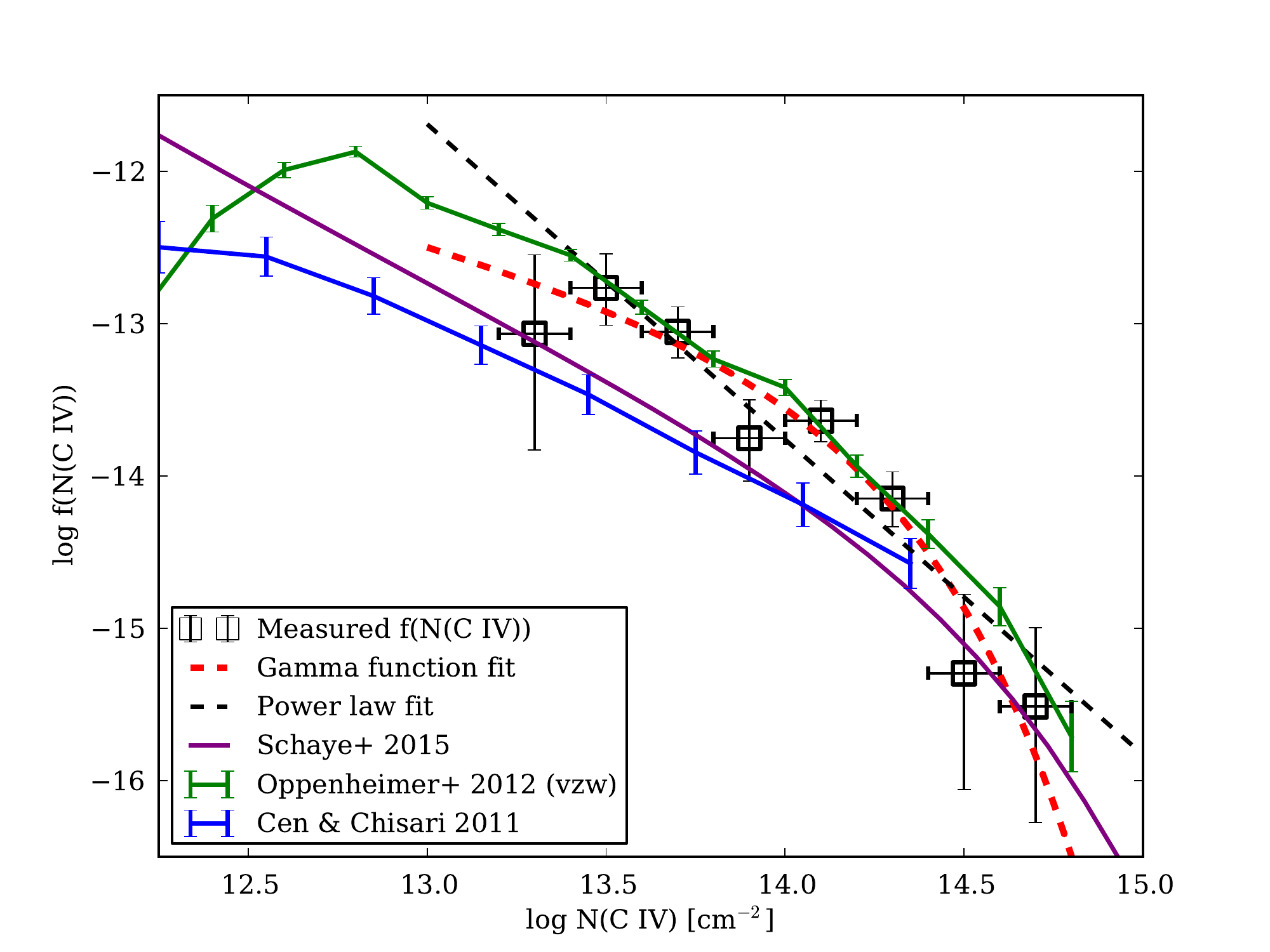}
\caption{Predictions of the column density distribution function from three cosmological hydrodynamic simulations [\citet[][green]{Oppenheimer:2012qy}, \citet[][purple]{Schaye:2015yg}, and \citet[][blue]{Cen:2011zl}] alongside our observational results. The black dashed line shows a single power law fit for f(N), while the red dashed line shows the preferred $\Gamma$-function fit.  All theoretical predictions exhibit qualitatively similar forms to the $\Gamma$-function we have adopted, while the single power law form overestimates both the measurements and predictions at low and high column densities.}
\label{fig:fnsims}
\end{figure}

As the \cfour\ doublet is among the most prominent metal-line spectral features, the \om\ and \fn\ derived from simulation data are routinely used as metrics of the self-consistency of sub-grid processes invoked to reproduce observations of galaxies and the IGM.  In Figure \ref{fig:fnsims}, we show our measured column density distribution function alongside predictions from three cosmological simulations.  Cosmological hydrodynamical simulations largely use yields from stellar population synthesis models to establish global metallicities, and the uncertainties from these models can introduce uncertainties of $\pm$0.3 dex in the column density distribution function. Given this uncertainty, both the \citet{Schaye:2015yg} and \citet{Oppenheimer:2012qy} models are largely consistent with our observations over the column density range probed, yet it is quite striking how well the \citet{Oppenheimer:2012qy} predictions seem to follow our measured distribution function.  Most importantly, both simulations predict  downturns relative to a single power law at low and high column densities, corroborating both the results of \citet{Cooksey:2013lr} that \fn\ deviates from a power law and our adoption of a $\Gamma$-function form.  The \citet{Cen:2011zl} results do not extend to to the highest column density regime covered by our observations and the other simulations, but they predict a smooth downturn (relative to a power law) at the lowest column densities.  The greatest dispersion among the models occurs at \logcfourcol\ $ \lesssim 13.0~\cmt$, but much higher S/N spectroscopy is required to constrain the behavior of \fn\ in this regime.

\begin{figure}[t!]
\centering
\includegraphics[width=1.05\linewidth]{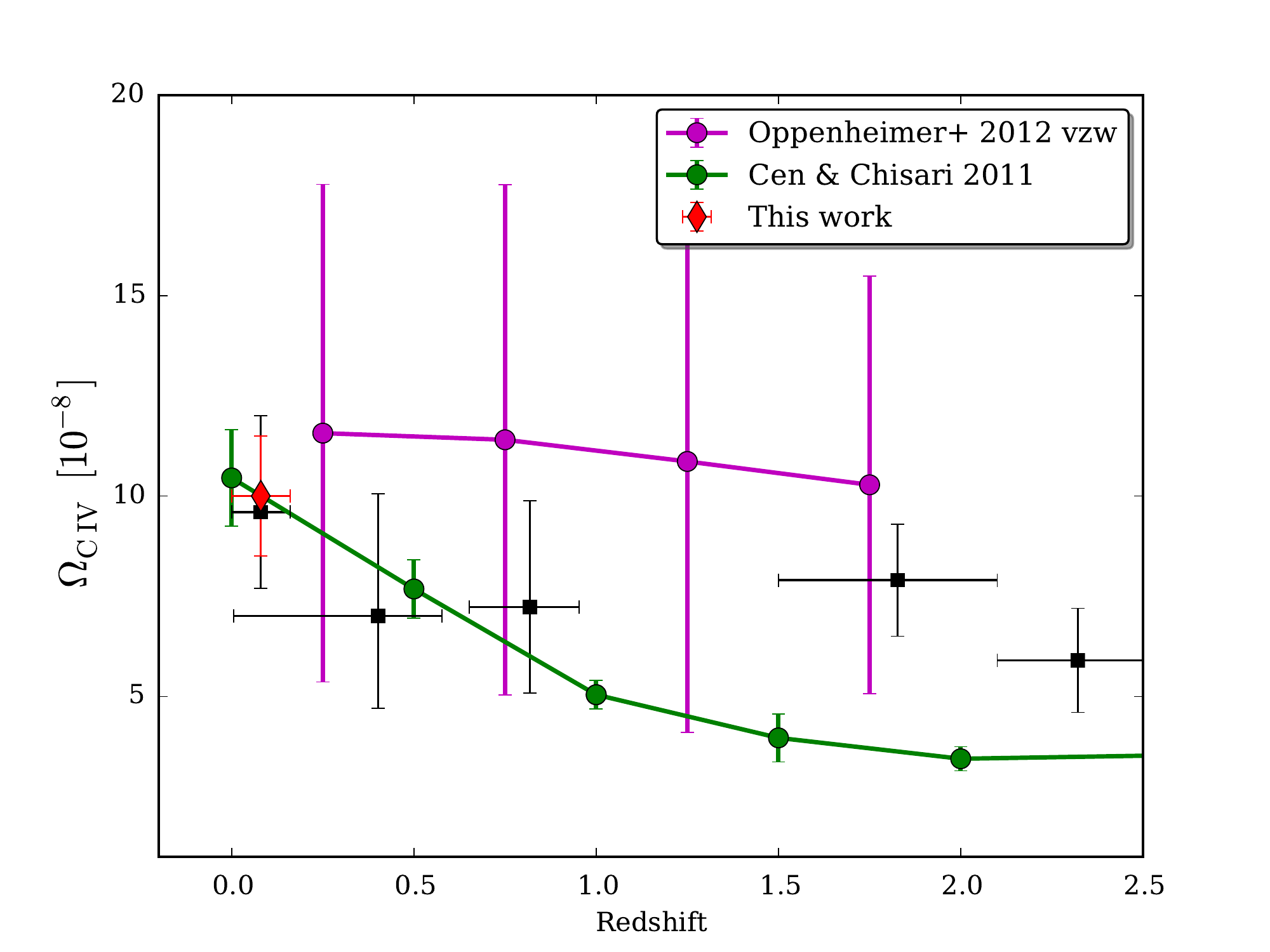}
\caption{The evolution of \om\ since $z=2.5$ as observed \citep[black squares;][]{Cooksey:2010lr, DOdorico:2010fj, Shull:2014zr} and as predicted by the cosmological hydrodynamical simulations of \citet{Oppenheimer:2012qy}  and \citet{Cen:2011zl} (purple and green lines and circles, respectively).  Our measurement is shown with a red diamond and is consistent with either theoretical result, although the evolutionary behaviors of these simulations differ from $z\sim2$ to $z\sim0.5$.}
\label{fig:omsims}
\end{figure}

Figure \ref{fig:omsims} shows the predicted evolution of \om\ by \citet{Oppenheimer:2012qy} and \citet{Cen:2011zl} from $z=2.5$ to the present alongside observational measurements.  The  \citet{Oppenheimer:2012qy} simulations predict minimal evolution in \om, but the uncertainties dwarf any differences between the observational constraints.  However, \citet{Cen:2011zl} predict a steady upward evolution in \om.  The behaviors of these predictions differ most significantly between $z\sim1.5$ and $z\sim0.5$, where the observational constraints are insufficient to favor one model over the other.  Regardless, the predicted \om\ values at $z\sim0$ are consistent with our reported measurement.

\begin{figure}[t!]
\centering
\includegraphics[width=1.05\linewidth]{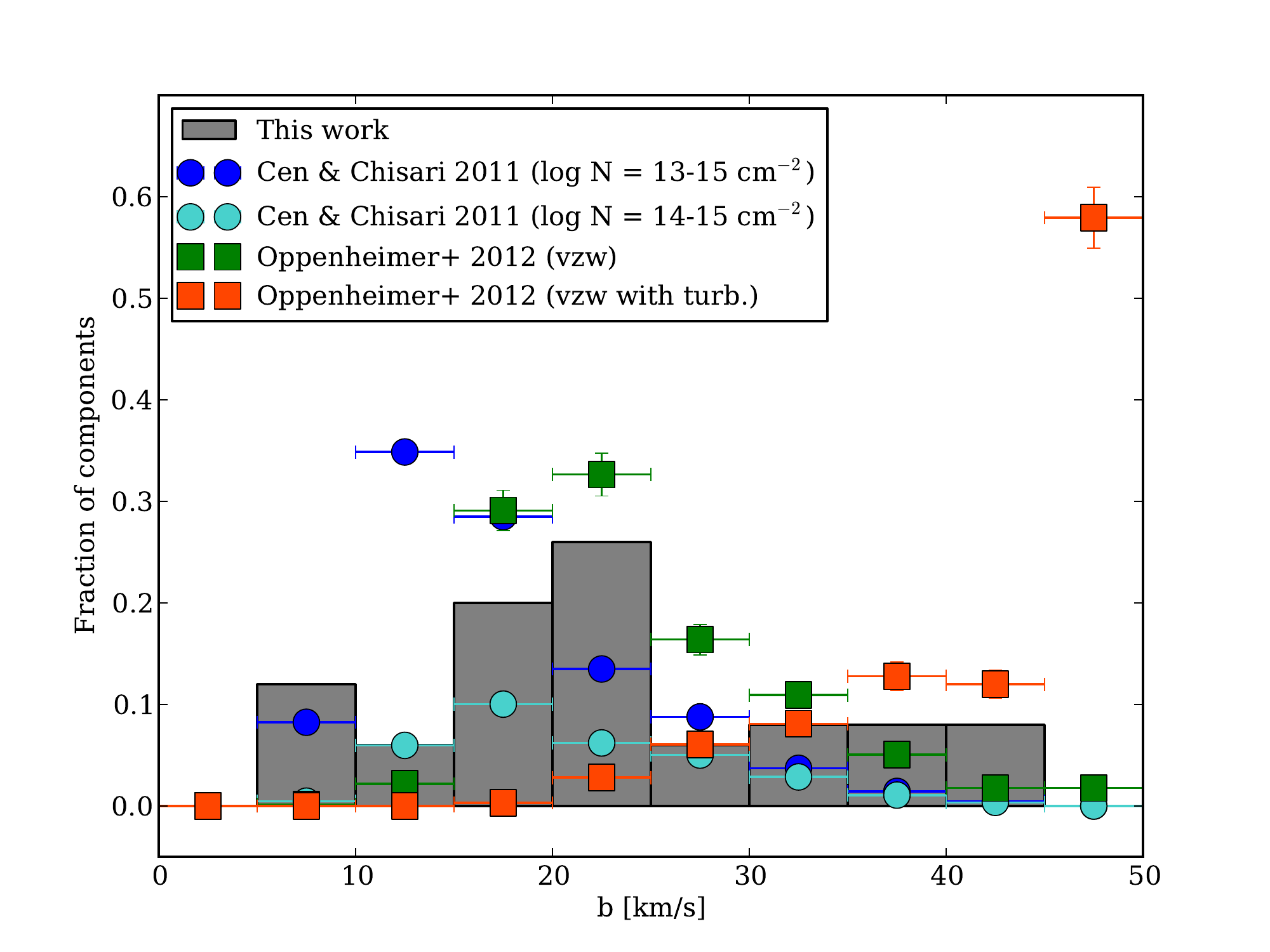}
\caption{The distribution of Doppler $b$ parameters measured from our \cfour\ absorber sample and from cosmological simulations.  The orange and green squares show predictions from the \citet{Oppenheimer:2012qy} simulations run with and without added turbulence, respectively.   Note the pronounced overprediction of \cfourb\ when including the turbulence model.  The blue and cyan circles show the $b$-value distributions from \citet{Cen:2011zl} for absorbers with \logcfourcol\ $ = 13-15~\cmt$ and \logcfourcol\ $ = 14-15~\cmt$, respectively.  Their measurement methods produce data that may not be compared with observations, and we present them here only for comparison between their two samples (see discussion).}
\label{fig:bsims}
\end{figure}

As an indicator of the physical conditions in the gas, the Doppler $b$ value provides another observable in the absorption line data by way of Voigt profile fitting. As \citet{Cen:2011zl} remark, directly comparing the $b$ values from any simulation versus those measured in observed absorbers is complicated by differences in measurement procedures (we derive \cfourb\ from Voigt profile fitting); the problem is exacerbated in systems with many components spread over a wide velocity range.  However, our sample is primarily composed of systems with simple velocity structures (one or two components spread over $<$ 100 km/s), so our data often provide reliable b-values that can be compared with simulation predictions.  We compare our sample with the predicted \cfourb\ distributions from \citet{Oppenheimer:2012qy}, who produced mock spectra for direct comparison with COS observations, in Figure \ref{fig:bsims}.   The distributions shown here from \citet{Oppenheimer:2012qy} employ their preferred feedback prescription including and excluding a model for turbulence.  While their turbulence model improves the reproduction of \osix\ absorber observations, the model overpredicts the line widths of observed \cfour\ absorbers.  Our data suggest that the media probed by \cfour\ absorption at $z\sim0$ may not be highly turbulent or that the turbulence model does not properly treat the media where \cfour\ is observed.  We return to the discussion of these predictions in Section \ref{sec:corrcoldens}.  

Also shown in Figure \ref{fig:bsims} are the predicted \cfourb\ distributions from \citet{Cen:2011zl} for two column density bins: \logcfourcol$\ = 13-15~\cmt$ and \logcfourcol$\ = 14-15~\cmt$, ranges over which our survey is sensitive.   Their data are not intended for direct comparison with observations, and we only present them here for discussion.  They report an increasing mean \cfourb\ with increasing \cfourcol, and as seen in Figure \ref{fig:bsims}, the $b$-value distribution for their \logcfourcol$\ = 14-15~\cmt$ bin (cyan circles in Fig. \ref{fig:bsims}) shows a higher mean \cfourb\ than that of the wider \cfourcol~range (blue circles).  \citet{Cen:2011zl} physically explain this trend by lower column \cfour\ absorbers residing in `quiescent' environments.  If this interpretation is valid, the dearth of low-\cfourcol, high-\cfourb\ absorbers in our sample (see Section \ref{sec:bvalcoldens}) may in fact be of physical origin rather than observational bias.

\section{Ion-ion Correlations}
We now explore relationships between the various metal species accessed by the COS data in addition to \textsc{C~iv}.  While \textsc{C~iv}~provides the most distinctive absorption-line tracer of metals within the COS G130M/G160M bandpass at very low redshift (\zabs $\lesssim 0.1$), other metal-line transitions also fall within this bandpass and redshift range.  As summarized in Table 1, we have identified and measured these additional lines occurring within our sample of \textsc{C~iv} absorbers.

\begin{figure*}[!t]
\centering
\includegraphics[width=.8\paperwidth]{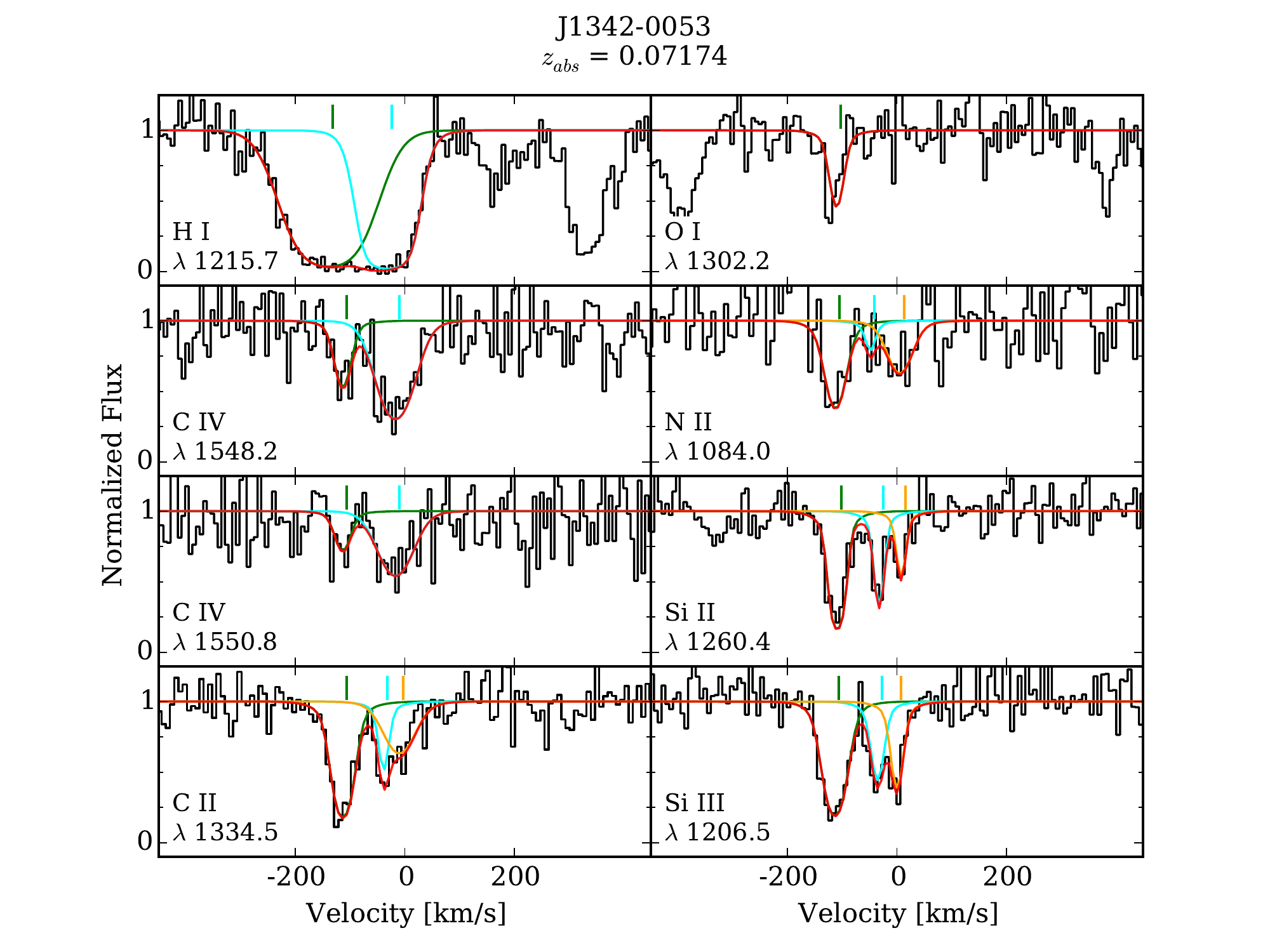}
\caption{A \cfour~absorber at $z_{abs}$=0.07174 towards J1342-0053 that shows several other metal lines as labeled at the bottom of each panel.  For each line, the individual component Voigt profile fits are colored green, cyan, and orange, while the composite profile resulting from a simultaneous fit of all components is shown in red. Interestingly, this system shows a neutral oxygen line well-aligned with a \cfour~component, which is typically associated with more highly-ionized gas.}
\label{fig:stackplot}
\end{figure*}

The additional metal-line species, Si~\textsc{ii}, Si~\textsc{iii}, \textsc{C~ii}, etc., can probe gas that is less ionized than \cfour\ (and possibly in different phases altogether), as the ionization potentials vary widely among them.  For example, the ionization energy from Si~\textsc{ii} to Si~\textsc{iii} is 16.35~eV, but the energy required to ionize \textsc{C~iii} to \textsc{C~iv} is 47.9~eV \citep{NIST_ASD}.  Consequently, species such as Si~\textsc{ii} and \cfour~are expected to exist in very different physical conditions, and one might intuitively anticipate that these ions would be located in entirely different locations with different kinematics (i.e., Si~\textsc{ii} and \textsc{C~iv} should exhibit different velocity centroids, line widths, etc.).  Indeed, this is predicted by cosmological simulations, which indicate that high ionization stages originate in large, low-density halos surrounding galaxies while lower ionization species such as \textsc{C~ii}, Si~\textsc{ii}, and Mg~\textsc{ii} are located in much smaller clumps embedded within the large hot halos \citep[e.g.,][]{Shen:2012nr,Ford:2013lr}.  

However, QSO absorber observations do not entirely conform to this expectation.  Many QSO absorbers show absorption lines of high-ionization species such as \textsc{C~iv}, \textsc{O~vi} and Ne~\textsc{viii} that have velocity centroids and line widths that are remarkably similar to those of low-ionizations stages such as \textsc{H~i}, \textsc{C~ii}, Si~\textsc{ii}, and Mg~\textsc{ii} or intermediate ionization stages including \textsc{C~iii}, \textsc{O~iii}, or Si~\textsc{iii} \citep{Tripp:2008lr,Tripp:2011wd,Meiring:2013fj,Savage:2014ty}, even in multicomponent systems spread over large velocity ranges.  Some absorbers are found to have \textsc{O~vi} profiles that clearly differ in detail from the profiles of low ions \citep[e.g.,][]{Werk:2013qy}, but even in these systems broad similarities in the shapes of the low-ion and high-ion profiles are evident.  This kinematical alignment of low- and high-ionization species suggests that the high ions could be located in a configuration such as the surface of a cool (low-ionization) cloud instead of the large gaseous halo origin predicted by simulations.  

As illustrated in Figure~\ref{fig:stackplot} and the Appendix, many of the blindly selected \textsc{C~iv} absorbers in this paper also display clear velocity alignment of lines of lower ionization species and higher ionization \textsc{C~iv} lines.  To explore the relationship between \textsc{C~iv} and lower ionization gas, we now consider the suite of additional metal lines covered in our blind survey.  Also, if future observations of the same QSOs cover shorter wavelengths and access unsaturated higher Lyman series \hone~lines, the lower ionization species may be employed to study abundances in these absorbers.

\subsection{Velocity alignment}
The spectral resolution of the G130M and G160M gratings coupled with Voigt profile fitting enable the velocity centroids of unsaturated individual lines to be localized within $< 10$ km/s, although the COS wavelength calibration introduces additional uncertainties.  Even where lines are blended, fitting the individual profiles can attempt to `deblend' the constituents, albeit with increased uncertainty.  All of the analyses in this section (this velocity comparison and the column density relationships to follow) utilize the individual components resulting from Voigt profile fitting.  Thus, we begin by examining the velocity differences between fitted components of various species identified within the same system.

We examined the component structure of all species in each system and grouped together components of different species with the smallest velocity separation without yet imposing a maximum velocity offset.  For example, in the absorber shown in Figure~\ref{fig:stackplot} (the $z_{abs} = 0.07174$ system in the spectrum of J1342-0053), we identified three components in the absorption profiles of \sitwo, \ctwo, N~\textsc{ii}, and \sithree, two components in the profiles of \textsc{C~iv} and \hone, and only one component in \oone.  For this absorber, we grouped together the \sitwo, \ctwo, N~\textsc{ii}, and \sithree~components at \dv = 8, -10, 5, and 0 km/s, respectively. The \ctwo, \cfour, \hone, \sitwo, N~\textsc{ii}, \oone, and \sithree~components at \dv = -114, -114, -139, -123, -110, -110, and -113 km/s, respectively, were placed in a second group.  We proceeded in this manner until all apparently related lines in the 42 absorbers in our sample had been assigned to a group. We note that in certain instances, the data are inadequate for this exercise; for example, we see that the \textsc{H~i} Ly$\alpha$ line in Figure~\ref{fig:stackplot} is strongly saturated, thus precluding reliable measurements of the quantities required for this analysis.

\begin{table*}[!t] 
\caption{Correlation Statistics of Ion Column Densities} 
\begin{center} 
\begin{threeparttable} 
\begin{tabular}{ccccccccc} 
\toprule 
Ion 1 & Ion 2  & $\tau_{ss}$\tnote{a}  & $P_{\tau ss}$\tnote{b}  & $\tau_{sd} $\tnote{c} & P$_{\tau sd}$\tnote{d} & $\Phi_{I}(1)$ (eV)\tnote{e} & $\Phi_{I}(2)$ (eV)\tnote{f} & $|\Delta \Phi|$ (eV) \\ 
\hline 
C IV & C II & 0.4286 & 0.0559 & 0.3744 & 0.1314 & 47.8 & 11.3 & 36.5 \\ 
C IV & O I & 0.0585 & 0.7698 & 0.0 & 1.0 & 47.8 & 0.0 & 47.8 \\ 
C IV & Si III & 0.4729 & 0.0242 & 0.3744 & 0.1353 & 47.8 & 16.3 & 31.5 \\ 
C IV & Si IV & 0.5867 & 0.0091 & 0.4933 & 0.0483 & 47.8 & 33.5 & 14.3 \\ 
C IV & O VI & 1.1667 & 0.0154 & 1.2778 & 0.0159 & 47.8 & 113.9 & 66.1 \\ 
C IV & N V & 0.3922 & 0.0244 & 0.4052 & 0.0224 & 47.8 & 77.5 & 29.7 \\ 
C II & Si III & 1.0292 & 0.0014 & 1.076 & 0.0009 & 11.3 & 16.3 & 5.0 \\ 
C II & Si IV & 0.8182 & 0.0172 & 0.8182 & 0.0172 & 11.3 & 33.5 & 22.2 \\ 
Si II & Si IV & 0.2821 & 0.4017 & 0.2821 & 0.4017 & 8.1 & 33.5 & 25.4 \\ 
Si III & Si IV & 0.8382 & 0.0057 & 0.8971 & 0.0034 & 16.3 & 33.5 & 17.2 \\ 
C II & Si II & 1.1048 & 0.0041 & 1.1048 & 0.0041 & 11.3 & 8.1 & 3.2 \\ 
Si II & Si III & 0.8235 & 0.014 & 0.8235 & 0.0145 & 8.1 & 16.3 & 8.2 \\ 
C IV & Si II & 0.2333 & 0.2427 & 0.1267 & 0.5993 & 47.8 & 8.1 & 39.7 \\ 
\hline 
\end{tabular} 
\begin{tablenotes} 
\item[a] {\scriptsize The Kendall tau correlation coefficient between the column densities of Ion 1 and Ion 2, assuming that lines flagged as saturated yield lower limits for their column densities.} 
\item[b] {\scriptsize Probablility that a correlation does not exist between the column densities of Ion 1 and Ion 2, assuming that lines flagged as saturated yield lower limits for their column densities.} 
\item[c] {\scriptsize The Kendall tau correlation coefficient between the column densities of Ion 1 and Ion 2, assuming that lines flagged as saturated yield reliable column densities.} 
\item[d] {\scriptsize Probablility that a correlation does not exist between the column densities of Ion 1 and Ion 2, assuming that lines flagged as saturated yield reliable column densities.} 
\item[e] {\scriptsize Energy to attain ionization state of Ion 1} 
\item[f] {\scriptsize Energy to attain ionization state of Ion 2} 
\end{tablenotes} 
\end{threeparttable} 
\end{center} 
\label{table:ionioncorr} 
\end{table*} 

\begin{figure*}[h!]
\centering
\includegraphics[width=0.6\paperwidth]{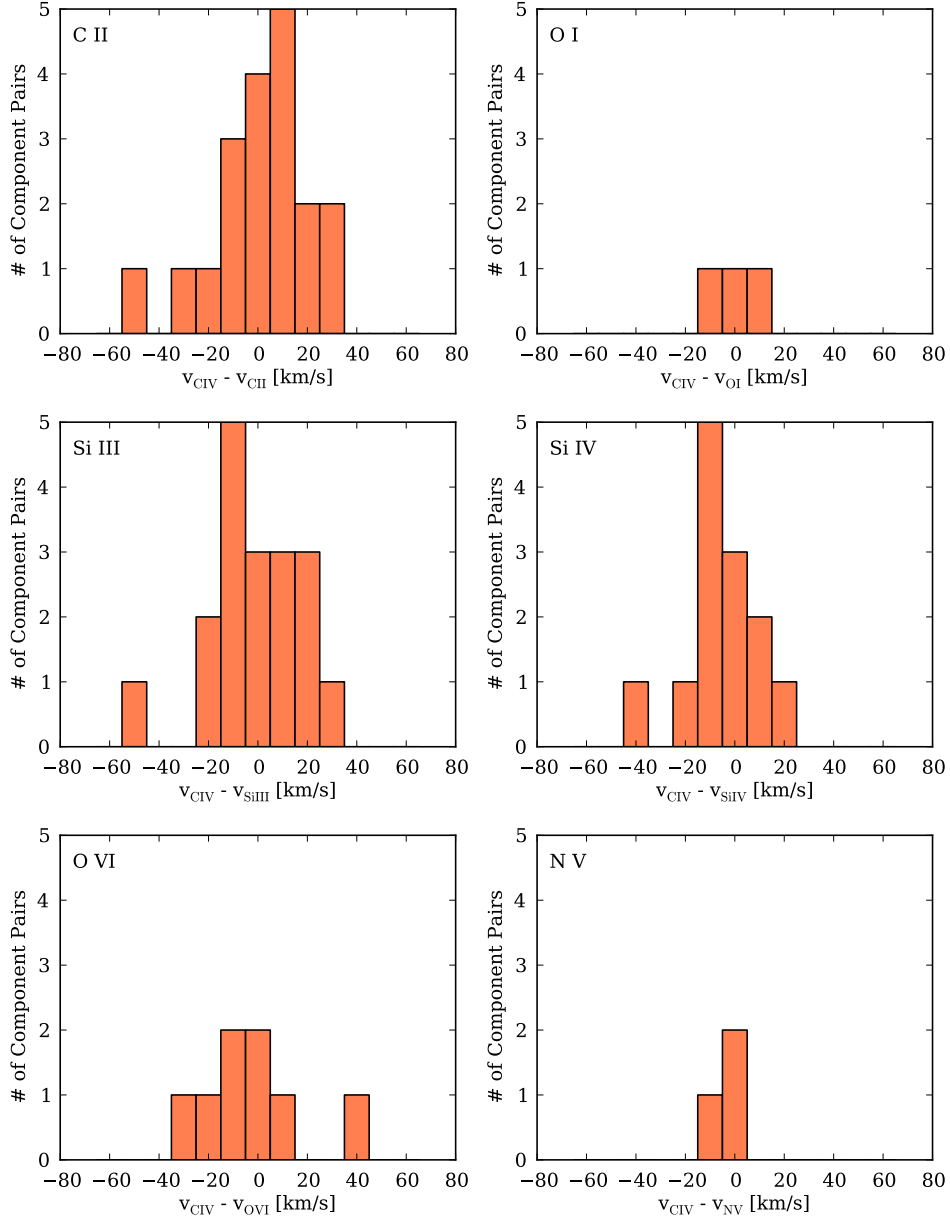}
\caption{Offsets in velocity between components of \cfour~and other ions detected in the data.  The plots are roughly arranged to show the  \cfour~offsets from low, intermediate, and high ions in the top, middle, and bottom rows, respectively. Close alignments in velocity space among components are typically interpreted as indicators of physical association, such as the species arising in the same kinematically connected structure (e.g., isolated gas cloud in a galaxy halo) or even in the same gas phase.  The many instances of offsets similar to the nominal resolution of 15 km/s indicate closely associated \cfour~components with a variety of ionization stages.}
\label{fig:velcomp}
\end{figure*}

Figure \ref{fig:velcomp} shows distributions of velocity offsets between \cfour~and a variety of ions as grouped by the above procedure; the species detected in our data cover a range of ionization stages, from \ctwo~and O~\textsc{i} (low ions) to \sithree~and \sifour~(intermediate ions) to \osix~and N~\textsc{v} (high ions).  In some instances, Figure~\ref{fig:velcomp} confirms expected outcomes, i.e., species that should be aligned (e.g., \textsc{C~iv} and Si~\textsc{iv}) are aligned. However, Figure~\ref{fig:velcomp} also reveals some surprising alignments. For example, the only 3 detections of O~\textsc{i} in our sample show close alignments with components of \cfour, even though the ionization potentials greatly differ: E(O~\textsc{I} $\rightarrow$ \textsc{II}) = 13.6 eV and E(C~\textsc{III} $\rightarrow$ \cfour) = 47.9 eV \citep{NIST_ASD}.  Likewise, the majority of the \textsc{C~ii} absorption lines are well aligned with the \textsc{C~iv} lines.  For \textsc{C~ii}, we have a larger sample, and we find that the distribution of velocity differences is centered on 0 km s$^{-1}$, and 79\% of the \textsc{C~ii} lines are aligned with \textsc{C~iv} to within $\pm$ 20 km s$^{-1}$.  Furthermore, our limited sample of \osix\ detections show decent alignment with \cfour\ components; in contrast, \citet{Lehner:2014kq} find that \osix\ and \cfour\ components often show quite distinct kinematics from one another in their $z\sim3$ sample of Lyman limit and damped Ly$\alpha$ systems.  Larger low-redshift samples at higher resolution would enable a further investigation of possibly evolving kinematic relationships between these species over the age of the Universe.

It is clear from Figure~\ref{fig:velcomp} that not all component pairs classified in the above manner are well aligned.  However, it is important to recognize that the spectral resolution, line-spread function, and well-known wavelength calibration problems of COS \citep{Wakker:2015rf} limit our ability to accurately measure velocity centroids. Therefore, we impose the following requirement for individual component pairs in order to include them in our subsequent analyses of aligned absorption lines:

\beq
\delta v_{XY} \leq \sqrt{ \sigma_v^2(\rm X) + \sigma^2_v(\rm Y) + \sigma^2_v({\rm COS})}
\label{eqn_aligned_groups}
\eeq
where $\delta v_{XY}$ is the velocity offset between components of species X and Y, and $\sigma_v(\rm X)$ and $\sigma_v(\rm Y)$ are the uncertainties in the component velocity centroids of species X and Y, respectively, from Voigt profile fitting.  The final term, $\sigma_v({\rm COS})$, accounts for known (but poorly understood) errors in the COS wavelength solution that well exceed the 15 km/s resolution of the instrument \citep{Wakker:2015rf}.  Efforts by several teams are underway to solve for corrections to these errors, but the effects appear to be highly wavelength dependent and nonlinear.  We have adopted  $\sigma_v({\rm COS}) = 25$ km/s, commensurate with offsets we have measured in our spectra between multiple transitions of the same ion that should be perfectly aligned but are not (Tripp et al., in preparation).

\begin{figure*}[!t]
\centering
\includegraphics[width=0.6\paperwidth]{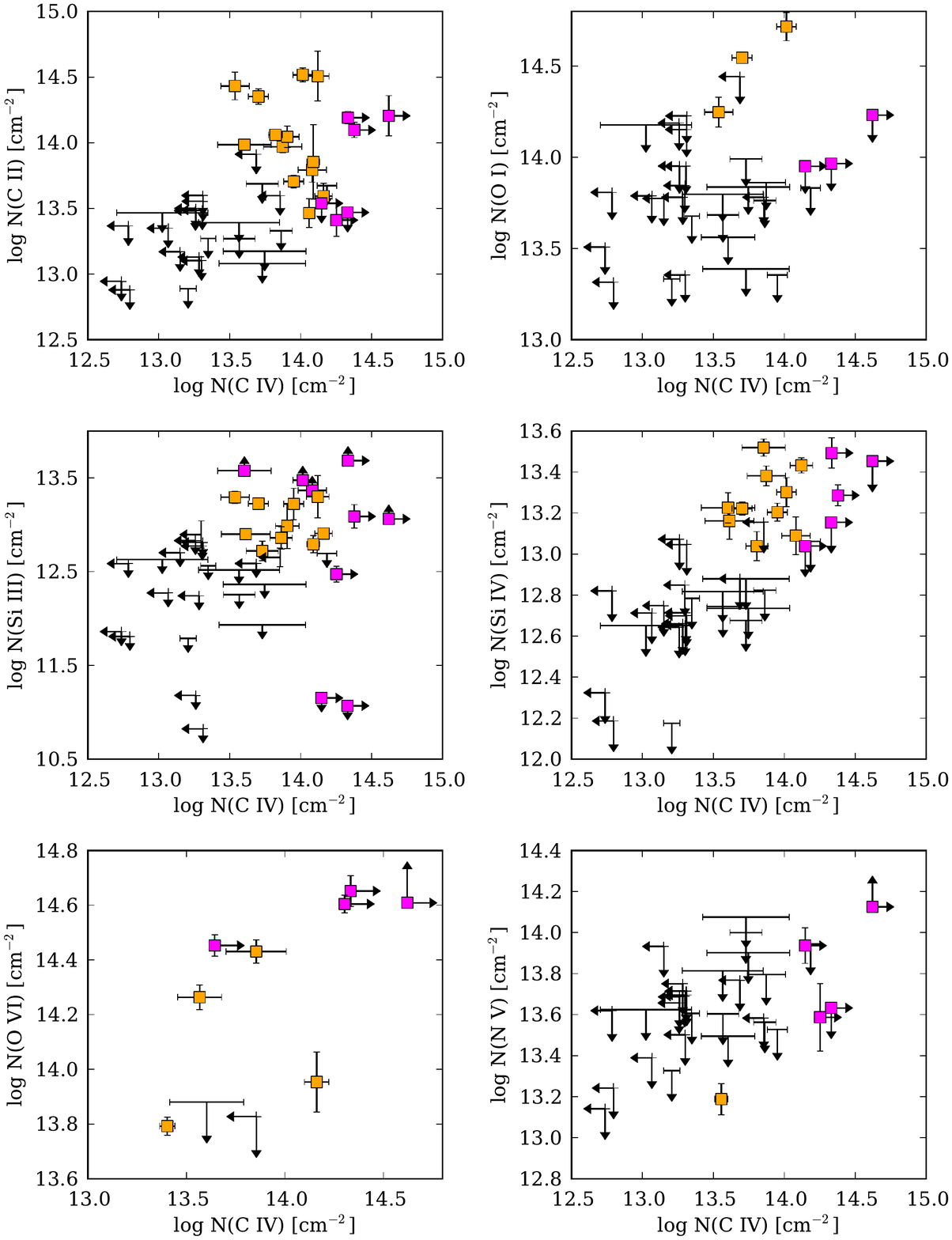}
\caption{Column densities of various species as a function of \cfour~column density.  Yellow squares represent unsaturated detections in both species plotted while magenta squares denote saturation in one or both species.  Nondetections are shown with 3-$\sigma$ upper limits.}
\label{fig:coldens6}
\end{figure*}

\subsection{Correlations of column densities between ions}
\label{sec:corrcoldens}

Using the components grouped in accord with Eq.~\ref{eqn_aligned_groups}, we next examine the correlations in column density of one species with another.  The aforementioned limitations in our \hone\ measurements (due to strong saturation of the only available \hone\ lines) inhibit our ability to analyze the absorbers with the usual photoionization or collisional ionization models, but we can nevertheless gain insights on the nature of these systems by comparing unsaturated species.  For example, if two given species generally are located in the same (cospatial) gas phase, then one might expect their column densities to be strongly correlated if the various clouds have the same relative abundance patterns.  Conversely, if two species have a physical relationship but are not necessarily cospatial (e.g., if the \cfour\ arises in an interface layer on the surface of a low-ionization phase), then the column densities of those species might be poorly correlated despite being kinematically well aligned.  We note that it has already been shown that the population of \osix~absorbers that are well aligned with \hone\ exhibit very weak correlation between $N$(\osix) and $N$(\hone)~\citep{Tripp:2008lr}, and we might anticipate a similar situation with \cfour.  

\begin{figure*}[!t]
\centering
\includegraphics[width=0.6\paperwidth]{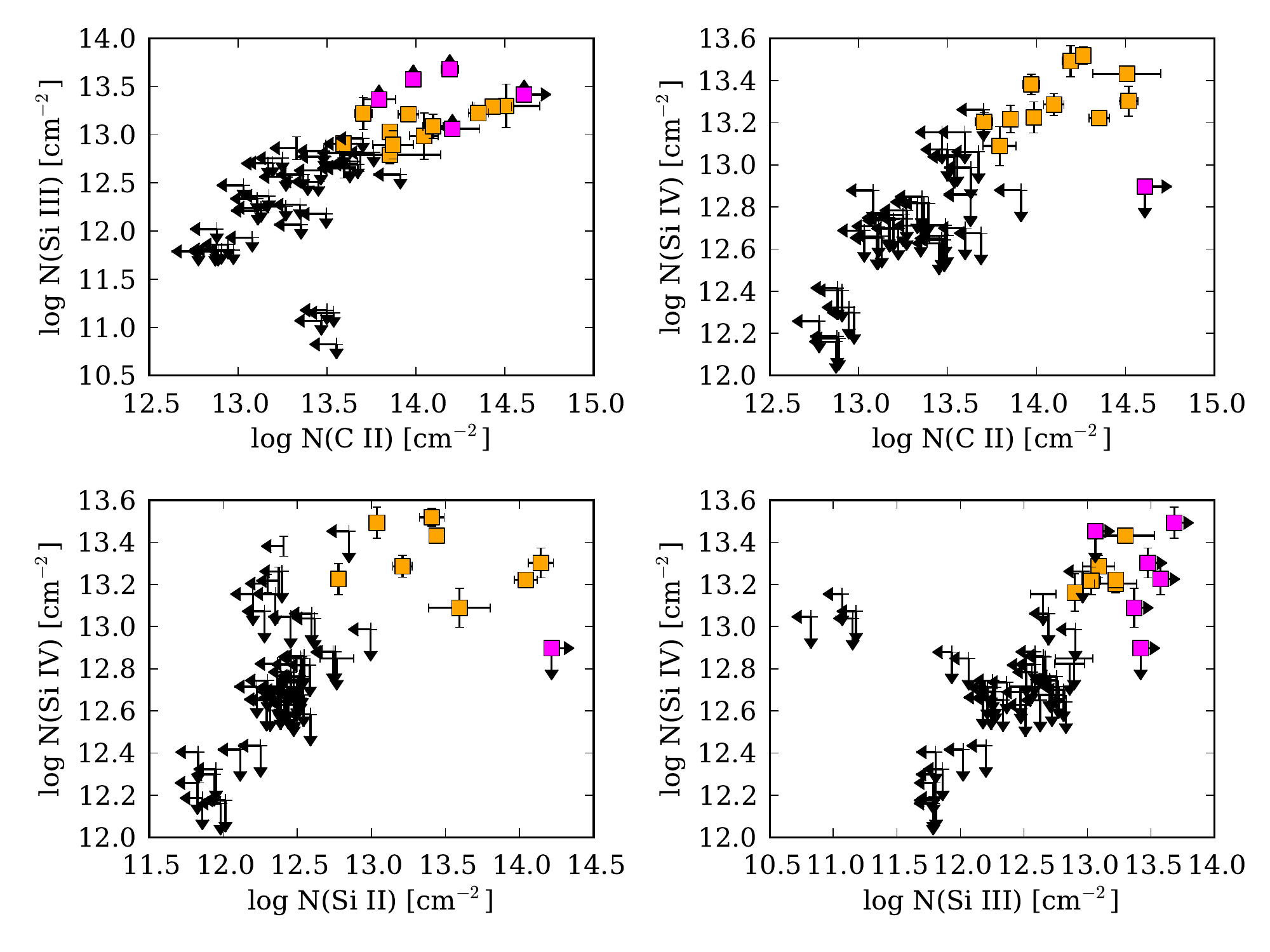}
\caption{Column densities of low- and mid-ionization species versus one another.  Yellow squares represent unsaturated detections in both species plotted while magenta squares denote saturation in one or both species.  Nondetections are shown with 3-$\sigma$ upper limits.}
\label{fig:coldensLowMid}
\end{figure*}

\begin{figure*}[!bth]
\centering
\includegraphics[width=0.85\paperwidth]{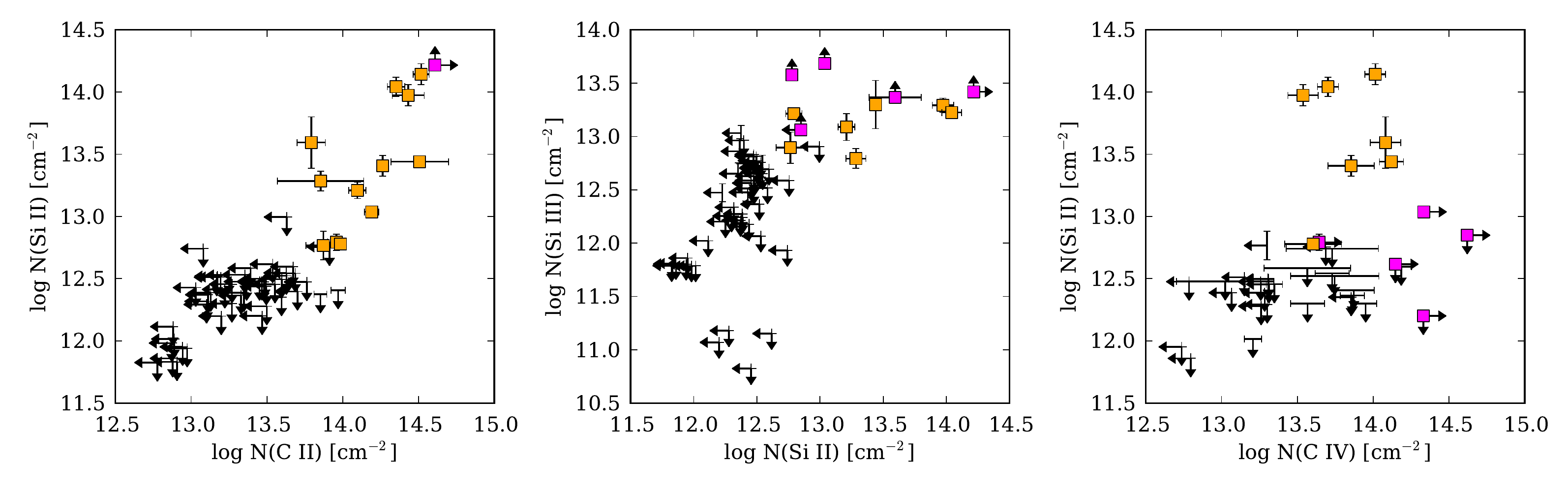}
\caption{Column densities of low-ionization species \sitwo~and \textsc{O I} as a function of \ctwo, another low ion. Yellow squares represent unsaturated detections in both species plotted while magenta squares denote saturation in one or both species.  Nondetections are shown with 3-$\sigma$ upper limits.}
\label{fig:coldensLow}
\end{figure*}

Figures \ref{fig:coldens6}, \ref{fig:coldensLowMid}, and \ref{fig:coldensLow} show the column density measurements of individual species components associated in velocity as described above.  The points marked by yellow squares indicate unsaturated detections of both species, and magenta squares indicate that one of the two species is saturated.  Here, we have flagged a line as saturated if $\geq$10\% of the pixels in the line profile have flux values that are less than their noise values.  Upper limits are indicated where one or both species was not detected at a coincident velocity to components identified in one or more other species.  Certain points in the ion-ion plot reflect only upper limits for both species, and we have excluded these cases from the statistical analysis below.

To quantitatively assess whether the species shown in Figures~\ref{fig:coldens6}, \ref{fig:coldensLowMid}, and \ref{fig:coldensLow} are correlated, Table~\ref{table:ionioncorr} shows the ion-ion column density correlation statistics calculated using the Kendall Tau rank correlation method, which incorporates censored data (upper/lower limits) to calculate the probability of the null hypothesis that no correlation exists between the two variables (smaller values of $P_{tau}$ in Table 2 correspond to \textit{greater} confidence of rejecting the null hypothesis that the quantities are not correlated).  We have imposed a criterion for saturation upon the column density measurements as described above.   However, these column densities may be better constrained than merely assigning lower limits, as Voigt profile fitting (used to measure column densities) is able to measure mildly saturated lines with adequate accuracy. Therefore, we calculate the Kendall tau correlation statistics flagging the measurements of the `saturated' lines as both lower limits and detections, which are reflected in $P_{\tau ss}$ and $P_{\tau sd}$, respectively.

With the exception of cases in which our sample is very small (e.g., \textsc{O~i}, \textsc{N~v}), most of the Kendall-Tau tests in Table~\ref{table:ionioncorr} suggest that the column densities of most of these species are correlated, albeit in some cases weakly.  However, we find that the likelihood of a correlation is generally very strong (i.e., small $P_{\tau}$) between species with small ionization energy differences ($|\Delta \Phi| < 20$ eV).  The most notable exception occurs for the \cfour-\osix~pair, which shows a $\sim 98 \%$ probability for the existence of a correlation even though these species have the greatest separation in ionization potential ($|\Delta \Phi| = 66.1$ eV).  This result should be interpreted cautiously as we only have a small sample of \osix\ lines, but this could occur if the \cfour\ and \osix\ arise in a collisionally ionized hot phase.  We note that our survey is \cfour~selected and therefore includes, by design, gaseous systems with ionization conditions conducive to maintaining sufficient quantities of \cfour~ions for detection.  Given the small sample sizes of certain ion pairs, we caution against overinterpreting these correlation results; some peculiarities clearly arise in Table~\ref{table:ionioncorr}, such as the apparent correlations for \ctwo-\sitwo\ and \ctwo-\sifour\ but the lack thereof for \sitwo-\sifour.  

The decreasing likelihood of correlation with increasing differences in ionization potential may indicate that \cfour\ has some type of physical relationship with lower ionization stages but nevertheless arises in a phase that is distinct from the low ions.  This would help to explain the poor agreement between the observed \textsc{C~iv} $b-$values and the predicted $b-$values from \citet{Oppenheimer:2012qy} when turbulence is included (purple points in Figure \ref{fig:bsims}): \citet{Oppenheimer:2012qy} were motivated to add turbulence to their model based on the nonthermal broadening required by aligned \textsc{O~vi} and \textsc{H~i} absorbers under the assumption that the $b-$values of \textsc{O~vi} and \textsc{H~i} can be jointly used to solve for the temperature and nonthermal broadening.  If the \textsc{O~vi} and \textsc{H~i} lines arise in physically distinct phases, then the nonthermal broadening found this way may not be valid.  Indeed, the addition of this ``turbulence'' is not supported by the measured \cfour\ $b$-values in our sample (see Figure \ref{fig:bsims}).  More sophisticated modeling of multiphase absorbers may be required for these CGM absorbers.  It would also be helpful to expand the sample of \cfour\ absorbers to improve the statistical significance of these analyses.

\section{N(\cfour)-$b$ relationship and physical origin of \cfour~absorbers}
\label{sec:bvalcoldens}
The larger aim of our survey is to place the gas traced by \cfour~absorption in context with the galaxies that may host the gas or have some past or present association.  While we may, in general, expect the gas to arise from a variety of astrophysical configurations, we may inquire whether the absorption data themselves suggest that the gas arises under some characteristic set of physical conditions.

\citet{Heckman:2002lr} found that \osix~absorbers observed in the Galactic disk, Galactic halo, Magellanic Clouds, and IGM all fall on a common N(\osix) vs. $b$-value relation.  They further interpreted this result to suggest that the gas is tracing radiatively cooling gas that is passing through the `warm-hot' regime at $T = 10^5-10^6$ K.  \citet{Tripp:2008lr} find that their large sample of \osix~absorbers do not follow the \citet{Heckman:2002lr} relation as closely, but their subsample of `intervening' absorbers (those with a greater velocity separation from the QSO itself) do show a marginal correlation between N(\osix) and $b$(\osix).  

\begin{figure}[!t]
\centering
\includegraphics[width=1.05\linewidth]{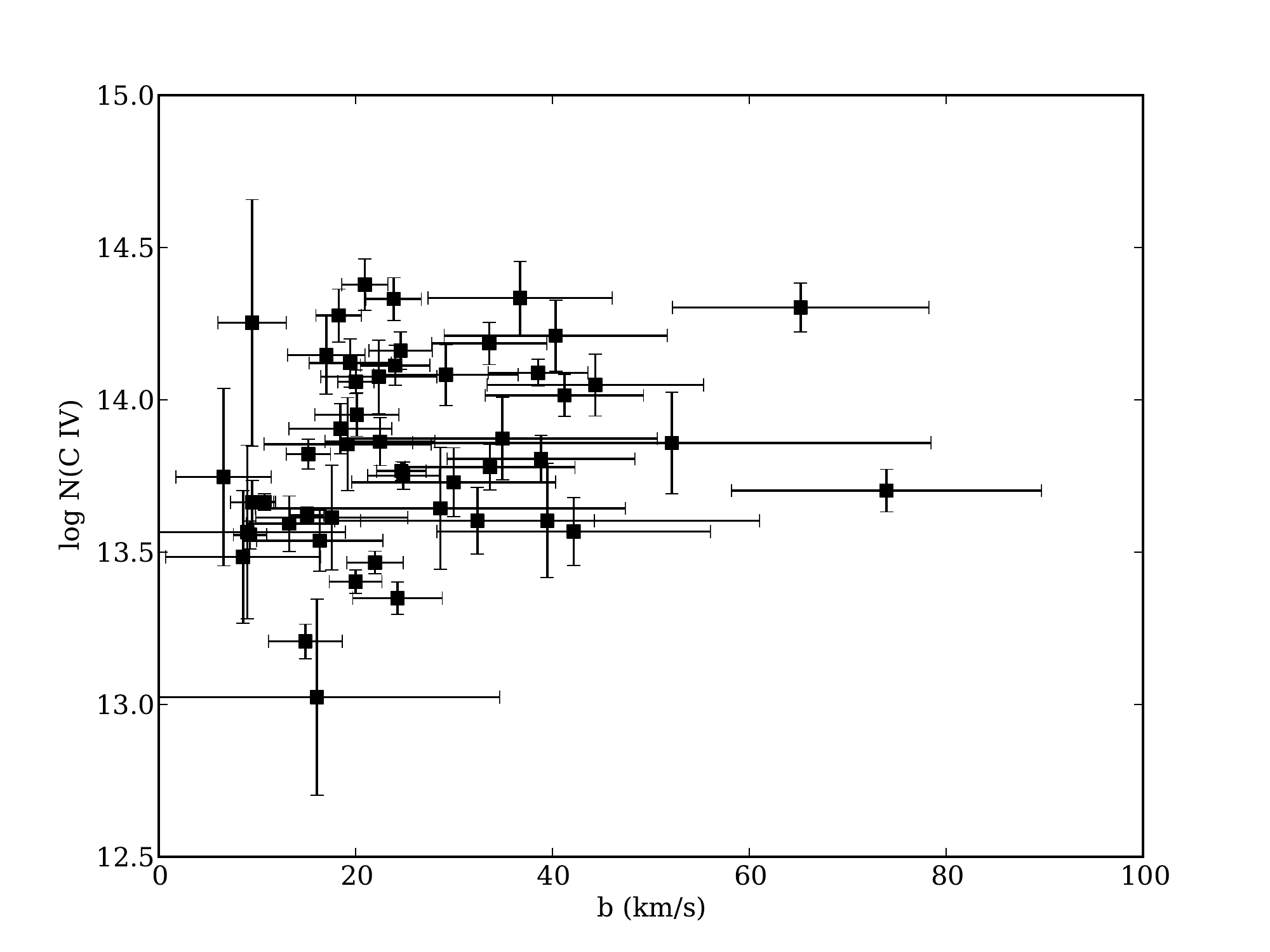}
\caption{Column densities of \cfour~as a function of the $b$-value for individual absorption components.  Both quantities and their errors are measured from Voigt profile fitting.}
\label{fig:coldensb}
\end{figure}

We present the N(\cfour)-$b$(\cfour) relationship for our absorber sample in Figure \ref{fig:coldensb}.   To assess whether a correlation exists between the two variables, we employ a nonparametric Spearman rank-order correlation test.  For the entire sample, we obtain a p-value for the null hypothesis (that no correlation exists) of 0.022, suggesting that a correlation exists at the 97.8\% confidence level.  However, the apparent envelope separating the region devoid of low-N(\cfour), high-$b$ points in Figure \ref{fig:coldensb} (the lower-right corner) is at least partly due to an observational bias.  Very broad, low-column density absorbers produce shallow line profiles, which are difficult to detect above the noise.  To account for this bias, we also consider only systems that have log N(\cfour) $>$ 13.5 $\cmt$, approximately above which we measure the full range of $b$-values.  A Spearman test for the correlation among these points yields a likelihood of 86.8\%, i.e., when we account for our inability to detect \cfour\ absorbers with large $b$-values and low column densities, we find that there is no compelling evidence that $b$(\cfour)  and \cfourcol\ are correlated in our CIV sample.

\section{Summary and Conclusions}
From a blind survey utilizing HST/COS spectra of 89 QSOs, we have presented our sample of 42 $z<0.16$ \cfour\ absorbers, their measurements, statistics, and correlations between the ions detected.  These absorbers compose the parent sample for a larger program that aims to study the state of cosmic heavy-element enrichment in the most recent epoch (as in this paper) and characterize the CGM with unprecedented sensitivity to faint dwarf galaxies (Papers I, III, and subsequent works), the latter enabled by the low redshift regime of our sample. 

We summarize our primary findings as follows:
\begin{enumerate}

\item At $z<0.16$, we measure \om~$= 10.0 \pm 1.5 \times 10^{-8}$ for \logcfourcol\ $> 13.0~\cmt$; therefore, the \cfour\ cosmic mass density has indeed increased over cosmic time, but marginally so since $z\sim1.5$.  
\item  At $z\leq0.16$, the frequency of \cfour\ absorbers per comoving path length \dNdX $= 7.5 \pm 1.1$, representing an increase relative to the value measured over the full $z<1$ epoch \citep[4.9; ][]{Cooksey:2010lr}.
\item Various cosmological hydrodynamic simulations that include galactic outflows qualitatively produce increased \dNdz, \dNdX, and \om\ since $z\sim4$, and their predictions are indeed consistent with our measurement of \om\ at $z\sim0$.  However, discrepancies arise among the evolutionary tracks of these theoretical predictions, likely due to `sub-grid' physics prescriptions that handle processes such as feedback and star formation.  Our results are consistent with \om\ remaining relatively constant at $z=0$ from $z=1$.
\item Close  alignments in velocity occur between species of varying ionization potential, but evidence for correlations between the column densities of several of these well-aligned ion pairs appears to weaken as the differences in their ionization potentials increase. This comparative analysis would benefit from larger samples of systems that show components of multiple ions; however, the close velocity alignments of certain species that show a decreased likelihood of a correlation suggests that many absorbers in our sample reside in multiphase material.
\item  Some evidence exists for a correlation between \cfourcol\ and \cfourb\, but the evidence is not strong over the entire sample when considering incompleteness to high-$b$, low-\cfourcol\ absorbers.  

\end{enumerate}

\section*{Acknowledgements}
The authors would like to thank George Becker for sharing software; Valentina D'Odorico, Hsiao-Wen Chen, and Charles Danforth for helpful discussions; and Ben Oppenheimer, Renyue Cen, Elisa Chisari, and Ali Rahmati for sharing their simulation results to include in this study. We also thank Kathy Cooksey for providing software and assistance with statistics calculations.  Support for this research was provided by NASA through grants HST-GO-11741, HST-GO-11598, HST-GO-12248, and HST-AR-13894 from the Space Telescope Science Institute, which is operated by the Association of Universities for Research in Astronomy, Incorporated, under NASA contract NAS5-26555.
  
\clearpage

\clearpage

 \section*{Appendix}
This section contains plots showing the spectra and Voigt profile fits of all species in our \cfour\ absorber sample.  The individual components are color coded, with the composite profile marked in dark red, and the error vector is plotted in light red.  In cases where a line was deemed too saturated to yield reliable column density measurements from profile fitting, no fit is plotted. Components marked with magenta crosses denote interloping lines blended with components attributed to the species labeled in each panel.  Certain spectral regions are affected by geocoronal emission and are marked with $\oplus$.
	
\maxdeadcycles=1000

\begin{figure*}[!b] 
\centering 
\includegraphics[width=0.9\textwidth,height=0.9\textheight,keepaspectratio]{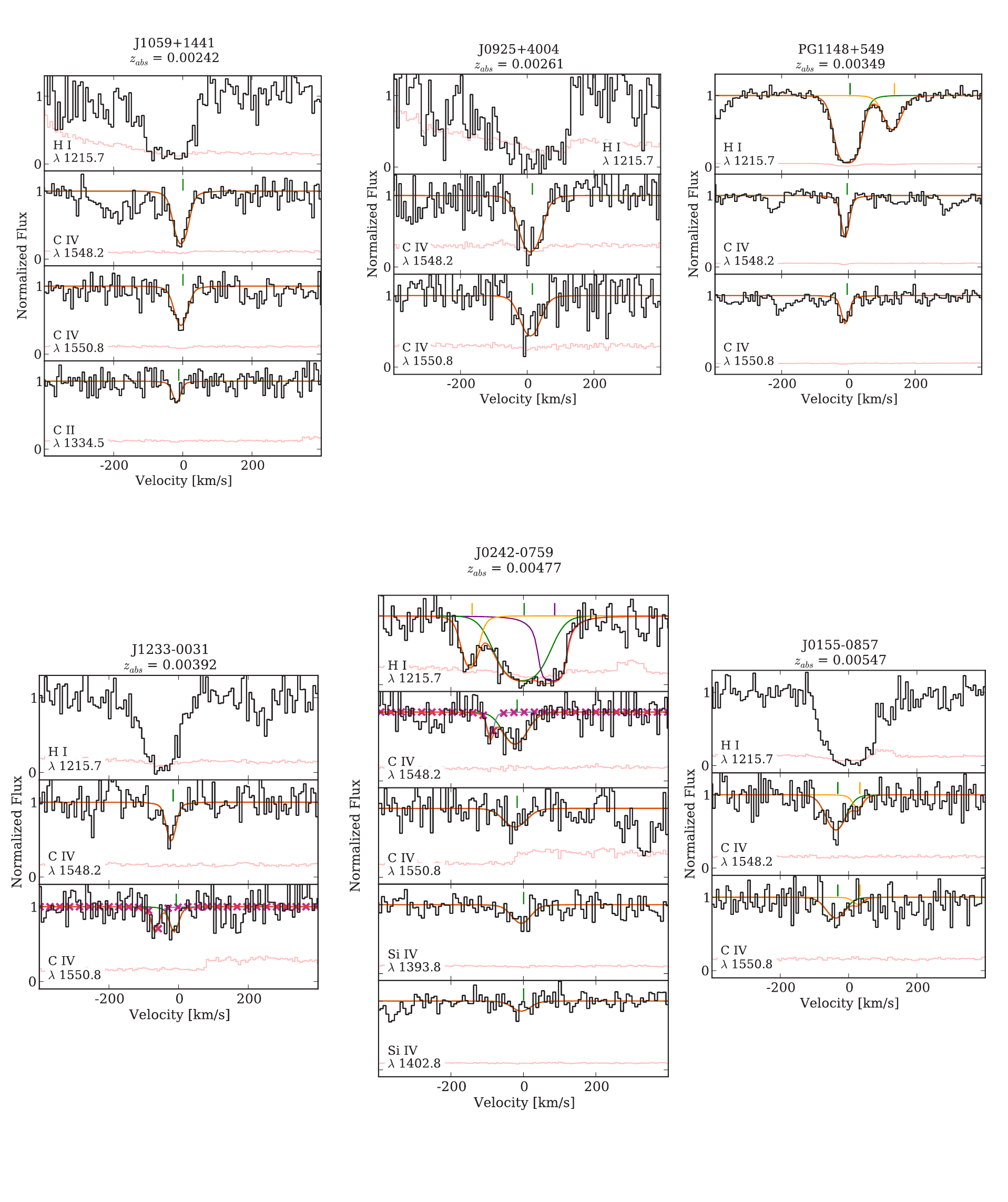} 
\caption{Spectra and fitted Voigt profiles for our \cfour~absorber sample.  For each system, at least one transition is shown for every ion detected, and each component is colored to match components of other species approximately aligned in velocity.  Profiles of interloping lines from other systems that were included in a fit are marked with purple x's, while the overalll fit is shown in maroon.  The error vector is plotted in light red.} 
\end{figure*} 

\setcounter{figure}{\value{figure}-1} 
\begin{figure*}[!h] 
\centering 
\includegraphics[width=\textwidth,height=\textheight,keepaspectratio]{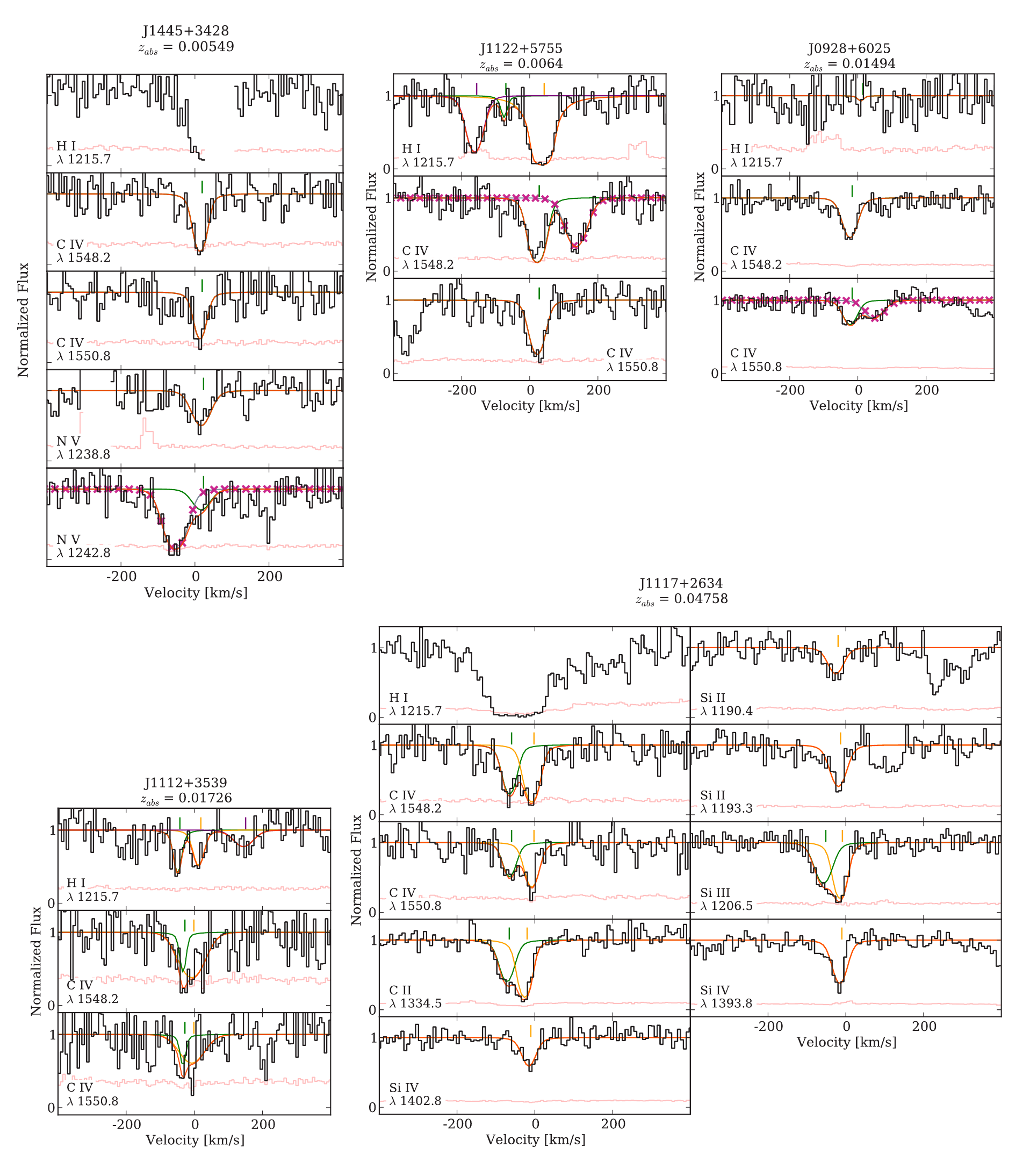} 
\caption{continued.} 
\end{figure*} 

\setcounter{figure}{\value{figure}-1} 
\begin{figure*}[!h] 
\centering 
\includegraphics[width=\textwidth,height=\textheight,keepaspectratio]{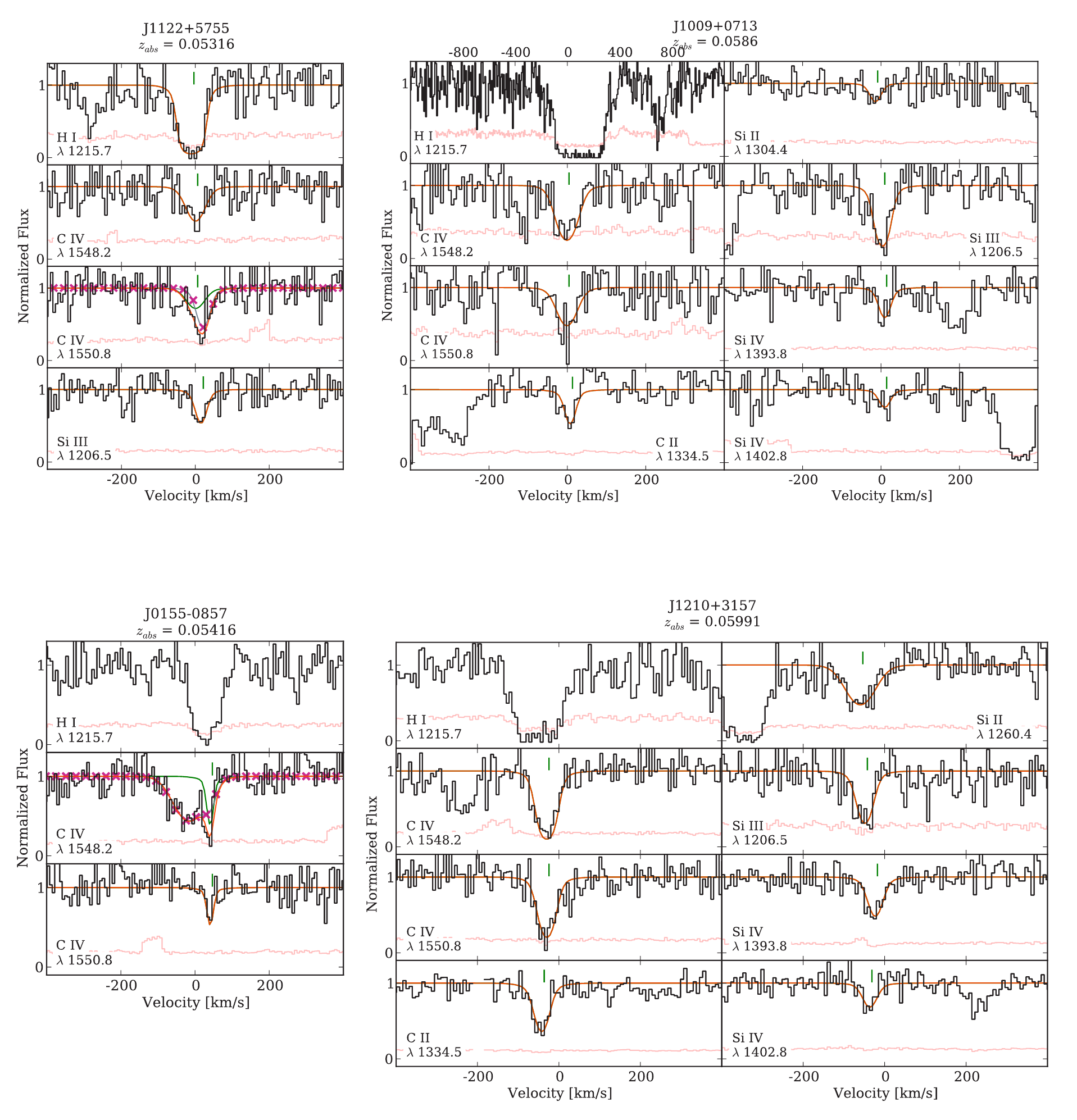} 
\caption{continued.} 
\end{figure*} 

\setcounter{figure}{\value{figure}-1} 
\begin{figure*}[!h] 
\centering 
\includegraphics[width=\textwidth,height=\textheight,keepaspectratio]{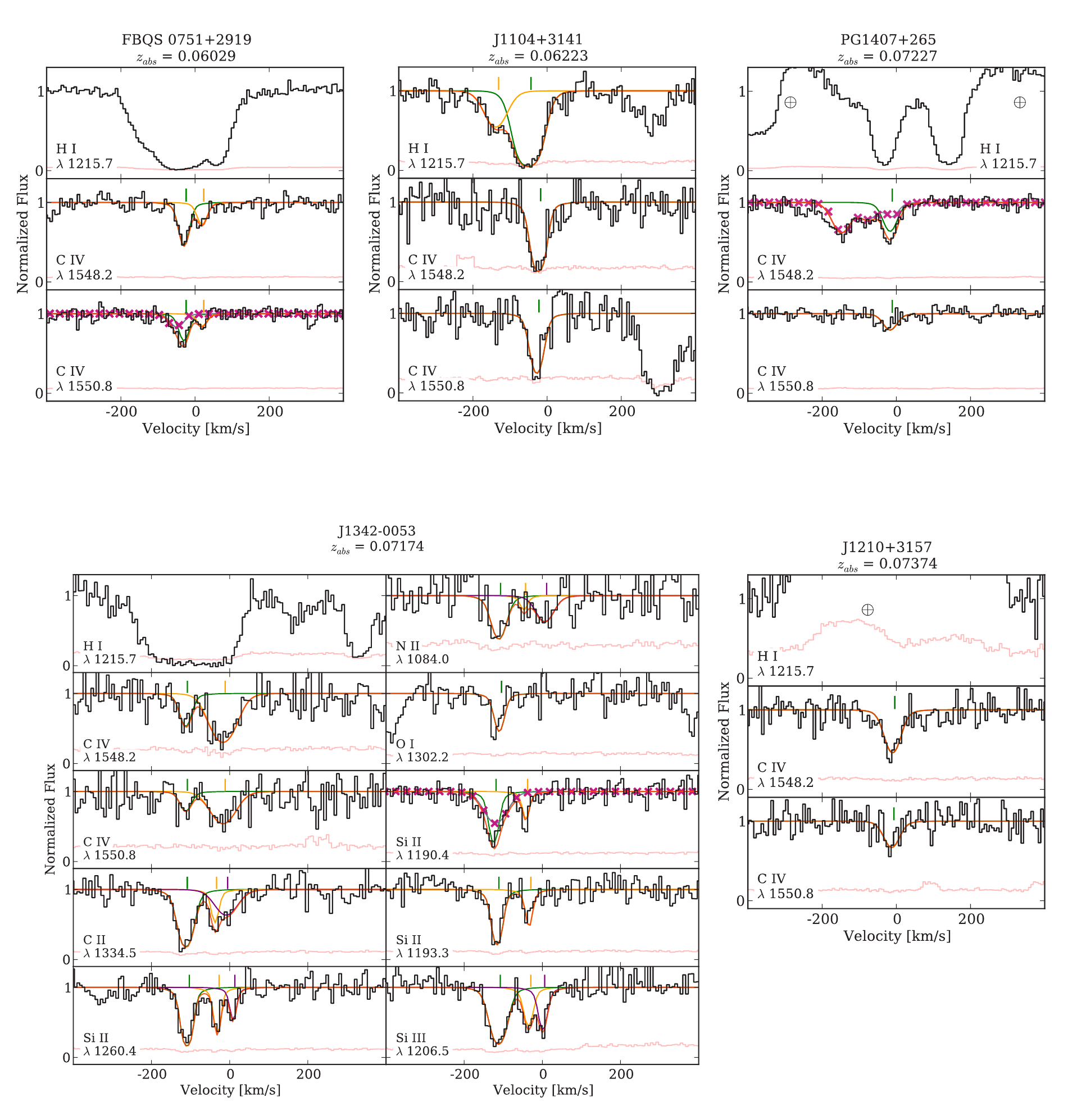} 
\caption{continued.} 
\end{figure*} 

\setcounter{figure}{\value{figure}-1} 
\begin{figure*}[!h] 
\centering 
\includegraphics[width=\textwidth,height=\textheight,keepaspectratio]{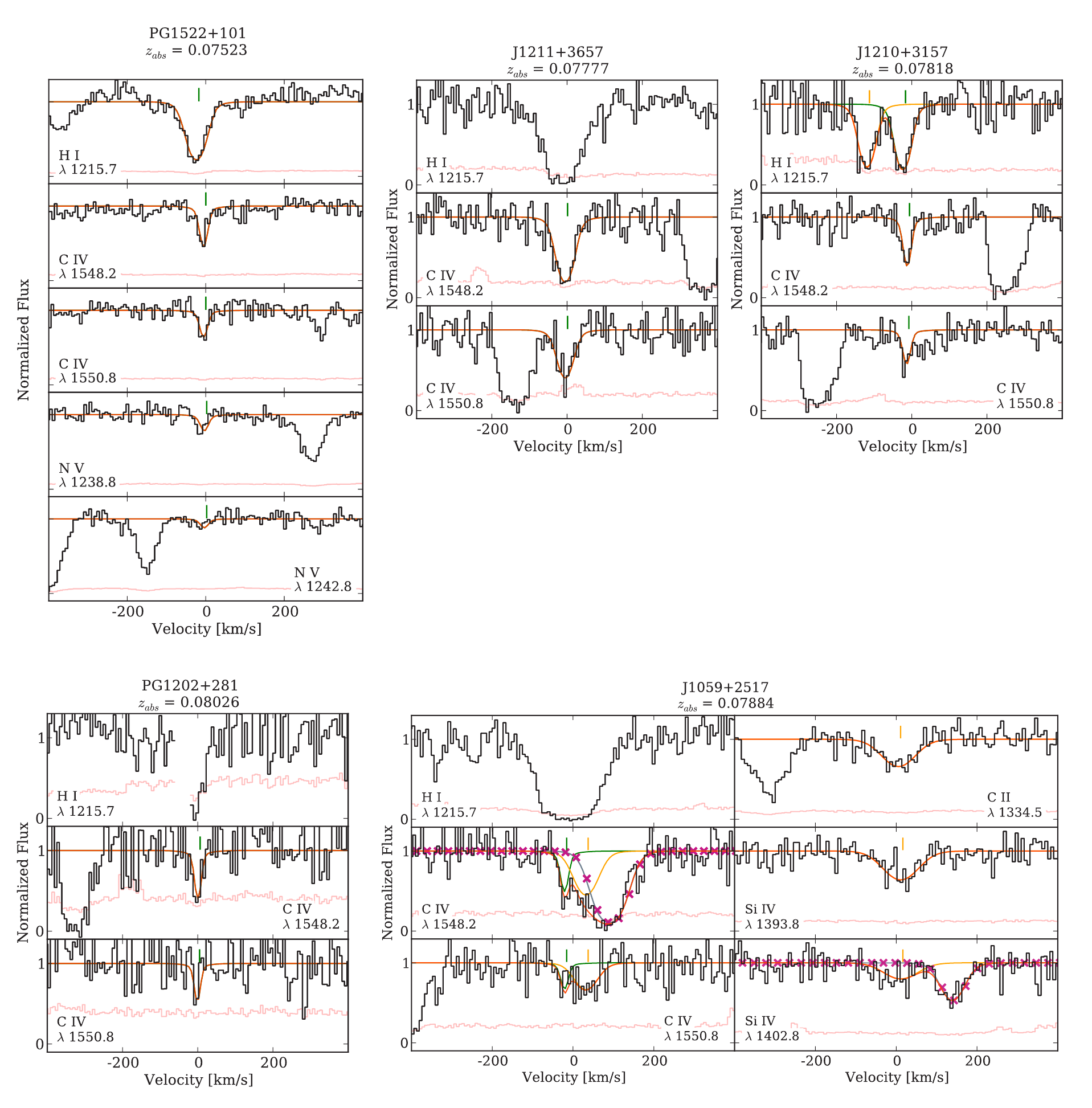} 
\caption{continued.} 
\end{figure*} 

\setcounter{figure}{\value{figure}-1} 
\begin{figure*}[!h] 
\centering 
\includegraphics[width=\textwidth,height=\textheight,keepaspectratio]{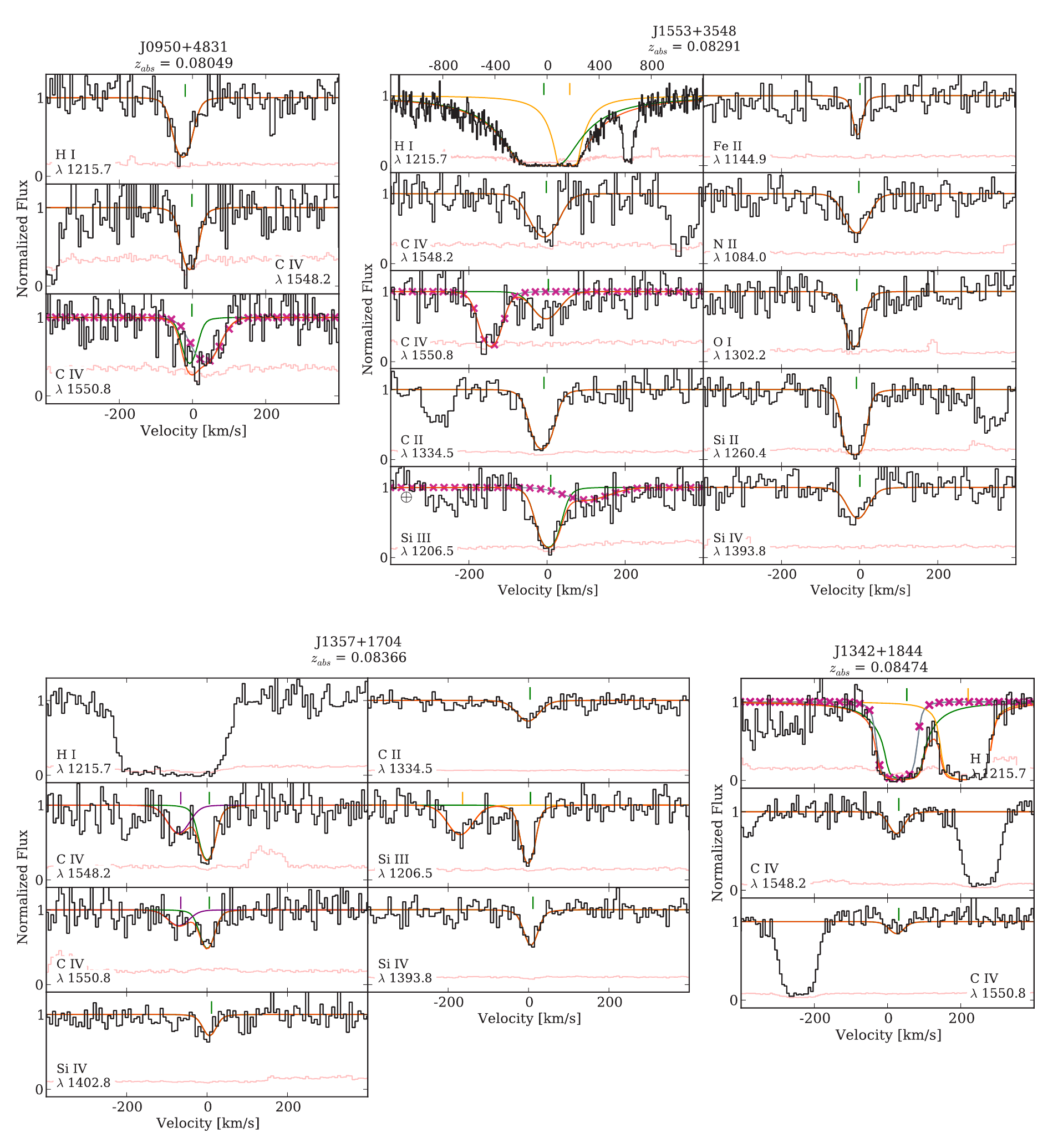} 
\caption{continued.} 
\end{figure*} 

\setcounter{figure}{\value{figure}-1} 
\begin{figure*}[!h] 
\centering 
\includegraphics[width=\textwidth,height=\textheight,keepaspectratio]{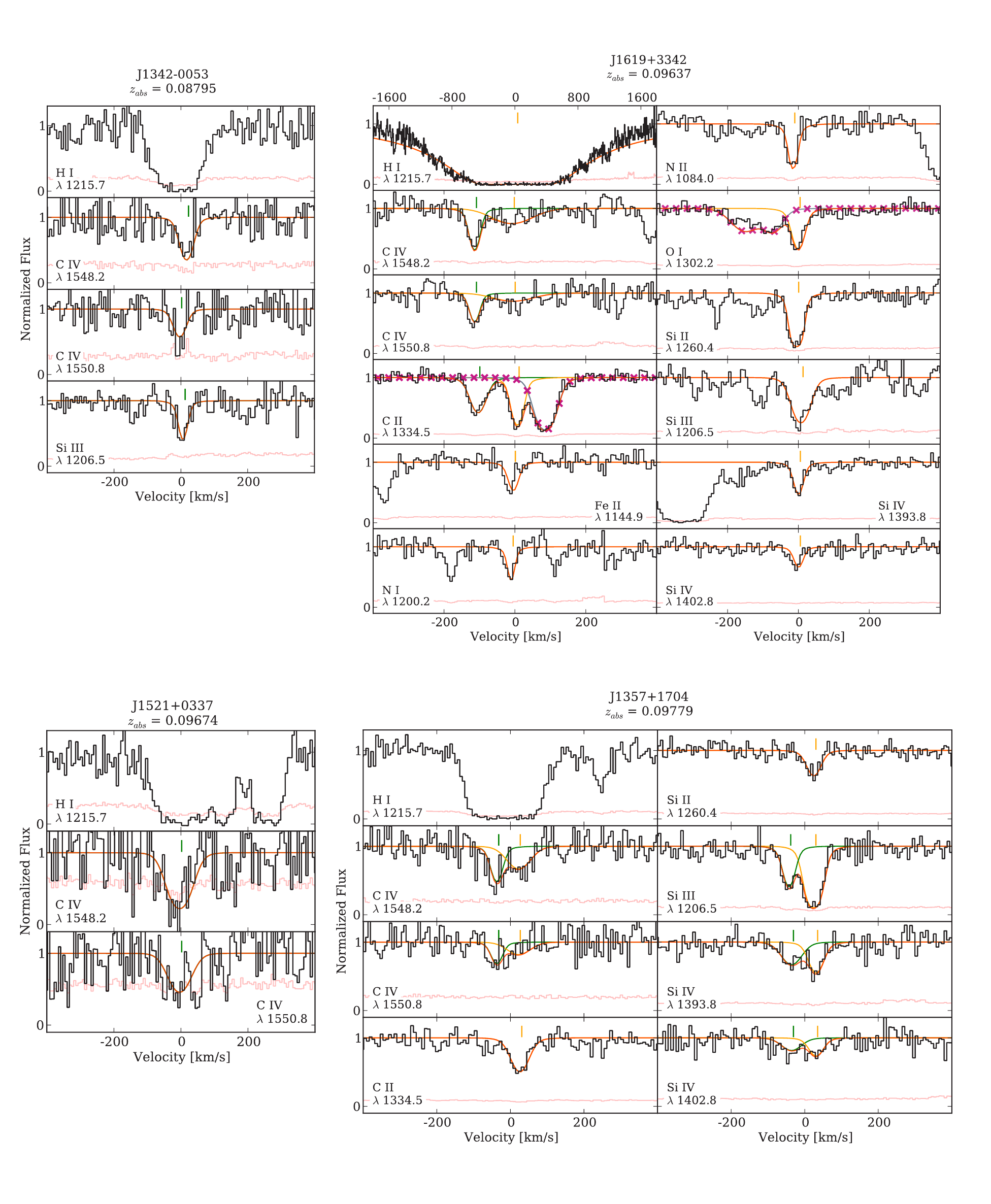} 
\caption{continued.} 
\end{figure*} 

\setcounter{figure}{\value{figure}-1} 
\begin{figure*}[!h] 
\centering 
\includegraphics[width=\textwidth,height=\textheight,keepaspectratio]{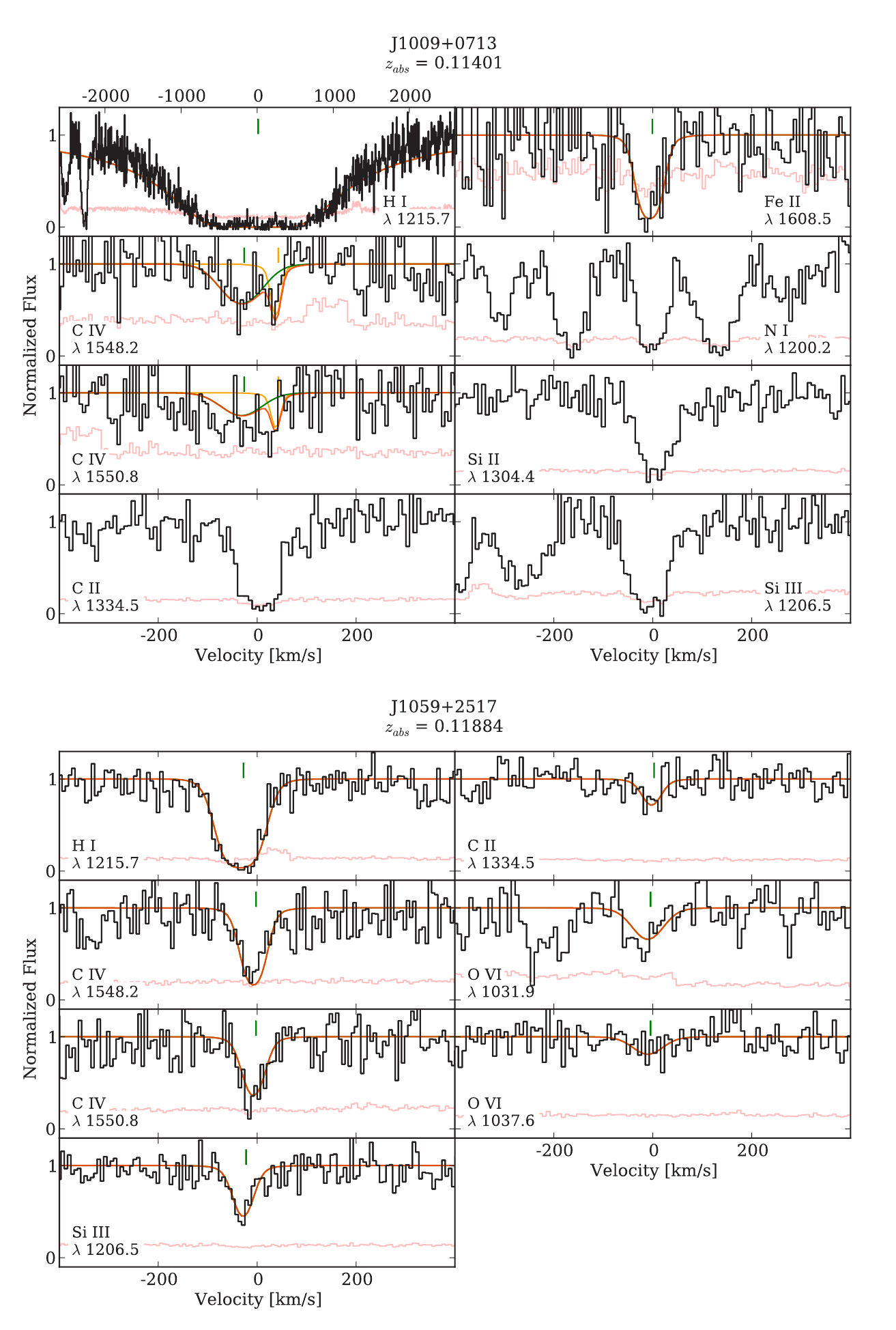} 
\caption{continued.} 
\end{figure*} 

\setcounter{figure}{\value{figure}-1} 
\begin{figure*}[!h] 
\centering 
\includegraphics[width=\textwidth,height=\textheight,keepaspectratio]{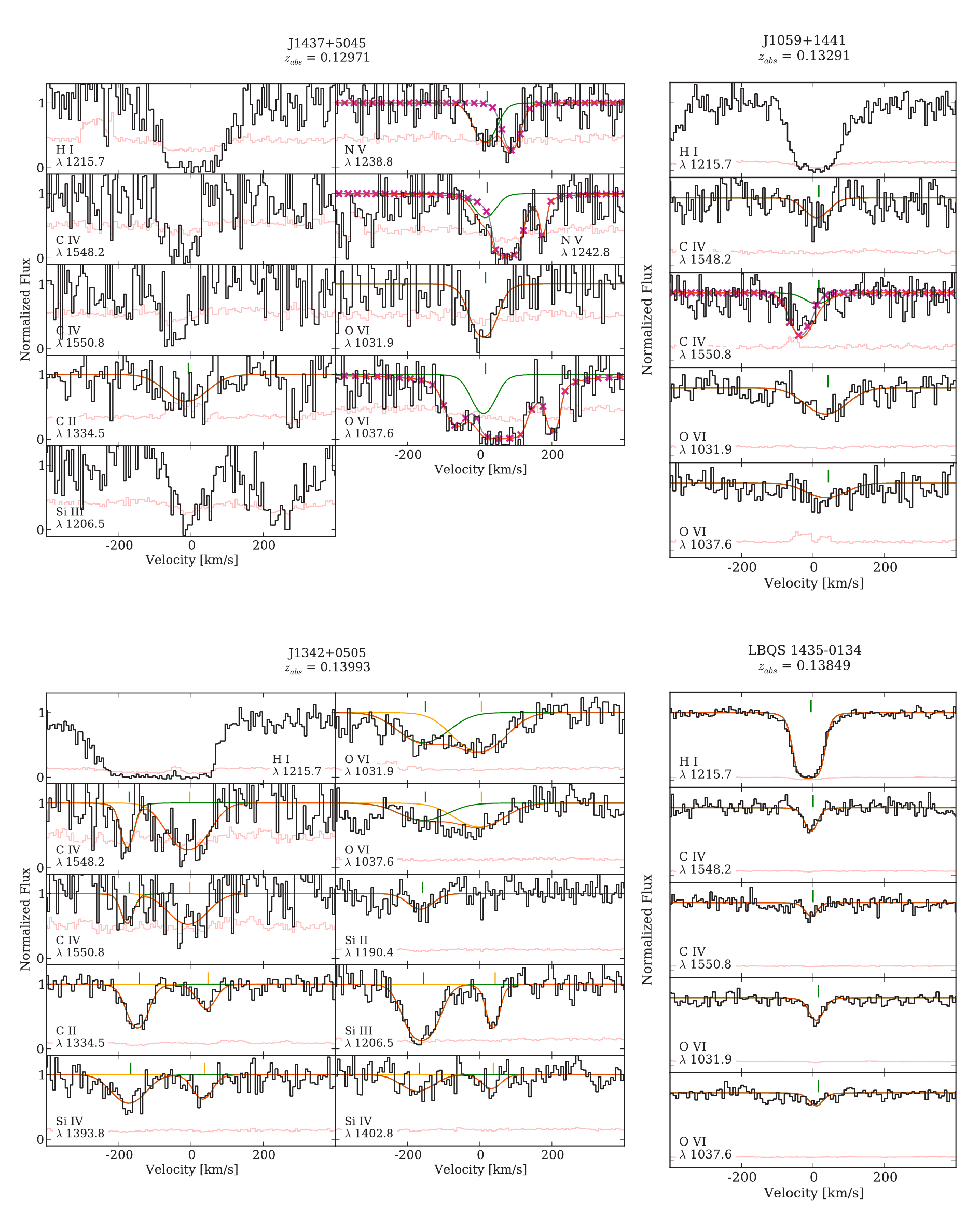} 
\caption{continued.} 
\end{figure*} 

\setcounter{figure}{\value{figure}-1} 
\begin{figure*}[!h] 
\centering 
\includegraphics[width=\textwidth,height=\textheight,keepaspectratio]{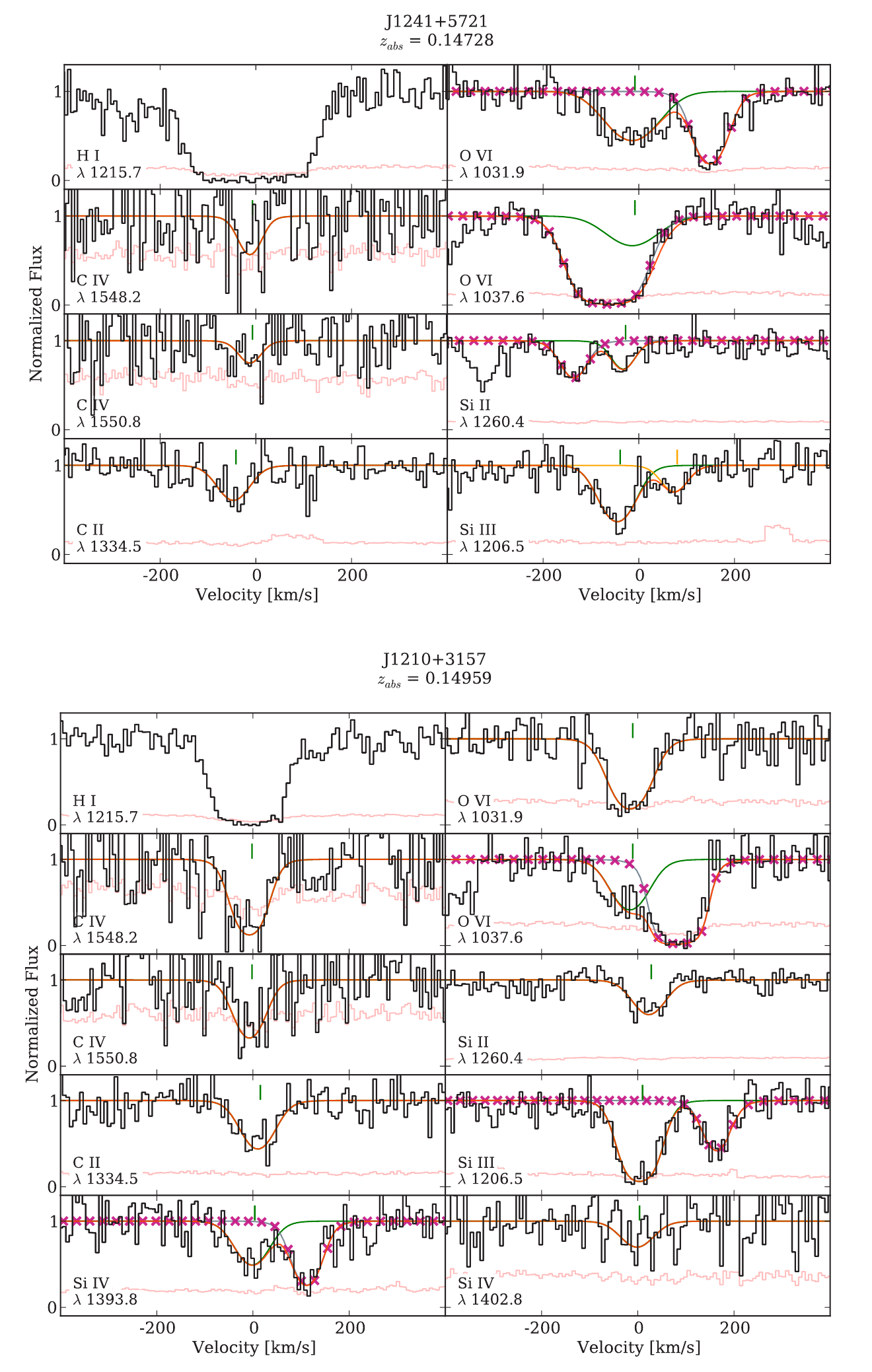} 
\caption{continued.} 
\end{figure*} 

\setcounter{figure}{\value{figure}-1} 
\begin{figure*}[!h] 
\centering 
\includegraphics[width=\textwidth,height=\textheight,keepaspectratio]{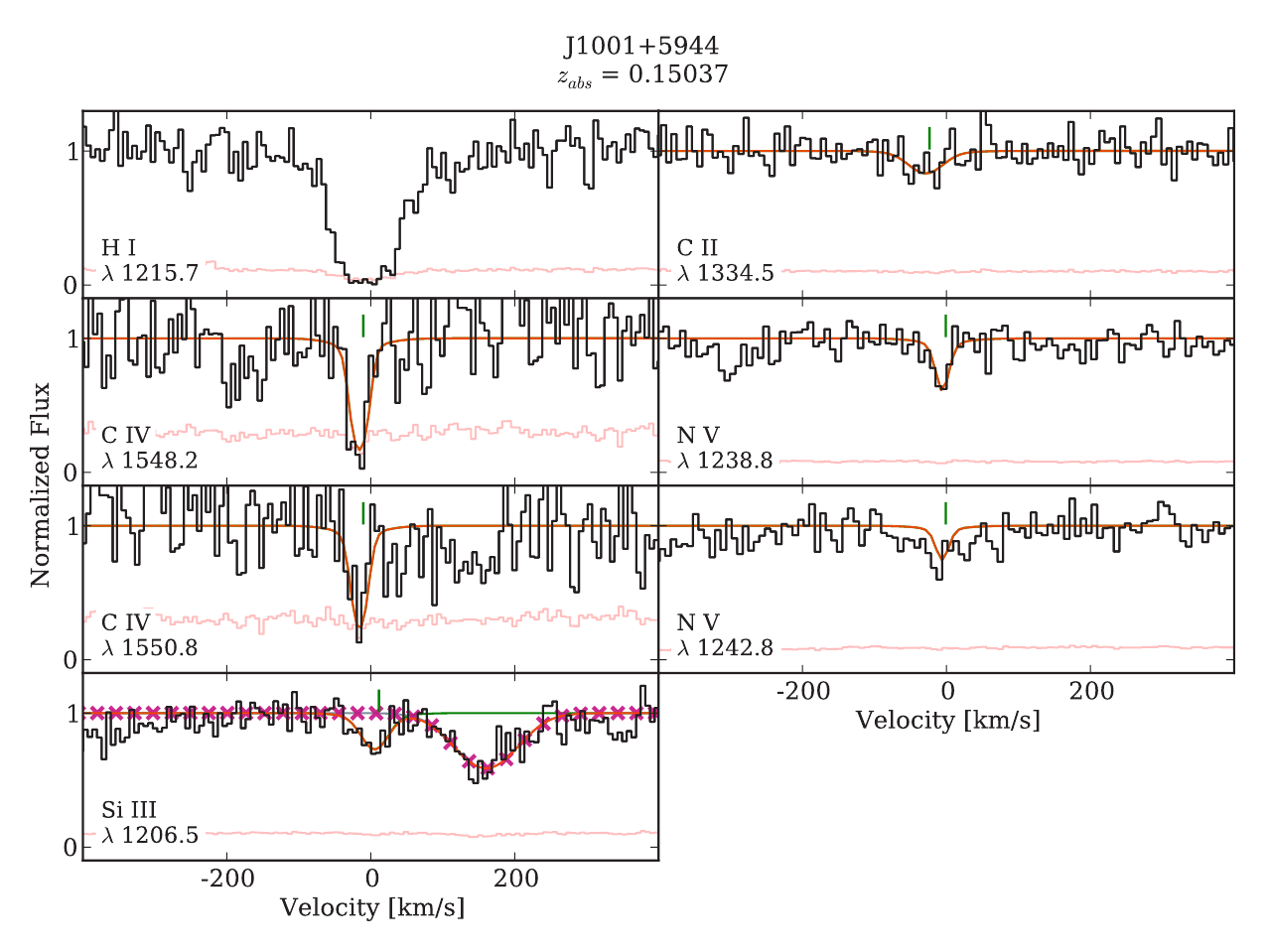} 
\caption{continued.} 
\end{figure*}

\clearpage

\end{document}